\documentclass{amsart}
\usepackage{comment,amsfonts, amsmath, amssymb, graphicx, latexsym,hyperref}

    \def\ds{\displaystyle}
    \def\tr{{\rm tr \,}}
    \def\Re{{\rm Re \,}}
    
    \def\Ai{{\rm Ai \,}}

    \def\Span{{\rm span}}
    \def\bigO{{\mathcal O}}

    \newcommand{\sgn}{{\operatorname{sgn}}}

{     \theoremstyle{plain}
        \newtheorem{theorem}{Theorem}[section]
         \newtheorem{lemma}[theorem]{Lemma} 
          \newtheorem{corollary}[theorem]{Corollary} 
          \newtheorem{proposition}[theorem]{Proposition} 
}
{       \theoremstyle{remark}
        \newtheorem{remark}[theorem]{Remark}
}
\makeatletter
\@addtoreset{equation}{section}
\makeatother

\begin{document}

\title{Universality for orthogonal and symplectic Laguerre-type ensembles}
\author[Deift]{P. Deift}
\address{Courant Institute of Mathematical Sciences, New York University, 251 Mercer St., New York, NY 10012, U.S.A.}
\email{deift@cims.nyu.edu}
\author[Gioev]{D. Gioev}
\address{Department of Mathematics, University of Rochester, Hylan Bldg., 
Rochester, NY 14627, U.S.A.}
\email{gioev@math.rochester.edu}
\author[Kriecherbauer]{T. Kriecherbauer}
\address{Fakult\"at f\"ur Mathematik, Ruhr--Universit\"at Bochum,
Universit\"atsstr. 150, 44780 Bochum, Germany}
\email{thomas.kriecherbauer@ruhr-uni-bochum.de}
\author[Vanlessen]{M. Vanlessen}
\address{Department of Mathematics,
Katholieke Universiteit Leuven, 3030 Leuven (Heverlee), 
Belgium}
\email{maarten.vanlessen@wis.kuleuven.ac.be}

\begin{abstract}
We give a proof of the Universality Conjecture for orthogonal ($\beta=1$) 
and symplectic ($\beta=4$) random matrix ensembles of Laguerre-type in the
bulk of the spectrum as well as at the hard and soft spectral edges. Our results
are stated precisely in the Introduction (Theorems \ref{theorem: universality hard edge}, 
\ref{theorem: universality soft edge}, \ref{theorem: universality bulk} and Corollaries
\ref{corollary:hard}, \ref{corollary:soft}, \ref{corollary:bulk}). They concern
the appropriately rescaled kernels $K_{n, \beta}$, correlation and cluster functions, gap 
probabilities and the distributions of the largest and smallest eigenvalues.
Corresponding results for unitary ($\beta=2$) Laguerre-type ensembles have been
proved by the fourth author
in \cite{v6}. The varying weight case at the hard spectral edge
was analyzed in \cite{KV} for $\beta=2$: In this paper we do
not consider varying weights.

Our proof follows closely the work of the first two authors who showed in 
\cite{DeiftGioev, DeiftGioev2} analogous results for Hermite-type ensembles.
As in \cite{DeiftGioev, DeiftGioev2} we use the version of the orthogonal polynomial method 
presented in \cite{Widom}, \cite{TracyWidom} to analyze the local eigenvalue 
statistics. The necessary asymptotic information on the Laguerre-type
orthogonal polynomials  is taken from \cite{v6}.
\end{abstract}

\maketitle

\section{Introduction}
\label{sec1}
In this paper we consider ensembles
of  matrices $\{M\}$ with invariant distributions
of Laguerre type
\begin{equation}
\label{eq1p1}
    d \mathbb P_{n, \beta} (M) = 
\mathcal{P}_{n,\beta}(M)\,dM 
  = \frac{1}{\mathcal{Z}_{n,\beta}}\, \det (W_\gamma(M)) e^{-\tr Q(M)}\,dM,
\end{equation}
for $\beta=1$, $2$ and $4$, the so-called Orthogonal, Unitary
and Symplectic ensembles, respectively (see \cite{M}).
For $\beta=1$, $2$, $4$, the ensemble consists of $n\times{}n$
real symmetric matrices, $n\times{}n$
Hermitian matrices, and $2n\times{}2n$
Hermitian self-dual matrices (see \cite{M}), respectively. 
The above terminology for $\beta=1$, $2$ and $4$
reflects the fact that (\ref{eq1p1})
is invariant under conjugation of $M$, $M\mapsto{}UMU^{-1}$,
by orthogonal, unitary and unitary-symplectic
matrices $U$.
Furthermore, 
in (\ref{eq1p1}), $dM$ denotes Lebesgue measure on the 
algebraically independent 
entries of $M$, 
$W_\gamma (x) = x^\gamma {\bf 1}_{\mathbb R_+} (x)$ with $\gamma > 0$,
$Q$ denotes any polynomial of positive degree and with positive
leading coefficient,
and $\mathcal{Z}_{n,\beta}$ is a normalization constant. Of course,
$\mathcal{P}_{n,\beta}$ and $\mathcal{Z}_{n,\beta}$ depend
not only on $n$ and $\beta$ which are
implicit in (\ref{eq1p1}) but also on the quantities $\gamma$ and
$Q$. For the sake of readability the dependence on $\gamma$ and $Q$
is suppressed in all of our notation. 

For ensembles (\ref{eq1p1}) the joint probability density 
function for the eigenvalues 
$x_1, x_2, \ldots, x_n$ of $M$ is given by (see \cite{M}) 
\begin{equation}
\label{eq_PN}
     P_{n,\beta}(x_1,\ldots,x_n) =  \frac{1}{{Z}_{n,\beta}}
         \prod_{1\leq{}j<k\leq{}n}
        |x_j-x_k|^\beta\prod_{j=1}^n w_\beta(x_j) \quad \textrm{on} \,\, 
        {\mathbb{R}}_+^n
\end{equation}
where again $Z_{n,\beta}$ denotes the corresponding normalization constant
and
\begin{equation}
\label{eq1p3}
   w_\beta(x)=\begin{cases}
                  x^\gamma e^{-Q(x)},&\beta=1,2\cr
                  \left( x^\gamma e^{-Q(x)} \right)^2,&\beta=4.
               \end{cases}
\end{equation}
The second power appearing in $w_{\beta=4}$ simply reflects the fact 
that the eigenvalues of self-dual Hermitian matrices come in pairs.

Our main results stated below
show that the appropriately 
rescaled local eigenvalue statistics for ensembles (\ref{eq1p1})
are universal (i.e. independent of $Q$) in the limit $n \to \infty$,
where for $\beta=1$ only even values for $n$ are 
considered\footnote{For $\beta=1$, $n$ odd, see the discussion
following equation (1.13) in \cite{DeiftGioev}.}.
Consequently, the limiting local eigenvalue statistics agree for all
ensembles (\ref{eq1p1}) with the corresponding limiting statistics
in the well studied classical cases of linear $Q$ (see e.g. 
\cite{NW, Fo, TW, NF, Forr}
and references therein). Ensembles (\ref{eq1p1}) with linear $Q$
are called Laguerre ensembles because $w_\beta$ in (\ref{eq1p3}) is then a
Laguerre weight. More generally, all matrix ensembles
with eigenvalue probability density 
function of the form (\ref{eq_PN}), (\ref{eq1p3}) and with linear
$Q$ are called Laguerre ensembles irrespective of whether they
arise from matrix ensembles
of the form (\ref{eq1p1}). In fact, Laguerre ensembles appeared first
in statistics and in physics and these were not of type (\ref{eq1p1}).
In statistics, for example, Wishart ensembles $\{ M \}$ with $M = X^t X$
and $X$ being a random $N \times n \, (N \geq n)$ rectangular matrix with
real entries that are independently distributed standard Gaussian variables,
have an eigenvalue probability density 
function of the form (\ref{eq_PN}), (\ref{eq1p3}) with
$\beta = 1$, $\gamma = (N-n-1)/2$ and $Q(x) = x/2$ (see e.g. \cite{Muirhead}).
In physics, Laguerre ensembles emerge e.g. in the study
of Dirac operators in quantum chromodynamics and in
the study of disordered superconductors in mesoscopic physics,
see e.g. \cite{Altland,Verbaarschot}.
Here we encounter not only Wishart ensembles but also random
matrices with a $2 \times 2$ block structure which lead again
to an eigenvalue probability density 
function of the form (\ref{eq_PN}), (\ref{eq1p3}). For example,
random Dirac operators in the chiral gauge are modelled by 
$\begin{pmatrix} 0&X\\X^t&0 \end{pmatrix}$ where
$X$ is a rectangular  $N \times n$ random matrix. Choosing again
the entries of $X$ to be independently distributed real standard Gaussian 
variables one obtains a density function for (the squares of) the 
eigenvalues which is of the form (\ref{eq_PN}), (\ref{eq1p3}) with
$\beta = 1$, $\gamma = (N-n-1)/2$ and $Q(x) = x/2$.

In \cite{DeiftGioev,DeiftGioev2} the authors proved 
universality in the bulk \cite{DeiftGioev} and at the spectral
edge \cite{DeiftGioev2} for {\em Hermite-type ensembles,\ }i.e. for
ensembles (\ref{eq1p1}) with $W_\gamma (x) = 1$ for all $x \in
\mathbb R$ and with $Q(x)$ denoting any polynomial of even positive
degree and with positive leading coefficient. 
To the best of our knowledge, universality results for Laguerre-type 
ensembles have so far only been proved for unitary ($\beta = 2$)
ensembles in \cite{KV} (varying weights) and in
\cite{v6} where
the author showed universality for unitary 
ensembles of the form (\ref{eq1p1}). All the results
regarding $\beta = 2$ stated in the present paper can be found already
in \cite{v6} and we only include them here for the sake of 
completeness. Moreover, a number of formulae and estimates proved in 
\cite{v6} play a key role in our proof of universality for 
$\beta = 1$, $4$. 
Universality for Laguerre-type ensembles, for all three
cases $\beta=1$, $2$ and $4$, has been considered in the 
physics literature  (see e.g. \cite{ADMN, SV} and 
references therein).
More information on the history of universality
for matrix ensembles can be found the introductions 
of \cite{DeiftGioev, DeiftGioev2} and in \cite{Deift}. 

The basic structure of the proof in this paper is similar
to \cite{DeiftGioev,DeiftGioev2} and relies on the 
orthogonal polynomial method developed in 
\cite{TracyWidom} and 
\cite{Widom}. A detailed 
description of the strategy of proof can be found in the 
Introductions of \cite{DeiftGioev} and \cite{DeiftGioev2}.
We now introduce some further notation
that is needed to state our main results.

Following \cite{Widom},
\cite[Remark 1.3]{DeiftGioev} 
we define 
weights of the form
\begin{equation} \label{def:w1}
    w(x)=x^\alpha e^{-V(x)},\qquad\ \mbox{for $x \in \mathbb R_+$,}
\end{equation}
with
\begin{equation} \label{def:w2}
    \alpha := \begin{cases}
                  \gamma,&\beta=2\cr
                  2\gamma,&\beta=1, 4
               \end{cases} \ ; \qquad
    V := \begin{cases}
                  Q,&\beta=2\cr
                  2 Q,&\beta=1, 4
               \end{cases}
\end{equation}
($\gamma$, $Q$ as in (\ref{eq1p1})) in order to be able to use the same
set of orthogonal polynomials in all three cases $\beta = 1$, $2$, $4$.
By the assumptions made on $\gamma$ and $Q$ we will assume that
\begin{equation} \label{def:V}
\qquad  \alpha > 0 \qquad  \qquad \textrm{and} \qquad \qquad 
    V(x) = \sum_{j=0}^m q_j x^j 
\end{equation}
where the polynomial $V$,
known as the external field,
 has positive degree $m$ and positive leading 
coefficient $q_m$. The orthogonal polynomials $p_k$ with respect to the
weight $w$ are uniquely defined by the conditions
\[
    \int_0^\infty p_k(x)p_l(x)w(x)dx=\delta_{k,l}\qquad\mbox{for $k,l\in \mathbb
    N_0$,}
\]
and $p_k(x)=\gamma_kx^k+\ldots$ is a polynomial of degree $k$ with positive
leading coefficient $\gamma_k>0$. The functions
\begin{equation} \label{def:phi_k}
    \phi_k(x):= p_k(x) \sqrt{w(x)}
\end{equation}
then form an orthonormal system in $L_2(\mathbb R_+)$. The statement of
our main results involves several quantities that arise in
the asymptotic analysis of
the orthogonal polynomials $p_k$, viz., the
Mhaskar--Rakhmanov--Saff numbers $\beta_n$, 
the densities $\omega_n$ of the
equilibrium measures in the
presence of the rescaled external field $V_n(x)=\frac{1}{n}V(\beta_n x)$, and
numbers $c_n$, $\tilde c_n$ related to the behavior of the equilibrium
measure at the soft, hard edges respectively.
 The definition and relevant
properties of all these quantities are summarized in 
equations (\ref{defrel: MRS})--(\ref{definition: cntilde auxiliary results}) 
of Section \ref{subsection: relevant result from v6} below where one
can also find references to \cite{v6} for their respective derivations.

As mentioned above our proof relies on the
orthogonal polynomial method for invariant matrix ensembles. 
This method is based
on the observation that the eigenvalue statistics (e.g. correlation and 
cluster functions, gap probabilities, distributions of smallest and largest eigenvalues) 
can be analyzed using functions
$K_{n, \beta}$ of two variables which can be expressed 
in terms of the orthogonal
polynomials $p_k$ (see \cite{TracyWidom}). More
precisely, let $\varepsilon$ 
denote the integral operator with kernel $\varepsilon(x,y)=\frac{1}{2}\sgn(x-y)$
where $\sgn = {\bf 1}_{\mathbb R_+} - {\bf 1}_{\mathbb R_-}$ 
is the standard sign-function. We then define 
\begin{align}
    \label{definition: Kn2}
    & K_{n,2}(x,y) := K_n(x, y) := \sum_{k=0}^{n-1}\phi_k(x)\phi_k(y) 
       \quad \mbox{(Christoffel--Darboux kernel)}\\[2ex]
    \label{definition: Kn1}
    & K_{n,1}(x,y)=
    \begin{pmatrix}
        S_{n,1}(x,y) & -\frac{\partial}{\partial y}S_{n,1}(x,y) \\[1ex]
        (\varepsilon S_{n,1})(x,y)-\frac{1}{2}\sgn(x-y) &
        S_{n,1}(y,x)
    \end{pmatrix}, 
\qquad \mbox{for $n$ even\footnotemark[\value{footnote}],}\\[2ex]
    \label{definition: Kn4}
    & K_{n,4}(x,y)=
    \frac{1}{2}\begin{pmatrix}
        S_{n,4}(x,y) & -\frac{\partial}{\partial y}S_{n,4}(x,y) \\[1ex]
        (\varepsilon S_{n,4})(x,y) & S_{n,4}(y,x)
    \end{pmatrix}.
\end{align}
Here $S_{n,\beta}$ $(\beta=1,4)$ are certain
specific scalar functions which will be
discussed in detail in Section~\ref{sec2}. 
The analysis in the 
present paper depends critically
on the formulae of Widom \cite[Theorem 2]{Widom} that express the
functions $S_{n, \beta}$
in terms of the orthogonal polynomials $p_k$.

We will prove the convergence of 
$K_{n, \beta}$ for $n \to  \infty$ to a universal limit that is independent of
$V$. In proving the convergence
one needs to rescale the arguments $x$ and $y$
appropriately. Since the (1,2)-entry of $K_{n, \beta}$
for $\beta=1$, $4$  
contains differentiation with respect to $y$,
and the (2,1)-entry of $K_{n, \beta}$ 
contains integration with respect to $x$,
these two entries behave differently
under rescaling.
In order to take this into account
it is convenient to introduce the following notation for $\beta =1$, $4$:
\begin{equation}\label{definition: conjugate Kn}
    K_{n,\beta}^{(\lambda)}=
    \begin{pmatrix}
        \lambda^{-1} & 0 \\[1ex]
        0 & \lambda
    \end{pmatrix} K_{n,\beta}
    \begin{pmatrix}
        \lambda & 0 \\[1ex]
        0 & \lambda^{-1}
    \end{pmatrix}=
    \begin{pmatrix}
        (K_{n,\beta})_{11} & \lambda^{-2}(K_{n,\beta})_{12} \\[1ex]
        \lambda^2 (K_{n,\beta})_{21} & (K_{n,\beta})_{22}
    \end{pmatrix},\qquad \lambda>0.
\end{equation}

We now are ready to state our main results. 
Since the statistical behavior
is different for eigenvalues in the bulk of the spectrum and 
at the spectral edges, we need to distinguish these cases.
Moreover, for Laguerre-type ensembles the lower and upper spectral
edges have a different character. The lower edge at the origin is called a
{\em hard edge}, because no eigenvalue can be less than zero by 
definition of the ensemble. For the upper edge, on the other hand,
there is no apriori upper bound for the eigenvalues. The
existence of the upper spectral edge is due to the fact that the 
probability for an eigenvalue to be bigger than a certain $n$-dependent 
threshold value is exponentially small: This threshold value
is known as the {\em soft edge} of the spectrum. 
Both the rescaling and 
the limit of $K_{n, \beta}$ are different for the bulk, for the soft edge and 
for the hard edge. In \cite{DeiftGioev, DeiftGioev2} the authors
proved universality for Hermite-type ensembles in the bulk and at the soft edge,
respectively. We state the analogous results for Laguerre-type
ensembles in Theorems \ref{theorem: universality bulk}, 
\ref{theorem: universality soft edge} below. 
Note that another manifestation of universality
is seen in the fact 
that the limits of the appropriately rescaled $K_{n, \beta}$ 
are the same for Hermite-type and Laguerre-type ensembles both 
in the bulk and at the soft edge. 

We start by stating our results for the {\em hard edge}, a case
which is not 
present in Hermite-type ensembles \cite{DeiftGioev, DeiftGioev2}. 

{\bf Notational remark.\ }In Theorem \ref{theorem: universality hard edge}
and also in other situations where we consider the hard edge, we will use the
notation that an estimate holds
{\em uniformly for $\xi,\eta$ in bounded subsets of $(0,\infty)$.\ }By
this we mean that the estimate holds for $\xi,\eta$ in any
set of the form $(0,L)$, $0<L<\infty$. By {\em uniformly\ }we
mean that the constant in the $\bigO$-term in \eqref{1.13} below,
for example, depends only on $L$.
This somewhat unusual notation is necessitated by the actual
form of the error estimates for the correlation kernel near $0$,
see e.g.~\eqref{1.13} and the proof of Corollary \ref{corollary:hard}(b)
in Subsection \ref{sec6.1} below.

\begin{theorem}\label{theorem: universality hard edge} {\bf(hard edge).} \ \
    Let $\beta=1,2$ or $4$ and introduce the notation
    \[
        \nu_n=\left(\frac{\beta_n}{4\tilde c_n n^2}\right)^{-1/2},\qquad
        \tilde x^{(n)}=\frac{1}{\nu_n^2}x=\frac{\beta_n}{4\tilde c_n n^2}x.
    \]
    Then, as $n\to\infty$ ($n$ even for the cases $\beta=1,4$) the following 
    holds uniformly for $\xi,\eta$ in bounded subsets of $(0,\infty)$.

    \medskip

    \noindent (i) The case $\beta=2$:
    \begin{equation}
        \frac{1}{\nu_n^2} K_n(\tilde\xi^{(n)} , \tilde\eta^{(n)})=K_J(\xi,\eta)+
            \bigO\left(\frac{\xi^{\frac{\alpha}{2}}\eta^{\frac{\alpha}{2}}}{n}\right).
    \end{equation}
    where $K_J$ denotes the Bessel kernel,
    \[
        K_J(\xi,\eta)=\frac{J_\alpha(\sqrt\xi)\sqrt\eta J_\alpha'(\sqrt\eta)
        -
        J_\alpha(\sqrt\eta)\sqrt\xi J_\alpha'(\sqrt\xi)}{2(\xi-\eta)}.
    \]

    \medskip

    \noindent (ii) The case $\beta=4$:
    \begin{equation}\label{1.13}
        \frac{1}{\nu_n^2}K_{\frac{n}{2},4}^{(\nu_n)}(\tilde\xi^{(n)},\tilde\eta^{(n)})=
            K^{(4)}(\xi,\eta)
        +\bigO\left(\frac{\xi^{\frac{\alpha}{2}}\eta^{\frac{\alpha}{2}}}{n}\right)
        \begin{pmatrix}
            \xi^{-1} & \xi^{-1}\eta^{-1} \\[1ex]
            1 & \eta^{-1}
        \end{pmatrix},
    \end{equation}
    where
    \begin{align*}
        2(K^{(4)})_{11}(\xi,\eta) &=
            2(K^{(4)})_{22}(\eta,\xi) \\[2ex]
            & = K_J(\xi,\eta)+\frac{1}{4}
            \left(\frac{J_{\alpha+1}(\sqrt\xi)}{\sqrt\xi}-\frac{2\alpha}{\xi}J_\alpha(\sqrt\xi)\right)
            \int_0^{\sqrt\eta}J_{\alpha+1}(s)ds, \\[3ex]
        2(K^{(4)})_{12}(\xi,\eta) &= -\frac{\partial}{\partial \eta}K_J(\xi,\eta) -\,
            \frac{1}{8}\left(\frac{J_{\alpha+1}(\sqrt\xi)}{\sqrt\xi}-\frac{2\alpha}{\xi}J_\alpha(\sqrt\xi)\right)
            \frac{J_{\alpha+1}(\sqrt\eta)}{\sqrt\eta}, \\[3ex]
        2(K^{(4)})_{21}(\xi,\eta) &=\int_0^\xi K_J(s,\eta)ds+\frac{1}{2}
            \int_0^{\sqrt\xi}\left(J_{\alpha+1}(s)-\frac{2\alpha}{s}J_\alpha(s)\right)ds
            \int_0^{\sqrt\eta}J_{\alpha+1}(s)ds.
    \end{align*}

    \medskip

    \noindent (iii) The case $\beta=1$: there exists $0<\tau=\tau(m,\alpha)<1$ such that
    \begin{equation}
        \frac{1}{\nu_n^2}K_{n,1}^{(\nu_n)}(\tilde\xi^{(n)} , \tilde\eta^{(n)})=
            K^{(1)}(\xi,\eta)
        +\bigO(n^{-\tau})\begin{pmatrix}
            \xi^{\frac{\alpha}{2}} &
            \xi^{\frac{\alpha}{2}}\eta^{\frac{\alpha}{2}-1} \\[1ex]
            1 & \eta^{\frac{\alpha}{2}}
        \end{pmatrix},
    \end{equation}
    where
    \begin{align*}
        (K^{(1)})_{11}(\xi,\eta) &=
            (K^{(1)})_{22}(\eta,\xi) \\[2ex]
            & =K_J(\xi,\eta)-\frac{1}{4}\frac{J_{\alpha+1}(\sqrt\xi)}{\sqrt\xi}
            \int_{\sqrt\eta}^\infty\left(J_{\alpha+1}(s)-\frac{2\alpha}{s}J_\alpha(s)\right)ds,
        \\[3ex]
         (K^{(1)})_{12}(\xi,\eta)&= -\frac{\partial}{\partial \eta}K_J(\xi,\eta) -\,
            \frac{1}{8}\frac{J_{\alpha+1}(\sqrt\xi)}{\sqrt\xi}
            \left(\frac{J_{\alpha+1}(\sqrt\eta)}{\sqrt\eta}-\frac{2\alpha}{\eta}J_\alpha(\sqrt\eta)\right),
            \\[3ex]
         (K^{(1)})_{21}(\xi,\eta)
        &= -\int_\xi^\eta
        K_J(s,\eta)ds+\frac{1}{2}\int_{\sqrt\xi}^{\sqrt\eta}J_{\alpha+1}(s)ds
        \int_{\sqrt\eta}^\infty\left(J_{\alpha+1}(s)-\frac{2\alpha}{s}J_\alpha(s)\right)ds\\[1ex]
        &\phantom{=\,} -\frac{1}{2}\sgn(\xi-\eta).
    \end{align*}
\end{theorem}

As in \cite{DeiftGioev, DeiftGioev2} we now present two consequences of Theorem
\ref{theorem: universality hard edge} which demonstrate the relevance of the theorem
for the understanding of the local eigenvalue statistics in the limit $n \to \infty$.
Here we consider the distribution of the lowest eigenvalue as well as the $l$-point
correlation functions. The latter are obtained from the probability density function 
$P_{n, \beta}$ essentially by integrating out the last $n-l$ variables,
\begin{equation} \label{def:correlation}
R_{n, \beta, l} (x_1, \ldots x_l) := \binom{n}{n-l} \int_{\mathbb R^{n-l}} P_{n, \beta}(x_1, \ldots, x_n)\, dx_{l+1}
\ldots dx_n.
\end{equation}
\begin{corollary} \label{corollary:hard}
With the notation of Theorem \ref{theorem: universality hard edge} and (\ref{def:correlation}) and
$\lambda_1(M)$ denoting the smallest eigenvalue of $M$ we have for
$l \in \mathbb N$, $\xi$, $\xi_i \in (0,\infty)$ that the following limits
\begin{itemize}
\item[(a)] $\displaystyle \lim_{n \to \infty} \frac{1}{\nu_n^{2l}} R_{n, \beta, l}( \frac{\xi_1}{\nu_n^{2}}, 
\ldots, \frac{\xi_l}{\nu_n^{2}})$
for $\beta=1$, $2$; \quad
$\displaystyle \lim_{n \to \infty}  \frac{1}{\nu_n^{2l}}
R_{n/2, 4, l}(\frac{\xi_1}{\nu_n^{2}}, \ldots, \frac{\xi_l}{\nu_n^{2}})$,
\item[(b)]
$\displaystyle \lim_{n \to \infty} \mathbb P_{n, \beta} (\{M \colon \lambda_1(M) \le 
\frac{\xi}{\nu_n^{2}} \})$ for $\beta=1$, $2$; \quad
$\displaystyle \lim_{n \to \infty} \mathbb P_{\frac{n}{2}, 4} (\{M \colon \lambda_1(M) \le 
\frac{\xi}{\nu_n^{2}} \})$
\end{itemize}
exist (with $n$ even for $\beta =1$, $4$) and are independent of $Q$ (cf. (\ref{eq1p1})). 
\end{corollary}
Existence and universality of the limits appearing in statement (a) of the 
Corollary follow from the 
convergence of the cluster functions and the relation between cluster and correlation functions 
(see \cite[Section 2]{TracyWidom}). The convergence of the cluster functions is immediate from Theorem
\ref{theorem: universality hard edge} together with the formulae in \cite[Section 3]{TracyWidom}
which express the cluster functions in terms of the kernels $K_{n, \beta}$. For $\beta = 1$, $4$ one needs to
observe in addition that the formulae do not change if one replaces $K_{n, \beta}$ by
$K_{n,\beta}^{(\lambda)}$. The proof of existence and universality of the limits
in statement (b) of the corollary is slightly more involved and will be presented at the end of 
Subsection \ref{sec6.1}.
\begin{remark}
It is also possible to give explicit formulae for the limits considered in Corollary \ref{corollary:hard}
in terms of the kernels $K_J$, $K^{(1)}$ and $K^{(4)}$ for $\beta =2$, $1$, $4$ respectively. These
limits are easy to derive for the correlation functions (a), using the
determinantal formula for $\beta = 2$ and using the relation with cluster functions for $\beta=1$, $4$.

In contrast, the dependence of the limiting distribution of the 
smallest eigenvalue (b) on the limiting kernels
$K_J$, $K^{(1)}$ and $K^{(4)}$ is given via Fredholm determinants 
(cf. (\ref{t6.1}), (\ref{t6.10}), (\ref{t6.15})) and therefore 
is far more complicated. However, 
our universality result stated in Corollary \ref{corollary:hard} implies that
it suffices to understand the limiting distribution in the classical Laguerre case where the polynomial
$Q$ in (\ref{eq1p1}) has degree 1. Fortunately, this case 
has already been studied in the literature
and it was found that the limiting distributions of the smallest eigenvalue can be expressed in terms of
certain Painlev\'e functions (see \cite{TW} for $\beta =2$ and \cite{Forr} for $\beta =1$, $4$).
\end{remark}

Next we state our main result for the upper spectral edge.

\begin{theorem} {\bf(soft edge)} 
    \label{theorem: universality soft edge}
    {\rm (cf.~\cite[Theorem 1.1]{DeiftGioev2})}. \ \
    Let $\beta=1,2$ or $4$ and introduce the notation
    \[
        \lambda_n=\left(\frac{\beta_n}{c_n n^{2/3}}\right)^{-1/2},\qquad
        x^{(n)}=\beta_n + \frac{x}{\lambda_n^2}=\beta_n \left( 1 + \frac{x}{c_n n^{2/3}} \right).
    \]
    Fix a number $L_0$. Then, there exists $c=c(L_0)$ and $0<\tau=\tau(m,\alpha)<1$ such that as $n\to\infty$
    ($n$ even for the cases $\beta=1,4$) the following holds uniformly for $\xi,\eta\in[L_0,+\infty)$.

    \medskip

    \noindent (i) The case $\beta=2$:
    \begin{equation}
        \frac{1}{\lambda_n^2} K_n(\xi^{(n)} , \eta^{(n)})=K_\Ai(\xi,\eta)+
            \bigO\left(n^{-1/3}\right)e^{-c\xi}e^{-c\eta}.
    \end{equation}
    where $K_\Ai$ denotes the Airy kernel,
    \[
        K_\Ai(\xi,\eta)=\frac{\Ai(\xi)\Ai'(\eta)-\Ai(\eta)\Ai'(\xi)}{\xi-\eta}.
    \]

    \medskip

    \noindent (ii) The case $\beta=4$:
    \begin{equation}
        \frac{1}{\lambda_n^2}K_{\frac{n}{2},4}^{(\lambda_n)}(\xi^{(n)},\eta^{(n)})=
            K^{(4)}(\xi,\eta)
        +\bigO\left(\frac{e^{-c\xi} e^{-c\eta}}{n^{\tau}}\right)
         \begin{pmatrix}
            1& 1 \\[1ex]
            1 & 1
        \end{pmatrix} ,
    \end{equation}
    where
    \begin{align*}
        2(K^{(4)})_{11}(\xi,\eta) &=
            2(K^{(4)})_{22}(\eta,\xi) = K_\Ai(\xi,\eta)-\frac{1}{2}\Ai(\xi)\int_\eta^\infty\Ai(s)ds, \\[1ex]
        2(K^{(4)})_{12}(\xi,\eta) &= -\frac{\partial}{\partial \eta}K_\Ai(\xi,\eta) -
            \frac{1}{2}\Ai(\xi)\Ai(\eta), \\[1ex]
        2(K^{(4)})_{21}(\xi,\eta) &=-\int_\xi^\infty K_\Ai(s,\eta)ds+\frac{1}{2}
            \int_\xi^\infty\Ai(s)ds\int_\eta^\infty\Ai(s)ds.
    \end{align*}

    \medskip

    \noindent (iii) The case $\beta=1$:
    \begin{equation}
        \frac{1}{\lambda_n^2}K_{n,1}^{(\lambda_n)}(\xi^{(n)} , \eta^{(n)})=
            K^{(1)}(\xi,\eta)
        +\bigO(n^{-\tau})
          \begin{pmatrix}
            e^{-c\xi}& e^{-c\xi} e^{-c\eta}\\[1ex]
            e^{-c \min (\xi, \eta)} & e^{-c\eta}
        \end{pmatrix},
    \end{equation}
    where
    \begin{align*}
        (K^{(1)})_{11}(\xi,\eta) &=
            (K^{(1)})_{22}(\eta,\xi) =K_\Ai(\xi,\eta)+\frac{1}{2}\Ai(\xi)\int_{-\infty}^\eta\Ai(s)ds,
        \\[1ex]
         (K^{(1)})_{12}(\xi,\eta)&= -\frac{\partial}{\partial \eta}K_\Ai(\xi,\eta) - \frac{1}{2}\Ai(\xi)\Ai(\eta),
            \\[1ex]
         (K^{(1)})_{21}(\xi,\eta)
        &= -\int_\xi^\infty
        K_\Ai(s,\eta)ds-\frac{1}{2}\int_\xi^\eta\Ai(s)ds+\frac{1}{2}\int_\xi^\infty\Ai(s)ds\int_\eta^\infty\Ai(s)ds
        \\[1ex]
        &\phantom{=\,} -\frac{1}{2}\sgn(\xi-\eta).
    \end{align*}
\end{theorem}
As above we now state the consequences of this result for 
the $l$-point correlation functions and for the distribution of the 
largest eigenvalue.
\begin{corollary} \label{corollary:soft}
With the notation of Theorem \ref{theorem: universality soft edge} and (\ref{def:correlation}) and
$\lambda_n(M)$ denoting the largest eigenvalue of $M$ we have for
$l \in \mathbb N$, $\xi$, $\xi_i \in \mathbb R$ that the following limits
\begin{itemize}
\item[(a)] $\displaystyle \lim_{n \to \infty} \frac{1}{\lambda_n^{2l}} R_{n, \beta, l}
(\beta_n + \frac{\xi_1}{\lambda_n^{2}}, 
\ldots, \beta_n + \frac{\xi_l}{\lambda_n^{2}})$
for $\beta=1$, $2$; \quad
$\displaystyle \lim_{n \to \infty}  \frac{1}{\lambda_n^{2l}}
R_{n/2, 4, l}(\beta_n + \frac{\xi_1}{\lambda_n^{2}}, \ldots, \beta_n + \frac{\xi_l}{\lambda_n^{2}})$,
\item[(b)]
$\displaystyle \lim_{n \to \infty} \mathbb P_{n, \beta} 
(\{M \colon \lambda_n(M) \le \beta_n + \frac{\xi}{\lambda_n^2} \})$ 
for $\beta=1$, $2$; \quad
$\displaystyle \lim_{n \to \infty} \mathbb P_{\frac{n}{2}, 4} 
(\{M \colon \lambda_n(M) \le \beta_n + \frac{\xi}{\lambda_n^2} \})$,
\end{itemize}
exist (with $n$ even for $\beta =1$, $4$) and are independent of $Q$ (cf. (\ref{eq1p1})). 
\end{corollary}
This Corollary can be shown to be true in exactly the same way as Corollaries 1.2 and 1.3 were 
proven in \cite{DeiftGioev2} and we will not repeat the arguments here. Comparing the statements
of Theorem 1.1 in
\cite{DeiftGioev2} with Theorem \ref{theorem: universality soft edge} above 
shows that the 
limits in Corollary  \ref{corollary:soft} are exactly the same as the ones stated in 
Corollaries 1.2 and 1.3 of \cite{DeiftGioev2}. This implies in particular that the limits
in statement (b) are given by the celebrated Tracy--Widom distributions.
(Observe also that in \cite{DeiftGioev2}
the results were stated for cluster functions rather than for correlation functions.) 

We finally turn to the spectral statistics in the bulk.

\begin{theorem}\label{theorem: universality bulk} {\bf(bulk)}
    {\rm (cf.~\cite[Theorem 1.1]{DeiftGioev})}. \ \
    Let $\beta=1$,$2$ or $4$, $x\in(0,1)$ and define
    \begin{equation}\label{notation: bulk}
        q_n=\left(\frac{\beta_n}{n\omega_n(x)}\right)^{-1/2}, \qquad
        q_{n, 2}=q_{n,1}=q_n,\qquad q_{n,4}^2=\frac{1}{2}q_n^2,
    \end{equation}
    Then, for
    $n\to\infty$ ($n$ even for $\beta=1,4$) the following holds uniformly for
    $\xi,\eta$ in compact subsets of $\mathbb{R}$ and $x$ in compact subsets of
    $(0,1)$.

    \medskip

    \noindent (i) The case $\beta=2$:
    \begin{equation}\label{theorem: universality bulk: case beta=2}
        \frac{1}{q_{n,2}^2}K_n\left(\beta_n x+\frac{\xi}{q_{n, 2}^2},\beta_n
        x+\frac{\eta}{q_{n, 2}^2}\right)=K_\infty(\xi-\eta)+\bigO\left(\frac{1}{n}\right),
    \end{equation}
    where
    \begin{equation}
        K_\infty(t)=\frac{\sin\pi t}{\pi t}.
    \end{equation}

    \medskip

    \noindent (ii) The cases $\beta=1$ and $4$:
    \begin{align}
        & \frac{1}{q_{n,1}^2}K_{n,1}^{(q_{n,1})}\left(\beta_n x+\frac{\xi}{q_{n,1}^2},\beta_n
        x+\frac{\eta}{q_{n,1}^2}\right)=K_{\infty,1}(\xi,\eta)+
        \begin{pmatrix}
            \bigO(n^{-1/2}) & \bigO(n^{-1}) \\[1ex]
            \bigO(n^{-1}) & \bigO(n^{-1/2})
        \end{pmatrix}, \\[3ex]
        & \frac{1}{q_{n,4}^2}K_{\frac{n}{2},4}^{(q_{n,4})}\left(\beta_n x+\frac{\xi}{q_{n,4}^2},\beta_n
        x+\frac{\eta}{q_{n,4}^2}\right)=K_{\infty,4}(\xi,\eta)+
        \begin{pmatrix}
            \bigO(n^{-1/2}) & \bigO(n^{-1}) \\[1ex]
            \bigO(n^{-1}) & \bigO(n^{-1/2})
        \end{pmatrix},
    \end{align}
    where
    \begin{align}
        & K_{\infty,1}(\xi,\eta)=
        \begin{pmatrix}
            K_\infty(\xi-\eta) & \frac{\partial}{\partial \xi}
            K_\infty(\xi-\eta) \\[1ex]
            \int_0^{\xi-\eta}K_\infty(s)ds-\frac{1}{2}\sgn(\xi-\eta) & K_\infty(\eta-\xi)
        \end{pmatrix}, \\[3ex]
        & K_{\infty,4}(\xi,\eta)=
        \begin{pmatrix}
            K_\infty(2(\xi-\eta)) & \frac{\partial}{\partial \xi}
            K_\infty(2(\xi-\eta)) \\[1ex]
            \int_0^{\xi-\eta}K_\infty(2s)ds & K_\infty(2(\eta-\xi))
        \end{pmatrix}.
    \end{align}
\end{theorem}

Again we state the consequences of this theorem for 
the $l$-point correlation functions and for gap probabilities.
\begin{corollary} \label{corollary:bulk}
With the notation of Theorem \ref{theorem: universality bulk} and (\ref{def:correlation}) we have for
$l \in \mathbb N$, $x \in (0, 1)$, $\xi$, $\xi_i \in \mathbb R$ that the following limits
\begin{itemize}
\item[(a)] $\displaystyle \lim_{n \to \infty} \frac{1}{q_{n,\beta}^{2l}} R_{n, \beta, l}
(\beta_n x + \frac{\xi_1}{q_{n, \beta}^{2}}, 
\ldots, \beta_n x + \frac{\xi_l}{q_{n, \beta}^{2}})$
for $\beta=1$, $2$; \\
$\displaystyle \lim_{n \to \infty}  \frac{1}{q_{n, 4}^{2l}}
R_{n/2, 4, l}(\beta_n x + \frac{\xi_1}{q_{n, 4}^{2}}, \ldots, \beta_n x + \frac{\xi_l}{q_{n, 4}^{2}})$,
\item[(b)]
$\displaystyle \lim_{n \to \infty} \mathbb P_{n, \beta} ( 
\{M \colon \mbox{ no eigenvalue of } M \mbox{ lies in }
( \beta_n x - \frac{\xi}{q_{n, \beta}^2}\, , \, \beta_n x + \frac{\xi}{q_{n, \beta}^2} ) \} )$ 
for $\beta=1$, $2$; \\
$\displaystyle \lim_{n \to \infty} \mathbb P_{\frac{n}{2}, 4} ( 
\{M \colon \mbox{ no eigenvalue of } M \mbox{ lies in }
( \beta_n x - \frac{\xi}{q_{n, 4}^2}\, , \, \beta_n x + \frac{\xi}{q_{n, 4}^2} ) \} )$
\end{itemize}
exist (with $n$ even for $\beta =1$, $4$) and are independent of $Q$ (cf. (\ref{eq1p1})). 
\end{corollary}
For a proof and a description of the limits, see the corresponding results, Corollaries 1.2 and
1.3, in \cite{DeiftGioev}. We would like to stress again that the limiting local spectral statistics
of Hermite-type ensembles as considered in \cite{DeiftGioev, DeiftGioev2} agree in the bulk
and at the soft spectral edge exactly with
those for Laguerre-type ensembles considered in the present paper.

We conclude the Introduction with a brief outline of the remaining parts of this paper.
In Section \ref{sec2} we derive formulae (see Theorem \ref{thm:2.6}, 
Lemma \ref{lemma: simplification Widom}, 
Corollary \ref{S final}) for the scalar functions 
$S_{n, \beta}$, $\beta=1, 4$,
appearing in the definition of the matrix kernels $K_{n, \beta}$ in
(\ref{definition: Kn1}), (\ref{definition: Kn4}), in terms of orthogonal
polynomials. Here we follow mostly \cite{Widom} and \cite{DeiftGioev, DeiftGioev2}.
The precise form of the relation \eqref{2.40}
 in Proposition \ref{proposition: BAC}
below
and the skew symmetry of $G_{11}$ and $\widehat G_{11}$ 
reported in Lemma
\ref{lemma: simplification Widom}(ii),
 are extremely useful in proving precise error
estimates at various points in this paper. Relation \eqref{2.40}
and the skew symmetry in Lemma \ref{lemma: simplification Widom}(ii),
can also be used to improve some of the error estimates in
 \cite{DeiftGioev,DeiftGioev2}
(cf. Remark 4.1 in \cite{DeiftGioev2}).
At the end of Section \ref{sec2} we have
all the necessary ingredients to formulate the strategy for proving our main results
(see Remark \ref{remark:important}).

As in \cite{DeiftGioev, DeiftGioev2} one crucial step in the analysis
is to show the invertibility of a certain $m \times m$ matrix
(see $T_m$ in \eqref{definition: Tm} below), 
where $m$ denotes the
degree of the polynomial $Q$. 
This will be done in Section \ref{section: invertibility Tm}.
Here  estimates (essentially) derived
in \cite{DeiftGioev, CostinDeiftGioev}
are very useful (see Propositions \ref{proposition: u}, 
\ref{proposition: wq}, \ref{proposition: Wqhat}).
However, the proof of the invertibility
of the $m\times m$ matrix $T_m$ in the present situation,
is considerably more complicated
than the analogous situation in \cite{DeiftGioev, DeiftGioev2},
and new ingredients, over and above the estimates in 
\cite{DeiftGioev, CostinDeiftGioev}, are needed.

Sections 
\ref{section: asymptotics polynomials} 
and \ref{section: asymptotics B12 matrix} 
provide all the
asymptotic information on the orthogonal polynomials needed in this paper. 
We start the analysis from the pointwise asymptotic
 results derived in \cite{v6} by a Riemann--Hilbert (RH) 
steepest-descent analysis. 
In Section \ref{section: asymptotics polynomials} we reformulate 
these asymptotic results in such a way that
they can be conveniently used in the subsequent sections. 
Note that our splitting of
$\mathbb R_+$ into intervals with different leading asymptotics,
 differs from the 
one used in \cite{DeiftGioev}, and leads to improved error estimates,
in particular see Lemma \ref{lemma: B12matrix} below.
In Section 
\ref{section: asymptotics B12 matrix} we then derive
 asymptotic formulae for 
integrals of the functions $\phi_k$ defined in (\ref{def:phi_k}) and of
various related
functions. Most of these calculations are needed
 to determine the leading order behavior 
of the matrix $B$ which appears in Widom's 
formalism discussed in Section \ref{sec2}.

Our final Section \ref{sec6} combines all auxiliary results and provides proofs for our
main results. Here we give all details for the hard edge case which was not present in 
\cite{DeiftGioev, DeiftGioev2}. For the soft edge and the bulk we do not repeat those arguments
which can already be found in \cite{DeiftGioev, DeiftGioev2}.

{\bf Remark. }Throughout this paper, $D$ denotes differentiation 
and $\varepsilon$ 
denotes the integral operator with 
kernel $\varepsilon(x,y) = \frac{1}{2} \sgn(x-y)$. Furthermore,
by $\varepsilon f(x)$ we always mean the following,
$$
        \varepsilon f(x) = \frac{1}{2}\int_0^\infty \sgn(x-y)f(y)\,dy,
          \qquad x>0.
$$
The property $D\varepsilon f(x) = f(x)$ is clearly true
for all continuous and integrable functions $f$ on ${\mathbb R}_+$.
However, the relation $\varepsilon D f(x)=f(x)$ is only true
if $f(0)=0$. In what follows, the relevant function $f$
will always have this property, and we will use the relation
$\varepsilon D f(x)=f(x)$ without further comment.

\medskip

\noindent
{\bf Acknowledgments.}
The work of the first author was supported in part by
the NSF grant DMS--0500923.
While this work was being completed, the first author
was a Taussky--Todd and Moore Distinguished Scholar at Caltech,
and he thanks Professor Tombrello for his sponsorship
and Professor Flach for his hospitality.

The work of the second author was supported in part by
the NSF grant DMS--0556049.
The third author would like to thank
the Courant Institute
and Caltech for hospitality.

The forth author is a Postdoctoral Fellow
of the Fund for Scientific Research---Flaunders (Belgium). 

The first, third and forth author acknowlege support received from
the DFG within the program of the SFB/TR 12.

\section{Widom's formalism}
\label{sec2}

Following \cite{Widom} and \cite{DeiftGioev, DeiftGioev2} we will derive
in this section formulae for the scalar functions $S_{n, \beta}$, $\beta=1, 4$
appearing in the definition of the matrix kernels $K_{n, \beta}$ in
(\ref{definition: Kn1}), (\ref{definition: Kn4}). Furthermore, we will
present all properties of the terms appearing in the formulae needed
to prove our main theorems, except for the asymptotic results on the 
orthogonal polynomials. Those results will be provided in Section \ref{sec6}.    

Recall first (see \cite{TracyWidom}) the following representations 
for $S_{n, \beta}$ corresponding to probability density functions of the 
form (\ref{eq_PN}), (\ref{eq1p3}).  
Let $\{r_k(x)\}_{k\geq0}$ be any sequence of polynomials with $r_k$ having 
exact degree $k$.
For $k=0,1,2,\ldots$, set
\begin{equation}
\label{eq3p1}
        \psi_{k,\beta}(x)=\begin{cases}
                 r_k(x) w_1(x),&\beta=1\cr
                 r_k(x) (w_4(x))^{1/2},&\beta=4.
                                     \end{cases}
\end{equation}
Let $M_{n,1}$ denote the $n\times{}n$ matrix with entries
\begin{equation}
\label{eq3p2}
       (M_{n,1})_{jk} = \langle \psi_{j,1},\varepsilon \psi_{k,1} \rangle,\qquad 0\leq j,k\leq n-1,
\end{equation}
where we recall that $\varepsilon$ denotes the integral operator with 
kernel $\varepsilon(x,y) = \frac{1}{2} \sgn(x-y)$ and
$\langle f,h\rangle=\int_0^\infty f(x)h(x)\, dx$
is the standard real inner product on $\mathbb R_+$. Furthermore, denote by $M_{n,4}$ 
the $2n\times{}2n$ matrix with entries
\begin{equation}
\label{eq3p3}
       (M_{n,4})_{jk} = \langle \psi_{j,4},\psi'_{k,4}\rangle,\qquad 0\leq j,k\leq 2n-1,
\end{equation}
The matrices $M_{n,1}$ and $M_{n,4}$ are skew symmetric and invertible
(see e.g.~\cite[(4.17), (4.20)]{AvM}).
Let $\mu_{n,1}$, $\mu_{n,4}$ denote the inverses of $M_{n,1}$, $M_{n,4}$
respectively.
With this notation we have the following formulae (see \cite{TracyWidom})
for $S_{n,\beta}$ 
\begin{align}
\label{eq3pp1}
     & S_{n,1}(x,y) = -\sum_{j,k=0}^{n-1}
             \psi_{j,1}(x)\,(\mu_{n,1})_{jk}\,(\varepsilon \psi_{k,1})(y), \quad n \mbox{ even }, \\
\label{eq3pp2}
     & S_{n,4}(x,y) = \sum_{j,k=0}^{2n-1}
             \psi_{j,4}^\prime(x)\,(\mu_{n,4})_{jk}\,\psi_{k,4}(y).
\end{align}
As noted in \cite[(1.49), (1.50)]{DeiftGioev2} the following representations of $\varepsilon
S_{n, \beta}$ that are convenient for the study of the (2,1)-entries of $K_{n, \beta}$ are 
immediate from (\ref{eq3pp1}) and (\ref{eq3pp2}).
\begin{proposition} \label{proposition: 2.0}
\begin{align}
&(\varepsilon S_{n,1})(x,y)  \;=\; -\int_{x}^y S_{n,1}(t,y)\,dt\, , \quad n \mbox{ even },   \label{es1}\\
&(\varepsilon S_{n,4})(x,y)  \;=\; -\int_{x}^y S_{n,4}(t,y)\,dt\; = \; 
-\int_{x}^\infty S_{n,4}(t,y)\,dt \; = \; \int_{0}^x S_{n,4}(t,y)\,dt \label{es4}
\end{align}
\end{proposition}
\begin{proof}
The first equation follows from (\ref{eq3pp1}) and the skew symmetry of $\mu_{n,1}$ 
which implies in turn the skew symmetry of
$\varepsilon S_{n, 1}$: 
In particular $\varepsilon S_{n, 1}(y,y)=0$ for all $y > 0$.
The first relation of (\ref{es4}) follows from (\ref{eq3pp2}) in a similar way, using the skew symmetry of
$\mu_{n,4}$ and $\varepsilon \psi_{j,4}^\prime = \psi_{j,4}$. The remaining two equalities are consequences 
of $(\varepsilon S_{n,4})(+\infty,y)=0$ for all $y > 0$ together with the trivial relations
$\varepsilon f(x) = \int_0^x f(t)\,dt\, - \varepsilon f(+\infty) =  \varepsilon f(+\infty)-
\int_x^\infty f(t)\,dt\,$, which hold for integrable functions $f$.
\end{proof}

An essential feature of formulae (\ref{eq3pp1}), (\ref{eq3pp2}) 
is that the polynomials $\{r_k\}$ are arbitrary
and we are free to choose them conveniently to facilitate the
asymptotic analysis of \eqref{definition: Kn1}, \eqref{definition: Kn4} as $n\to\infty$
(see discussion in \cite[below (1.18)]{DeiftGioev}).
Widom \cite{Widom} found that the choice
 of orthogonal polynomials for $\{r_k\}$
leads to particularly convenient expressions for $S_{n, \beta}$ in cases where
$w'_{\beta} / w_{\beta}$ is a rational function. In \cite{DeiftGioev, DeiftGioev2}
it was then shown how these formulae together with detailed asymptotic information on
the orthogonal polynomials lead to universality results.

In order to be able to use the same set of orthogonal polynomials for
$\beta=1$, $4$ (and $2$) we have defined $w=w_1^2=w_4$($=w_2$) in 
(\ref{def:w1}), (\ref{def:w2}). The role of $r_k$, $\psi_{k, \beta}$ above
is then played by $p_k$ and $\phi_k$ defined in (\ref{def:phi_k}) above. 
The simultaneous treatment of $\beta = 1$ and $4$ is further 
facilitated by assuming $n$ to be even and by considering $S_{n, 1}$ together with 
$S_{\frac{n}{2}, 4}$. 

Consequently, let $n$ be an \textit{even} integer where we assume in addition that $n\ge
m$ (recall from (\ref{def:V}) that $m$ denotes the degree of the polynomial 
$V(x)=\sum_{j=0}^m q_j x^j$). Following Widom \cite{Widom} we denote
\begin{equation}
    \mathcal H:= \Span (\phi_0,\phi_1,\ldots ,\phi_{n-1}).
    \label{eqn:2.1}
\end{equation}
Following \cite[(3.3) and (3.4)]{Widom} we introduce the
$2m$-dimensional space
\[
    \mathfrak g:= \Span \left(\left\{ x^j \phi_n(x),x^j\phi_{n-1}(x)\mid -1\leq j\leq m-2\right\}\right).
\]
From the standard three-term recurrence relation satisfied by the
orthonormal functions $\phi_j$ (see \cite{Sz}), it follows directly that
\[
    \mathfrak g= \Span \left(\{ \phi_k\mid n-m+1\leq k\leq n+m-2\}\cup
    \left\{ \frac{\phi_n(x)}{x},
    \frac{\phi_{n-1}(x)}{x}\right\}\right).
\]
Define
\[
    \mathfrak g^{(1)}:=\mathfrak g \cap \mathcal H,\qquad
        \mbox{and}\qquad
        \mathfrak g^{(2)}:= \{ f\in \mathfrak g\mid \langle f,h\rangle=0,
        \mbox{ for all } h \in \mathcal H\}.
\]

Our first task is to construct a basis for $\mathfrak g^{(1)}$
and $\mathfrak g^{(2)}$. Define
\begin{align}
    \label{definition: psi1tilde}
    \tilde\psi_1(x) & := \frac{\gamma_{n-1}}{\gamma_n}\left[p_{n-1}(0)
        \frac{\phi_n(x)}{x}-p_n(0)\frac{\phi_{n-1}(x)}{x}\right], \\[1ex]
    \label{definition: psi2tilde}
    \tilde\psi_2(x) & := 2\pi i\frac{\gamma_{n-1}}{\gamma_n}\left[C(p_{n-1}w)(0)
        \frac{\phi_n(x)}{x}- C(p_nw)(0)\frac{\phi_{n-1}(x)}{x}\right],
\end{align}
where $C$ denotes the Cauchy transformation, i.e.
\[
    C(p_jw)(0)=\frac{1}{2\pi i}\int_0^\infty\frac{p_j(y)w(y)}{y}dy.
\]
Let $\beta_n$ be the Mhaskar--Rakhmanov--Saff number as defined in Subsection
\ref{subsection: relevant result from v6} below, let $d_n$ be some negative
number specified in (\ref{definition: dn}) below, and define
\begin{equation}
    \psi_1:=\alpha d_n\sqrt{\frac{\beta_n}{n}}\tilde\psi_1,\qquad
    \mbox{and}\qquad
    \psi_2:=\frac{1}{d_n}\sqrt{\frac{\beta_n}{n}}\tilde\psi_2.
    \label{th0}
\end{equation}
Furthermore, let $\Phi := (\Phi_1,\Phi_2)$ with
\[
    \Phi_1 := (\phi_{n-1},\phi_{n-2},\ldots ,
    \phi_{n-m+1},\psi_1),
    \qquad \Phi_2 := (\phi_n,\phi_{n+1},\ldots ,
    \phi_{n+m-2},\psi_2).
\]
With this notation we can prove the following Lemma.

\begin{lemma}\label{lemma: basis g}
    $\Phi_j$ is a basis of $\mathfrak g^{(j)}$ for $j=1,2$.
\end{lemma}

\begin{proof}
    Our approach to proving the Lemma is as follows.
    Assume that the following four statements are true:
    \begin{align*}
        & \mbox{(i) $\Span\, \Phi_1\subseteq \mathfrak g^{(1)}$}\\[1ex]
            & \mbox{(ii) $\Span\, \Phi_2\subseteq \mathfrak g^{(2)}$} \\[1ex]
        & \mbox{(iii) the $m$ functions in $\Phi_1$ 
                      are linearly independent}\\[1ex]
            & \mbox{(iv) the $m$ functions in $\Phi_2$ are linearly independent.}
    \end{align*}
    Then it only remains to be seen that $\dim (\Span\, \Phi_1)=\dim \mathfrak g^{(1)}$ and
    $\dim (\Span\, \Phi_2)=\dim \mathfrak g^{(2)}$.
    Since $\mathfrak
    g^{(1)}\cap \mathfrak g^{(2)}=\{0\}$, this follows from
    \[
        2m=\dim \mathfrak g  \geq \dim \mathfrak g^{(1)} +\dim \mathfrak g^{(2)}
             \geq \dim (\Span\, \Phi_1)+\dim (\Span\, \Phi_2)=2m.
    \]
    We now turn to verifying the four statements (i)--(iv).

    (i) One only needs to show that $\tilde\psi_1\in\mathfrak g^{(1)}$.
        Applying the Christoffel--Darboux formula (see \cite{Sz})
 to equation (\ref{definition: psi1tilde}) we have
        \begin{equation}
            \tilde\psi_1(x)=\sum^{n-1}_{k=0} p_k(0)\phi_k(x).
            \label{th1}
        \end{equation}
        This shows that $\tilde\psi_1$ is in $\mathcal H$ and
        hence in $\mathfrak g^{(1)}$.

    (ii) We need to prove that $\int_0^\infty\phi_k(x)\tilde\psi_2(x)dx=0$ for
        all $0\leq k\leq n-1$. Write
        \[
            \frac{\phi_k(x)}{x}=\left (q_{k-1}(x)+\frac{p_k(0)}{x}\right) \sqrt{w(x)}
        \]
        for some polynomial $q_{k-1}$ of degree $k-1$ (resp. $q_{-1}\equiv 0$ for $k=0$).
        From orthogonality we obtain
        for $0\leq k \leq n-1$,
        \begin{align*}
            \int_0^\infty \phi_k(x)\frac{\phi_{n-1}(x)}{x}dx
                &= \int_0^\infty\left(q_{k-1}(x)+\frac{p_k(0)}{x}\right)p_{n-1}(x) w(x) dx
                \\[1ex]
                &= p_k(0)\int_0^\infty \frac{p_{n-1}(x)w(x)}{x} dx =2\pi i p_k(0)
                C(p_{n-1}w)(0),
        \end{align*}
        and similarly
        \[
            \int_0^\infty \phi_k(x)\frac{\phi_n(x)}{x} dx = 2\pi i p_k(0) C(p_n
            w)(0).
        \]
        This implies that for $0\leq k\leq n-1$,
        \begin{multline*}
            \int_0^\infty \phi_k(x) \tilde \psi_2(x)dx
                \\[1ex] = (2\pi i)^2\frac{\gamma_{n-1}}{\gamma_n}
                    p_k(0)[C(p_{n-1}w)(0)C(p_nw)(0)-C(p_nw)(0)C(p_{n-1}w)(0)]
                    = 0.
        \end{multline*}

    (iii) It suffices to prove that $\tilde \psi_1\notin \Span(\phi_{n-1},\phi_{n-2},\ldots
        ,\phi_{n-m+1})$. This follows again from equation (\ref{th1})
        as $p_0(0)\not= 0$ and $n-m+1>0$.

    (iv) We prove by contradiction that $\tilde\psi_2\notin
        \Span\ (\phi_n, \ldots ,\phi_{n+m-2})$. Assume otherwise. Then
        $\lim_{x\to 0} \frac{x}{\sqrt{w(x)}}\tilde\psi_2(x)=0$. On the other
        hand, using the Christoffel--Darboux formula and the orthogonality
        relations for $p_k$ we have
        \begin{align*}
            \lim_{x\to 0}\frac{x}{\sqrt{w(x)}}\tilde \psi_2(x)
                & = \frac{\gamma_{n-1}}{\gamma_n}\int_0^\infty
                    \left(\frac{p_{n-1}(y)w(y)p_n(0)}{y}-\frac{p_n(y)w(y)p_{n-1}(0)}{y}\right) dy
            \\[1ex]
                & = -\sum^{n-1}_{k=0}p_k(0)\int_0^\infty p_k(y)w(y)dy
            \\[1ex]
                & = -p_0(0)^2\int_0^\infty w(y)dy = -1.
        \end{align*}
    This proves the Lemma.
\end{proof}

Next we consider the operator $[D,K]=DK-KD$ which plays a central
role in \cite{Widom}. Recall that $D$ denotes differentiation and $K$ denotes
the orthogonal projection onto $\mathcal H$, i.e.
\[
    (Kf)(x)=\int K(x,y)f(y)dy,\qquad\mbox{with } K(x,y)=\sum^{n-1}_{k=0} \phi_k(x)\phi_k(y).
\]
It follows from \cite{Widom} that the kernel of the operator $[D,K]$ can be expressed
in terms of functions in $\mathfrak g$ (in fact this motivates the definition of $\mathfrak g$).
More precisely, it is shown in \cite{Widom} that there 
exists a $2m\times2m$ real matrix $A$ such that
\begin{equation}
    [D,K]f=\Phi A\langle f, \Phi^t\rangle,\qquad
    \mbox{ for all $f\in C^1(\mathbb R_+)$ with $f'\in L^1(\mathbb
    R_+)$.}
    \label{th1.5}
\end{equation}
Moreover $A$ has the form
\begin{equation}
    A=\begin{pmatrix}
        0 & A_{12} \\
        A_{21} & 0
    \end{pmatrix},\qquad\mbox{where $A_{12}=A_{21}^t$ 
              is of size $m\times m$.}
    \label{th1.6}
\end{equation}
Here $\langle f,\Phi^t\rangle$ denotes the (column) vector $\int_0^\infty
f(x)\Phi^t(x)dx$. In order to determine the entries of $A$ we first prove the
following Proposition.

\begin{proposition}\label{proposition: entries A21-matrix}
    For all integers $\ell$ with $0\leq \ell\leq n-1$ we have
    \[
        [D,K]\phi_\ell
        = \sum^{n+m-2}_{k=n} \left(-\frac{1}{2}\langle V'\phi_\ell , \phi_k\rangle\right)\phi_k
            + \left(-\frac{n}{2\beta_n}\right)\langle \phi_\ell,\psi_1\rangle\psi_2.
    \]
\end{proposition}

\begin{proof}
    Let $0\leq \ell\leq n-1$. Then, since $K\phi_\ell=\phi_\ell$, we obtain
    \begin{align}\label{proposition: entries A21matrix: eq1a}
        \nonumber
        [D,K]\phi_\ell &= D\phi_\ell-KD\phi_\ell=(I-K)\phi_\ell' \\
            &= (I-K)(p_\ell'\sqrt{w})+\frac{\alpha}{2}(I-K)\left(\frac{\phi_\ell}{x}\right)
                -\frac{1}{2}(I-K)(V'\phi_\ell).
    \end{align}
    Let $\tilde w(x)=\frac{1}{x}\sqrt{w(x)}$. Observe that
    \[
        p_\ell'\sqrt w\in\mathcal H,\qquad
        \frac{\phi_\ell}{x}\in p_\ell(0)\tilde w+\mathcal
        H,\qquad\mbox{and}\qquad V'\phi_\ell\in\sum_{k=n}^{n+m-2}\langle
        V'\phi_\ell,\phi_k\rangle\phi_k+\mathcal H.
    \]
    Here the last formula follows from the fact $V'\phi_\ell\in \Span (\phi_0,\phi_1,\ldots
    ,\phi_{n+m-2})$. Since $(I-K)f=0$ for $f\in\mathcal H$, and since
    $(I-K)\phi_k=\phi_k$ for $k\geq n$, we then obtain from (\ref{proposition: entries A21matrix: eq1a})
    \begin{equation}\label{proposition: entries A21matrix: eq1}
        [D,K]\phi_\ell=\frac{\alpha}{2}p_\ell(0)(I-K)(\tilde w)
            +\sum_{k=n}^{n+m-2}\left(-\frac{1}{2}\langle V'\phi_\ell,\phi_k\rangle\right)\phi_k.
    \end{equation}
    It now remains to determine $(I-K)(\tilde w)$. Note that
        \begin{align*}
            \tilde\psi_2(x)
                & = \frac{\gamma_{n-1}}{\gamma_n}\int_0^\infty
                    \frac{\sqrt{w(y)}}{xy}\left(\phi_{n-1}(y)\phi_n(x)-\phi_n(y)\phi_{n-1}(x)\right) dy
                \\[1ex]
                    & = \int_0^\infty\frac{\sqrt{w(y)}}{xy} (x-y)
                        \sum^{n-1}_{k=0}\phi_k(x)\phi_k(y)dy
                \\[1ex]
                    & = \int_0^\infty K(x,y)\tilde w(y)dy-
                \frac{1}{x}\int_0^\infty K(x,y)\sqrt{w(y)}dy=K(\tilde w)
                    -\frac{1}{x}K(\sqrt w).
        \end{align*}
    Since $\sqrt{w} \in \mathcal H$, we have $K(\sqrt{w})=\sqrt{w}$. We
    then obtain $\tilde \psi_2=(K-I)(\tilde w)$, so that by
    (\ref{th0}),
    \[
        (I-K)\left(\tilde w\right)=-\tilde\psi_2=-d_n\sqrt{\frac{n}{\beta_n}}\psi_2.
    \]
    Inserting this relation into (\ref{proposition: entries A21matrix: eq1}) 
we obtain
    \begin{equation}\label{proposition: entries A21matrix: eq2}
        [D,K]\phi_\ell=-\frac{1}{2}\alpha d_n p_\ell(0)\sqrt{\frac{n}{\beta_n}}\psi_2
            +\sum_{k=n}^{n+m-2}\left(-\frac{1}{2}\langle V'\phi_\ell,\phi_k\rangle\right)\phi_k.
    \end{equation}
    Finally, observe that by (\ref{th0}) and (\ref{th1})
    \[
        \langle \phi_\ell,\psi_1\rangle =\alpha d_n\sqrt{\frac{\beta_n}{n}}
          \left\langle \phi_\ell,\sum^{n-1}_{k=0} p_k(0)\phi_k\right\rangle = \alpha d_n p_\ell(0)\sqrt{\frac{\beta_n}{n}}.
    \]
    The Proposition follows by inserting this
relation into (\ref{proposition: entries A21matrix: eq2}).
\end{proof}

Proposition \ref{proposition: entries A21-matrix} implies 
that for all $f\in\mathcal H$,
\[
     [D,K]f=-\sum^{n+m-2}_{k=n}\ \sum^{n-1}_{\ell=0} \phi_k
          \frac{1}{2}\langle V'\phi_\ell,\phi_k\rangle  \langle f,\phi_\ell\rangle -\psi_2
               \frac{n}{2\beta_n}\langle f,\psi_1\rangle.
     \label{th3}
\]
Note that $V'\phi_\ell\in\mathcal H$ for $\ell\leq n-m$: Hence
$\langle V'\phi_\ell,\phi_k\rangle=0$ for $\ell\leq n-m$
and $k\geq n$. Therefore,
\begin{align}
    \nonumber
    [D,K]f &= -\sum^{n+m-2}_{k=n}\sum^{n-1}_{\ell=n-m+1} \phi_k
          \frac{1}{2}\langle V'\phi_\ell,\phi_k\rangle  \langle f,\phi_\ell\rangle -\psi_2
               \frac{n}{2\beta_n}\langle f,\psi_1\rangle \\[1ex]
            &= \Phi_2\left[ -\frac{n}{\beta_n}
          \begin{pmatrix}
               Q_n & 0 \\
                    0 & \frac{1}{2}
           \end{pmatrix}
                \right] \langle f,\Phi_1^t\rangle, \qquad  \mbox{for $f\in\mathcal
                H$,}
\end{align}
where $Q_n$ is the $(m-1)\times (m-1)$ matrix given by
\[
     Q_n(i,j)=\frac{\beta_n}{2n}\langle V'\phi_{n-j},\phi_{n+i-1}\rangle, \qquad
        \mbox{for $1\leq i,j\leq m-1$.}
\]
On the other hand (\ref{th1.5}) and (\ref{th1.6}) imply
\[
     [D,K]f=\Phi_2 A_{21}\langle f,\Phi_1^t\rangle,\qquad
     \mbox{for $f\in\mathcal H$.}
\]
It is easy to see that the map $\mathfrak g^{(1)}\ni f\mapsto \langle
f,\Phi_1^t\rangle \in \mathbb R^m$ is a bijection. Since $\mathfrak
g^{(1)}\subseteq \mathcal H$ this shows that $\mathcal H\ni f\mapsto \langle
f,\Phi_1^t\rangle \in\mathbb R^m$ is onto, which in turn proves that the
matrix $A_{21}$ is given by
        \begin{equation}\label{determination A21}
            A_{21}=-\frac{n}{\beta_n}
                \begin{pmatrix}
                    Q_n & 0 \\
                    0 & \frac{1}{2}
                \end{pmatrix}.
        \end{equation}

\begin{remark}\label{remark: invertibility A}
  For $i+j>m$, $Q_n(i,j)=0$ and for $i+j=m$,
 $Q_n(i,j)=\langle V'\phi_{n+i-m},\phi_{n+i-1}\rangle$.
But by the orthogonality properties of the $\phi_j$'s, 
$\langle V'\phi_{n+i-m},\phi_{n+i-1}\rangle\neq 0$.
It follows that the matrix
  $A_{21}$, and hence also $A_{12}$, is invertible.
\end{remark}

\begin{lemma}{\rm (Asymptotics of the matrix $A$)} \label{lemma: A21matrix}
     The asymptotic behavior of the
     matrix  $A_{21}$ as $n\to\infty$, is given by
     \begin{equation}\label{definition: Y}
        A_{21}=-\frac{n}{\beta_n}\left(Y+\bigO(n^{-1/m})\right), \qquad
            \mbox{where }
            Y:= \begin{pmatrix}
                    Q & 0 \\
                    0 & \frac{1}{2}
                \end{pmatrix}.
     \end{equation}
     Here, $Q$ is an $(m-1)\times (m-1)$-matrix which is given by
     \begin{equation}
        Q(i,j):= c_{i+j-1}, \qquad \mbox{for $1\leq i,j\leq m-1$,}
     \end{equation}
     with
     \begin{equation}\label{definition: cl}
        c_\ell:=\frac{2^{2-2m}}{A_m} \binom{2m-2}{m-1-\ell},\qquad
            \mbox{and}\qquad A_m:=\prod_{j=1}^m \frac{2j-1}{2j}.
     \end{equation}
     Further, since $A_{12}=A_{21}^t$ and $Y=Y^t$, {\rm (\ref{definition: Y})} yields
     \begin{equation} \label{a12}
        A_{12}=A_{21}+\bigO\left(\frac{n}{\beta_n}n^{-1/m}\right).
     \end{equation}
\end{lemma}

\begin{proof}
    The proof uses the results in \cite{v6} on the asymptotics of the recurrence
    coefficients $b_{n-1}$ and $a_n$ appearing in the three-term 
recurrence relation
    \begin{equation}\label{three-term-recurrence-relation}
        x\phi_n(x)=b_n\phi_{n+1}(x)+a_n\phi_n(x)+b_{n-1}\phi_{n-1}(x),
    \end{equation}
    satisfied by the orthonormal functions $\phi_j$. The
    asymptotic behavior  of the recurrence coefficients as $n\to\infty$, is given by,
    cf.~\cite[Theorem 2.1]{v6}
    \begin{equation}\label{asymptotics recurrence coefficients}
        b_{n-1}=\frac{\beta_n}{4}\left[1+\bigO\left(\frac{1}{n}\right)\right],\qquad
        a_n=\frac{\beta_n}{2}\left[1+\bigO\left(\frac{1}{n}\right)\right].
    \end{equation}
    Here, $\beta_n$ is the Mhaskar--Rakhmanov--Saff number as defined
    in Subsection \ref{subsection: relevant result from v6} below,
    and has the following asymptotic behavior, cf.~\cite[Remark 2.2 and Proposition 3.4]{v6}
    \begin{equation}\label{asymptotics betan}
        \beta_n=\left(\frac{2n}{m q_m A_m} \right)^{1/m}
                 \left[1+\bigO(n^{-1/m})\right],\qquad
                 A_m=\prod_{j=1}^m \frac{2j-1}{2j},
    \end{equation}
    with $q_m$ the leading coefficient of the polynomial $V(x)=\sum_{k=0}^m q_k
    x^k$ (cf. (\ref{def:V})).

    Note first that for the case $i+j>m$ it is clear that $Q_n(i,j)=0$ as
    well as $c_{i+j-1}=0$ by the standard definition of binomials
    with negative second entry. Next, consider the case $i+j\leq m$. Since
    $\frac{\beta_k}{\beta_n}=1+
    \bigO\left(\frac{1}{n}\right)$ for $|k-n|$ bounded as $n\to\infty$ (see
    Proposition \ref{proposition1: double integrals out bulk}
    below), it follows from (\ref{asymptotics recurrence coefficients}) that
    \[
        \frac{b_k}{b_{n-1}}=1+\bigO\left(\frac{1}{n}\right),\qquad\mbox{and}\qquad
            \frac{a_k}{b_{n-1}}=2+{\mathcal O}\left(\frac{1}{n}\right)
    \]
    for $|k-n|$ bounded as $n\to \infty$. Using the three-term recurrence relation
    (\ref{three-term-recurrence-relation}), one can then prove
    by induction on $s$ that
    \[
        x^s\phi_\ell(x)=b^s_{n-1}\sum^{2s}_{r=0}
            \binom{2s}{r}
            \left[1+\bigO\left(\frac{1}{n}\right)\right] \phi_{\ell-s+r}(x),
    \]
    where the error bound $\bigO(1/n)$ does not depend on $x,s,\ell$ for
    $0\le s\le m-1$ and $n-m+1\le \ell\le n-1$. 
    It follows from this relation that for $i+j\le m$
    \begin{align*}
        Q_n(i,j) &= \frac{\beta_n}{2n}\langle
        V'\phi_{n-j},\phi_{n+i-1}\rangle = \frac{\beta_n}{2n}
        \sum^{m-1}_{s=0} (s+1)q_{s+1}\langle x^s\phi_{n-j},\phi_{n+i-1}\rangle
                \\[1ex]
                    & = \frac{\beta_n}{2n}
                        \sum_{s=i+j-1}^{m-1}(s+1)q_{s+1}b^s_{n-1}
                            \binom{2s}{s+(i+j-1)}
                            \left[1+\bigO\left(\frac{1}{n}\right)\right].
    \end{align*}
    Using (\ref{asymptotics recurrence coefficients}) and (\ref{asymptotics betan})
    we then arrive at the formula
    \begin{align*}
        Q_n(i,j)
                &= \frac{\beta_n}{2n}  mq_m b_{n-1}^{m-1}\binom{2m-2}{m-1 + (i+j-1)} \left[1+\bigO(n^{-1/m})\right]
                \\[1ex]
                &= \frac{2^{2-2m}}{A_m} \binom{2m-2}{m-1-(i+j-1)}
                    \left[1+\bigO(n^{-1/m})\right]
                \\[1ex]
                &= c_{i+j-1} + \bigO(n^{-1/m}).
     \end{align*}
     This completes the proof of the Lemma.
\end{proof}

Following \cite{Widom} we next define the real $2m\times2m$ matrix
\begin{equation} \label{definition: B12-matrix}
    B =
        \langle\varepsilon \Phi^t,\Phi\rangle =
        \begin{pmatrix}
            B_{11} & B_{12} \\
            B_{21} & B_{22}
        \end{pmatrix}.
\end{equation}
Observe that $B$ is skew symmetric so that
\begin{equation}\label{skew symmetrie B}
  B_{11}=-B_{11}^t,\qquad B_{21}=-B_{12}^t,\qquad \mbox{and}\qquad B_{22}=-B_{22}^t.
\end{equation}
For the convenience of the reader we display the entries of the
    matrix $B_{12}$, which is given by
    $B_{12}=\langle\varepsilon \Phi_1^t,\Phi_2\rangle$, more explicitly,
    \begin{equation}
        B_{12}(i,j)=\begin{cases}
            \langle\varepsilon\phi_{n-i},\phi_{n+j-1}\rangle,& 1\leq
            i,j\leq m-1, \\[1ex]
            \langle\varepsilon\psi_1,\phi_{n+j-1}\rangle, & i=m,1\leq j\leq
            m-1,\\[1ex]
            \langle\varepsilon\phi_{n-i},\psi_2\rangle, & 1\leq
            i\leq m-1,j=m, \\[1ex]
            \langle\varepsilon\psi_1,\psi_2\rangle, & i=j=m.
        \end{cases}
    \end{equation}

\begin{lemma}{\rm (Asymptotics of the matrix $B$)} \label{lemma: B12matrix}
    There exists $0<\tau=\tau(m,\alpha)<1$ such that:

    (i) As (even) $n\to\infty$,
    \begin{equation}\label{definition: X}
        B_{12}=\frac{\beta_n}{n}\left(X +\bigO(n^{-\tau})\right),
            \qquad \mbox{where }
            X = \begin{pmatrix}
                    R & v^t \\
                    v & 1-\frac{1}{\sqrt{2m-1}}
                \end{pmatrix}.
    \end{equation}
    Here, $R$ is an $(m-1)\times (m-1)$ matrix and $v$ is an $(m-1)$-dimensional
    row vector, which are given by
    \begin{equation}
        R(i,j)=\widehat I(i+j-1),\qquad
            v(j)=\sqrt{\frac{m}{2m-1}} I(j)-\frac{1}{2\sqrt m},
            \qquad \mbox{for $1\leq i,j\leq m-1$,}
    \end{equation}
    with
    \begin{align}
        \label{definition: Ihat}
        & \widehat
        I(q)=\frac{2}{\pi}\int_0^1\frac{\sin(q\arccos(2x-1))}{h(x)(1-x)}dx, \\[2ex]
        \label{definition: I}
        & I(q)=\frac{2}{\pi}\int_0^1\frac{\sin((q-1/2)\arccos(2x-1))}{h(x)x^{1/2}(1-x)}dx,
    \end{align}
    and $h(x)$ is expressed in terms of a particular hypergeometric
${}_2F_1$ function as follows:
    \begin{equation}\label{definition: h}
        h(x)=\sum_{k=0}^{m-1}2\frac{A_{m-1-k}}{A_m}x^k=\frac{4m}{2m-1}\,
        {}_2F_1(1,1-m,3/2-m;x).
    \end{equation}
    Further, since $B_{21}=-B_{12}^t$ and $X=X^t$, {\rm (\ref{definition: X})} yields
    \begin{equation} \label{b12}
        B_{21}=-B_{12}+\bigO\left(\frac{\beta_n}{n}n^{-\tau}\right).
    \end{equation}

    (ii) As (even) $n\to\infty$,
    \begin{equation} \label{b11}
        B_{11}=\bigO\left(\frac{\beta_n}{n}\right)=B_{22},\qquad
        B_{22}=-B_{11}+\bigO\left(\frac{\beta_n}{n}n^{-\tau}\right).
    \end{equation}
\end{lemma}

\begin{proof}
    The Lemma is immediate from the results
    (\ref{proof: lemma B12 matrix: innerproducts: eq1})--(\ref{proof: lemma B12 matrix: innerproducts: eq3})
    in Section \ref{section: asymptotics B12 matrix}. One should note that for
    the entries of the form $\langle\varepsilon\phi_{n-i},\psi_2\rangle$ we use the
    fact that $-I(-i+1)=I(i)$, which is true by definition.
\end{proof}

    \medskip

Finally we define the $2m\times2m$ matrix $C$  (see~\cite{Widom})
\begin{equation}\label{definition: C}
    C:=
        \begin{pmatrix}
            I & 0 \\
            0 & 0
        \end{pmatrix}+BA=
        \begin{pmatrix}
            I+B_{12}A_{21} & B_{11}A_{12} \\
            B_{22}A_{21} & B_{21}A_{12}
        \end{pmatrix}
        =\begin{pmatrix}
            C_{11} & C_{12} \\
            C_{21} & C_{22}
        \end{pmatrix},
\end{equation}
with $I$ the $m\times m$ identity matrix. We now have introduced all the
ingredients needed to state Widom's result \cite[Theorem 2]{Widom}
concerning the kernels $S_{n,1}$ and $S_{\frac{n}{2},4}$ 
(cf. \cite[(1.36), (1.37)]{DeiftGioev}). 

\begin{theorem} {\rm (Widom \cite{Widom})} 
    \label{thm:2.6}
    The kernels $S_{n,1}$ and
    $S_{\frac{n}{2},4}$ are given (for $n$ even) by
    \begin{align}
        \label{theorem: Widom: eq1}
        & S_{\frac{n}{2},4}(x,y)= K_n(x,y)-\Phi_2(x)A_{21}\varepsilon\Phi_1(y)^t-
            \Phi_2(x)A_{21}C_{11}^{-1}C_{12}\varepsilon\Phi_2(y)^t \\[1ex]
        \label{theorem: Widom: eq2}
        & S_{n,1}(x,y)= K_n(x,y)-(\Phi_1(x),0)\cdot (AC(I-BAC)^{-1})^t
            \cdot(\varepsilon\Phi_1(y),\varepsilon\Phi_2(y))^t.
    \end{align}
\end{theorem}
\begin{remark}\label{15p.1}
 The invertibility of $C_{11}$ in \eqref{theorem: Widom: eq1}
and of $I-BAC$ in \eqref{theorem: Widom: eq2}
is one of the assertions in \cite{Widom}
 (see also \cite[Remark~1.5]{DeiftGioev}).
\end{remark}
To simplify the analysis in the present paper we need a better understanding of
these kernels. We now establish the following interesting
and very useful relation.
\begin{proposition} \label{proposition: BAC}
    \begin{equation}\label{2.40}
        BAC=
            \begin{pmatrix}
                0 & 0 \\
                C_{21} & C_{22}
            \end{pmatrix}.
    \end{equation}
\end{proposition}

\begin{proof}
    Using $A=A^t$, and the fact that $\varepsilon f \in C^1(\mathbb R_+)$,
    $(\varepsilon f)^\prime = f \in L_1(\mathbb R_+)$
    for all $f\in\mathfrak g$, we conclude that
    \[
        DK\varepsilon f=KD\varepsilon f+[D,K]\varepsilon f =
            Kf+\Phi A\langle\varepsilon f,\Phi^t\rangle =
            Kf+\langle\varepsilon f,\Phi\rangle A \Phi^t,
    \]
            for all $f\in\mathfrak g$. Thus,
            \begin{align}\label{proof: proposition: BAC: eq1}
                & DK\varepsilon:\mathfrak g\to\mathfrak
                g,\qquad\mbox{with }
                (DK\varepsilon)\Phi^t=BA\Phi^t+\begin{pmatrix}
                    I & 0 \\ 0 & 0
                \end{pmatrix}\Phi^t=C\Phi^t,\\[1ex]
                & DK\varepsilon-K:\mathfrak g\to\mathfrak
                g,\qquad\mbox{with }
                (DK\varepsilon-K)\Phi^t=BA \Phi^t.
            \end{align}
            Using in addition that $\varepsilon D f = f$ for all $f \in \mathcal H$,
 we conclude
            \[
                (BAC)\Phi^t=DK\varepsilon(DK\varepsilon-K)\Phi^t=DK\varepsilon(I-K)\Phi^t
                    =\begin{pmatrix}
                        0 & 0 \\
                        0 & I
                    \end{pmatrix}C \Phi^t.
            \]
            Since $\Phi$ is a basis of $\mathfrak g$ we then have
            \[
                BAC=\begin{pmatrix}
                        0 & 0 \\
                        0 & I
                    \end{pmatrix}C=
                    \begin{pmatrix}
                        0 & 0 \\
                        C_{21} & C_{22}
                    \end{pmatrix},
            \]
            which proves the Proposition.
        \end{proof}
The above Proposition together with Lemma \ref{lemma: simplification Widom} 
below, restates Widom's result in a form 
which is particularly convenient for the asymptotic analysis in Section \ref{sec6}. 
Lemma \ref{lemma: simplification Widom} summarizes certain facts which were
already used in the analysis of \cite[Section 4]{DeiftGioev2}. Note, however,
that some of these facts were stated in \cite{DeiftGioev2} in a weaker 
form due to the use of a different version of Proposition \ref{proposition: BAC}.
\begin{lemma}\label{lemma: simplification Widom}
    (i)
    For $n$ even,
the kernels $S_{n,1}$ and $S_{\frac{n}{2},4}$ are given by,
    \begin{align}
        \label{proposition: kernels: eq1}
        & S_{\frac{n}{2},4}(x,y)=K_n(x,y)-\Phi_2(x)A_{21}\varepsilon\Phi_1(y)^t
        -\Phi_2(x)G_{11}\varepsilon\Phi_2(y)^t,
        \\[1ex]
        \label{proposition: kernels: eq2}
        & S_{n,1}(x,y)=
            K_n(x,y)
            - \Phi_1(x) A_{12} \varepsilon\Phi_2(y)^t - \Phi_1(x) \widehat G_{11} \varepsilon\Phi_1(y)^t,
    \end{align}
    where
    \[
        G_{11}=A_{21}C_{11}^{-1}C_{12},\qquad and \qquad
        \widehat G_{11}=-A_{12}B_{22}\widehat C_{22}^{-t}A_{21}\qquad
        \mbox{with $\widehat C_{22}=I-C_{22}$.}
    \]

    (ii) The matrices
    $G_{11}$ and $\widehat G_{11}$ are skew symmetric. Moreover,
    \[
        \widehat G_{11}=-A_{12}\widehat C_{22}^{-1}C_{21}.
    \]
\end{lemma}

\begin{proof}
    (i) Equation (\ref{proposition: kernels: eq1}) is precisely (\ref{theorem: Widom: eq1}).
    Next, consider the $2m\times 2m$ matrix
    $[AC(I-BAC)^{-1}]^t$ as a two by two block matrix with blocks of size $m\times m$ and denote
    the upper left and right blocks by $\widehat G_{11}$ and $\widehat G_{12}$, respectively.
    With this notation we have by (\ref{theorem: Widom: eq2}),
    \[
        S_{n,1}(x,y)=
            K_n(x,y) - \Phi_1(x) \widehat G_{11} \varepsilon\Phi_1(y)^t
            - \Phi_1(x) \widehat G_{12} \varepsilon\Phi_2(y)^t.
    \]
    In order to determine $\widehat G_{11}$ and $\widehat G_{12}$, observe that from
    Proposition \ref{proposition: BAC},
    \begin{equation}\label{proof: proposition: kernels: eq1a}
        \begin{pmatrix}
            \widehat G_{11}^t & * \\
            \widehat G_{12}^t & *
        \end{pmatrix}=AC
        \begin{pmatrix}
            I & 0 \\
            -C_{21} & \widehat C_{22}
        \end{pmatrix}^{-1}
        =AC
        \begin{pmatrix}
            I & 0 \\
            \widehat C_{22}^{-1}C_{21} & \widehat C_{22}^{-1}
        \end{pmatrix}.
    \end{equation}
    Note that the invertibility of $\widehat C_{22}$ is immediate from the invertibility of
    $I-BAC$.
    By (\ref{proof: proposition: kernels: eq1a}),
    \begin{align}\label{proof: proposition: kernels: eq1}
        \nonumber
        \widehat G_{11}^t
        &=(AC)_{11}+(AC)_{12}\widehat C_{22}^{-1}C_{21}=
            A_{12}(I+C_{22}\widehat C_{22}^{-1})C_{21}=A_{12}(\widehat C_{22}+C_{22})\widehat C_{22}^{-1}C_{21} \\[1ex]
        &=A_{12}\widehat C_{22}^{-1}B_{22}A_{21}.
    \end{align}
    Since $A_{12}=A_{21}^t$ and $B_{22}=-B_{22}^t$, see (\ref{th1.6}) and
    (\ref{skew symmetrie B}), this yields $\widehat G_{11}=-A_{12}B_{22}\widehat C_{22}^{-t}A_{21}$.
    Further, from (\ref{proof: proposition: kernels: eq1a}) we obtain,
    \begin{align}\label{proof: proposition: kernels: eq3}
        \nonumber
        \widehat G_{12}^t &=(AC)_{21}+(AC)_{22}\widehat C_{22}^{-1}C_{21}=A_{21}(C_{11}+C_{12}\widehat C_{22}^{-1}C_{21})
        \\[1ex]
        &= A_{21}+A_{21}(C_{11}-I+C_{12}\widehat C_{22}^{-1}C_{21}).
    \end{align}
    From Proposition \ref{proposition: BAC} it follows that
    \[
        \begin{pmatrix}
            C_{11}-I & C_{12} \\
            C_{21} & C_{22}
        \end{pmatrix}
        \begin{pmatrix}
            C_{11}\\
            C_{21}
        \end{pmatrix}=
        \begin{pmatrix}
            0 \\ C_{21}
        \end{pmatrix},
    \]
    which implies
    \[
        C_{11}-I=-C_{12}C_{21}C_{11}^{-1},\qquad \widehat C_{22}^{-1}C_{21}=C_{21}C_{11}^{-1}.
    \]
    Inserting the first relation
 into (\ref{proof: proposition: kernels: eq3}) 
we obtain $\widehat G_{12}^t=A_{21}=A_{12}^t$
    and the first part of the Lemma is proven.

    \medskip

    (ii)
    We will now prove that $G_{11}$ and $\widehat G_{11}$ are skew 
symmetric. Since $C_{11}^t=I-A_{12}B_{21}$, see (\ref{definition: C}), (\ref{th1.6}) and
    (\ref{skew symmetrie B}), and since
    $(C_{11}-I)C_{12}+C_{12}C_{22}=0$ (which follows from $(BAC)_{12}=0$) we have
    \[
        B_{11}C_{11}^tA_{12}=C_{12}-C_{12}C_{22}=C_{11}C_{12}=C_{11}B_{11}A_{12}.
    \]
    The invertibility of $A_{12}$ 
(see Remark \ref{remark: invertibility A})
yields $B_{11}C_{11}^{-t}=C_{11}^{-1}B_{11}$. Since
    $C_{12}^t=-A_{21}B_{11}$ we obtain
    \[
        \left(A_{21}C_{11}^{-1}C_{12}\right)^t=-A_{21}B_{11}C_{11}^{-t}A_{12}
            =-A_{21}C_{11}^{-1}B_{11}A_{12}=-A_{21}C_{11}^{-1}C_{12}.
    \]
    Hence $G_{11}=A_{21}C_{11}^{-1}C_{12}$ is skew symmetric.

    Next, since $\widehat C_{22}^t=I-C_{22}^t=I+A_{21}B_{12}$ and $C_{21}C_{11}+C_{22}C_{21}=C_{21}$ (which
    follows from $(BAC)_{21}=C_{21}$) we have,
    \[
        B_{22}\widehat C_{22}^t A_{21}=C_{21}C_{11}=\widehat C_{22} C_{21}= \widehat C_{22} B_{22}A_{21}.
    \]
    Since $A_{21}=A_{12}^t$ is invertible we therefore have
    $B_{22} \widehat C_{22}^{-t}=\widehat C_{22}^{-1}B_{22}$ and thus
    \begin{equation}\label{proof: proposition: kernels: eq2}
        \widehat G_{11}=-A_{12}\widehat C_{22}^{-1}B_{22} A_{21}=-A_{12}\widehat C_{22}^{-1}C_{21}.
    \end{equation}
    The skew symmetry of $\widehat G_{11}$
now follows from (\ref{proof: proposition: kernels: eq1}), and the Lemma
    is proven.
\end{proof}

As discussed in Remark \ref{remark:important} below, 
our universality results depend critically 
on bounds, uniform in $n$, 
for the inverse matrices $C_{11}^{-1}$ and $\widehat
C_{22}^{-1}$ which appear in the definitions of $G_{11}$ 
and $\widehat G_{11}$ given in the 
previous Lemma. In order to prove the existence of such 
 bounds we introduce
\begin{equation}\label{definition: Tm}
    T_m:=I-XY,
\end{equation}
where the $n$-independent matrices $X$ and $Y$ were defined in Lemmas
\ref{lemma: B12matrix} and \ref{lemma: A21matrix}, respectively.

\begin{theorem} \label{theorem: invertibility Tm}
    For all $m\geq 1$, the matrix $T_m$ is invertible.
\end{theorem}

\begin{proof}
   For $m=1$, the result is trivial as $X=0$ in this case
(see Lemma \ref{lemma: B12matrix}).
For $m\geq2$, the proof of the Theorem requires
considerable detailed analysis and occupies
all of Section \ref{section: invertibility Tm}.
\end{proof}

\begin{corollary}\label{corollary: control inverse}
    For all $m\geq 1$, there exists $N,L$ such that for all $n\geq N$,
    \begin{itemize}
        \item[(i)]
            $\|C_{11}^{-1}\|=\|(I + B_{12} A_{21})^{-1}\|\leq L$
        \item[(ii)]  $\|\widehat C_{22}^{-1}\|=\|(I - B_{21} A_{12})^{-1}\|\leq L$.
    \end{itemize}
\end{corollary}

\begin{proof}
    (i) It follows from Lemmas \ref{lemma: A21matrix} and \ref{lemma: B12matrix}
                    that $I + B_{12} A_{21}$ converges to $T_m$ as $n \to \infty$. The claim now follows from Theorem \ref{theorem: invertibility Tm}.

    (ii) Since $A_{12}^t = A_{21}$ and $B_{21}^t = - B_{12}$ we have that
             $(I - B_{21} A_{12})^t$ converges to
                    $I - YX$ as $n \to \infty$. Since $XY$ and $YX$ have the same 
(non-zero) eigenvalues,
                    the invertibility of $I - YX$ follows again from Theorem \ref{theorem: invertibility Tm},
                    leading to statement (ii).
\end{proof}

Lemma \ref{lemma: A21matrix}, Lemma \ref{lemma:
    B12matrix}(ii) together with Corollary \ref{corollary: control inverse} imply:

\begin{corollary}\label{corollary: asymptotics Gmatrices}
    The matrices $G_{11}$ and $\widehat G_{11}$ of Lemma \ref{lemma: simplification Widom}
    obey the following asymptotic bounds,
    \begin{equation}\label{corollary: asymptotics Gmatrices: eq1}
        G_{11}=\bigO\left(\frac{n}{\beta_n}\right),\qquad \widehat
        G_{11}=\bigO\left(\frac{n}{\beta_n}\right),\qquad n\to\infty.
    \end{equation}
\end{corollary}

Note that for $m=1$ it follows from the skew symmetry of $G_{11}$ and $\widehat G_{11}$, see
Lemma \ref{lemma: simplification Widom}(ii), that $G_{11}=\widehat G_{11}=0$.

The importance of the analog of the following 
observations for the proof of universality 
has already been noted in \cite[(1.46)]{DeiftGioev2}.

\begin{proposition}
\label{proposition: moreC} With the above notation,
 the following statements hold true.

\smallskip

\noindent
(i)  \, $A_{21}\varepsilon\Phi_1(+\infty)^t+G_{11}\varepsilon\Phi_2(+\infty)^t=0$. \\[1ex]
(ii)
There exists $0<\tau=\tau(m,\alpha)<1$ such that as $n\to\infty$\vspace{.1cm}\\ 
$\phantom{AAA}
A_{12}\varepsilon\Phi_1(+\infty)^t+\widehat G_{11}\varepsilon\Phi_2(+\infty)^t=
A_{12}\widehat C_{22}^{-1}
    \left[\bigO(n^{-\tau})\varepsilon\Phi_1(+\infty)^t+\bigO(n^{-\tau})\varepsilon\Phi_2(+\infty)^t\right].
$

\end{proposition}
\begin{proof}
(i)
Recall that for $f\in\mathcal H$ (see (\ref{eqn:2.1})) we have
$0=\frac{1}{2}\int_0^\infty f'(x)dx=\varepsilon(Df)(+\infty)$.
 Using (\ref{proof: proposition: BAC: eq1}) and
$K\varepsilon\Phi^t\in\mathcal{H}^{2m}$ we obtain
    $0=\varepsilon(DK\varepsilon\Phi^t)(+\infty)=\varepsilon\left(
    C\Phi^t\right)(+\infty)=C\varepsilon\Phi^t(+\infty).$
This implies by Lemma \ref{lemma: simplification Widom} that
\[
A_{21}\varepsilon\Phi_1(+\infty)^t+G_{11}\varepsilon\Phi_2(+\infty)^t=
A_{21} C_{11}^{-1}
    \left[C_{11} \varepsilon\Phi_1(+\infty)^t+C_{12} \varepsilon\Phi_2(+\infty)^t\right] = 0.
\]
(ii) Lemma \ref{lemma: simplification Widom} yields
\[
A_{12}\varepsilon\Phi_1(+\infty)^t+\widehat G_{11}\varepsilon\Phi_2(+\infty)^t=
A_{12} \widehat C_{22}^{-1}
    \left[\widehat C_{22} \varepsilon\Phi_1(+\infty)^t-C_{21} \varepsilon\Phi_2(+\infty)^t\right].
\]
Moreover, relations (\ref{a12}), (\ref{b12}), (\ref{b11}) together with (\ref{definition: Y}),
(\ref{definition: X}) imply 
$\widehat C_{22}=C_{11}+\bigO(n^{-\tau})$ and
        $C_{21}=-C_{12}+\bigO(n^{-\tau})$
for some suitable $0 < \tau < 1$. The claim then follows from (i).
\end{proof}
The above Proposition together with the simple observation 
that for integrable functions $f \in L^1(\mathbb R_+)$
\[
\varepsilon f (x) = \int_{0}^{x} f(s)ds -  \varepsilon f (+\infty)  = \varepsilon f (+\infty) -
\int_{x}^{\infty} f(s)ds
\]
allows us to convert (\ref{proposition: kernels: eq1}) and (\ref{proposition: kernels: eq2}) into a form
which is particularly
 suitable for the analysis both at the hard and the soft edge.
\begin{corollary} \label{S final}
 For $n$ even, and for some $0<\tau=\tau(m,\alpha)<1$,
  the kernels $S_{n,1}$ and $S_{\frac{n}{2},4}$ satisfy 
    \begin{align}
        \label{convenient formula S4: hard edge}
        S_{\frac{n}{2},4}(x,y)&=K_n(x,y)-\Phi_2(x)A_{21}\int_0^y\Phi_1(s)^tds
        -\Phi_2(x)G_{11}\int_0^y\Phi_2(s)^tds
        \\[1ex]
        \label{convenient formula S4: soft edge}
          &= K_n(x,y)+\Phi_2(x)A_{21}\int_y^\infty\Phi_1(s)^tds
           +\Phi_2(x)G_{11}\int_y^\infty\Phi_2(s)^tds \, ,
          \\[1ex]
        \nonumber
    S_{n,1}(x,y) &=
    K_n(x,y)-\Phi_1(x)A_{12}\left(\int_0^y\Phi_2(s)^tds
        -\varepsilon\Phi_2(+\infty)^t+\varepsilon\Phi_1(+\infty)^t\right)
    \\[1ex]
    \nonumber
    &\phantom{K_n(x,y)}-\Phi_1(x) \widehat G_{11}\left(\int_0^y\Phi_1(s)^tds
        -\varepsilon\Phi_1(+\infty)^t+\varepsilon\Phi_2(+\infty)^t\right)
    \\[1ex]
    \label{convenient formula S1: hard edge}
    &\phantom{K_n(x,y)}+\Phi_1(x)A_{12}\widehat C_{22}^{-1}
    \left[\bigO(n^{-\tau})\varepsilon\Phi_1(+\infty)^t+\bigO(n^{-\tau})\varepsilon\Phi_2(+\infty)^t\right]
          \\[1ex]
        \nonumber
    &=
     K_n(x,y)+\Phi_1(x)A_{12}\left(\int_y^\infty\Phi_2(s)^tds
        -\varepsilon\Phi_1(+\infty)^t-\varepsilon\Phi_2(+\infty)^t\right)
    \\[1ex]
    \nonumber
    &\phantom{K_n(x,y)}+\Phi_1(x) \widehat G_{11}\left(\int_y^\infty\Phi_1(s)^tds
        -\varepsilon\Phi_1(+\infty)^t-\varepsilon\Phi_2(+\infty)^t\right)
    \\[1ex]
    \label{convenient formula S1: soft edge}
    &\phantom{K_n(x,y)}+\Phi_1(x)A_{12}\widehat C_{22}^{-1}
    \left[\bigO(n^{-\tau})\varepsilon\Phi_1(+\infty)^t+\bigO(n^{-\tau})\varepsilon\Phi_2(+\infty)^t\right].
    \end{align}
\end{corollary}

\begin{remark}
\label{remark:important}
Corollary \ref{S final} allows us to indicate at this point which facts are essential for our
proof of universality. The details of the proofs can be found in Section \ref{sec6}.

\medskip

\noindent
(a) {\em Hard edge:} For the simpler case $\beta=4$ we see from 
(\ref{convenient formula S4: hard edge}) that $S_{\frac{n}{2}, 4}$ can be 
written as a sum of three terms. The first term is the Christoffel--Darboux kernel which we know to 
be universal from the analysis of the case $\beta = 2$ \cite{v6}. The key
 to understanding the second
term is the observation (cf. Propositions \ref{proposition: hard universality: Phi}, 
\ref{Pinthard}) that after rescaling $\Phi_2(x)$ 
and $\int_0^{y}\Phi_1(s)ds$ are both, to leading order,
 scalar multiples of the vector 
${\bf e}:=(0,\ldots,0,1)$ where the scalar factors can be expressed in terms of some Bessel functions which
only depend on $\alpha$. Moreover, ${\bf e} A_{21} {\bf e}^t$ just reproduces the $(m,m)$ entry of 
$A_{21}$, which by (\ref{determination A21}) is (universally)
 given by $-\frac{n}{2\beta_n}$. By similar
reasoning the leading order behavior of the last term of (\ref{convenient formula S4: hard edge})
is given by ${\bf e} G_{11} {\bf e}^t$ which is equal to $0$ by 
the skew symmetry of $G_{11}$. The vanishing of this term by skew
 symmetry is fortunate since an explicit evaluation 
of the asymptotics of the matrix
$G_{11}$ for general $m$ is a formidable problem. 
The heart of the problem is then to estimate the inverse
matrix $C_{11}^{-1}$ uniformly in $n$
(cf.~Corollary \ref{corollary: control inverse}).

For $\beta =1$ we use formula (\ref{convenient formula S1: hard edge}) which is a sum of four terms. 
The first term is the Christoffel--Darboux kernel and the last term is of lower order due to the 
$\bigO(n^{-\tau})$ estimate. As in the case $\beta = 4$ one can show by corresponding asymptotic
formulae for the expressions depending on $\Phi$, that 
the leading order behavior is given by
${\bf e} A_{12} {\bf e}^t$ and ${\bf e} \widehat G_{11} {\bf e}^t$. The latter term vanishes
by skew symmetry of $\widehat G_{11}$ and the first term equals $-\frac{n}{2\beta_n}$
since $A_{12} = A_{21}^t$ by (\ref{th1.6}).

\medskip

\noindent
(b) {\em Soft edge:} The arguments here are quite similar to the ones given for the hard edge
with (\ref{convenient formula S4: hard edge}), (\ref{convenient formula S1: hard edge}) replaced
by (\ref{convenient formula S4: soft edge}) and (\ref{convenient formula S1: soft edge}) respectively.
The most distinctive difference from the hard edge case is that the vector
${\bf e}$ is now replaced by ${\bf a}=\left(1,\ldots ,1,\sqrt\frac{m}{2m-1}\right)$. We still have the
vanishing of ${\bf a} G_{11} {\bf a}^t$ and ${\bf a} \widehat G_{11} {\bf a}^t$ by skew symmetry.
However, the universality result at the soft edge hinges on the relation
\[
{\bf a} A_{21} {\bf
        a}^t={\bf a}A_{12}{\bf a}^t=-\frac{n}{\beta_n}\left(\frac{m}{2}+\bigO(n^{-1/m})\right)
\] 
of Proposition \ref{proposition: algebraic formula soft edge}. This relation follows from the
leading order evaluation
\[
{\bf a} Y {\bf
        a}^t= \frac{m}{2}
\]
which by the defintion of $Y$ in (\ref{definition: Y}) is based for each $m$ on some identity 
for sums of binomial coefficients. It is somewhat surprising and maybe unsatisfactory
that the derivation of the universal Tracy--Widom distributions at the soft edge depends on 
such special identities. A similar situation already appeared in 
\cite[(4.13) and below]{DeiftGioev2}.

\medskip

\noindent 
(c) {\em Bulk:}
As in \cite{DeiftGioev} the proof of universality in the bulk is less subtle than at the edges, 
because one can show that the Christoffel--Darboux kernel $K_n$ dominates in  
(\ref{proposition: kernels: eq1}) and (\ref{proposition: kernels: eq2}) and the remaining two 
correction terms 
in each formula are of lower order as
$n \to \infty$.
\end{remark}

\section{Invertibility of $T_m$ for $m \ge 2$}  \label{section: invertibility Tm}

In this section we will always assume $m \ge 2$. Our objective is to
prove that for such $m$ the $m\times m$ matrices $T_m = I - XY$,
defined in (\ref{definition: Tm}), are invertible. A crucial step in the proof
of this result is provided by
 the estimates in Lemma \ref{lemma1: invertibility} for the 
entries of the matrix $X$. 
Our proof of the basic Lemma \ref{lemma1: invertibility} 
in Subsection \ref{sec3.2} follows 
closely the corresponding proofs in \cite{DeiftGioev, CostinDeiftGioev},
see in particular Proposition \ref{proposition: u} below.

However, as mentioned above, we face new
difficulties in the Laguerre-type case which are not present for
Hermite-type ensembles. 
In the Hermite-type case the authors show that, for any $m\geq1$,
as a map from $l_\infty$
to $l_\infty$, the analog of $T_m-I$
has the norm $<1$, and hence $T_m$ is invertible.
In the present situation, however, 
the last row and column in $X$ and $Y$, which have no 
analogue in the Hermite-type case, 
force the matrix $XY$ to have norm $\geq1$ for any 
operator norm on ${\mathbb{R}}^m$.
Thus we may not simply invert $T_m$ by a Neumann series and
one must take a different approach.
This approach is presented below and in
Subsection \ref{sec3.1}. 

We use the following
representation of $T_m$ which is immediate
 from (\ref{definition: Y}), (\ref{definition: X})
and (\ref{definition: Tm}):
\[
    T_m = I-XY = \begin{pmatrix}
                I-RQ & -\frac{1}{2}v^t \\[1ex]
                -vQ & \frac{1}{2}+\frac{1}{2}\frac{1}{\sqrt{2m-1}}
            \end{pmatrix},
\]
where $Q$ is defined in Lemma \ref{lemma: A21matrix}, and where
$R$ and $v$ are defined in Lemma \ref{lemma: B12matrix}. The
approach we follow to prove that $T_m$ is invertible is based on
the following fact. A matrix $T$ written in block form
\[
    T = \begin{pmatrix}
            a & b \\
            c & d
        \end{pmatrix}
\]
is invertible if both the matrices $a$ and $d-c a^{-1} b$ are invertible.
Therefore it suffices to prove that the following two conditions (1) and (2)
are satisfied.
\begin{itemize}
    \item[(1)] $I-RQ$ is invertible
    \item[(2)] $1+\frac{1}{\sqrt{2m-1}}-vQ(I-RQ)^{-1}v^t\neq 0$
\end{itemize}
In Subsection \ref{sec3.1} we will show how these two conditions follow from the
technical Lemmas \ref{lemma1: invertibility} and \ref{lemma2: invertibility}.
These Lemmas will then be proven in Subsections \ref{sec3.2} and 
\ref{subsection: proof invertibility: third part}.
  
\subsection{Proof of conditions (1) and (2)}
\label{sec3.1}
We introduce some convenient notation. Let $R_1$ and $U_0$ be the following
$(m-1)\times (m-1)$ matrices,
\begin{equation}
    R_1
        = R -
            \begin{pmatrix}
                \frac{1}{4} & 0 \\
                0 & 0
            \end{pmatrix},
    \qquad \mbox{and} \qquad
    U_0
        = I -
            \begin{pmatrix}
                \frac{1}{4} & 0 \\
                0 & 0
            \end{pmatrix} Q
        =   \begin{pmatrix}
                \gamma & -\frac{1}{4}u \\
                0 & I
            \end{pmatrix}.
\end{equation}
Here, $\gamma=1-\frac{c_1}{4}$ and $u=(c_2,\ldots ,c_{m-1})$
(cf. \eqref{definition: cl}). Further, define
$\hat Q=QU_0^{-1}$. It is clear that $U_0$ is invertible with inverse,
\[
    U_0^{-1}
        =   \begin{pmatrix}
                \frac{1}{\gamma} & \frac{1}{4\gamma}u \\[1ex]
                0 & I
            \end{pmatrix}.
\]
Then, since $1+\frac{c_1}{4\gamma}=\frac{1}{\gamma}$, we have
\begin{equation}
    \hat Q
        = Q U_0^{-1}
        = \frac{1}{\gamma}
            \begin{pmatrix}
                c_1 & u \\[1ex]
                u^t & \tilde Q
            \end{pmatrix},
        \qquad \tilde Q(i,j)=\frac{c_i c_j}{4}+\gamma c_{i+j-1},
        \qquad \mbox{for $2\leq i,j\leq m-1$.}
\end{equation}
With the above notation it is straightforward to check that
\begin{equation} \label{I-RQ}
    I-RQ
        = U_0 - R_1 Q
        = (I-R_1\hat Q) U_0.
\end{equation}
Hence condition (1) is equivalent to the invertibility of $(I-R_1\hat Q)$.
Assuming condition (1) and using in addition that $\hat Q$ is a symmetric
matrix and that $(I-R_1\hat Q)^{-1} = I  +  (I-R_1\hat Q)^{-1} R_1 \hat Q$ we
find
\begin{equation} \label{vQ(I-RQ)...}
    vQ(I-RQ)^{-1}v^t
        = (v\hat Q)[(I-R_1\hat Q)^{-1}R_1](v\hat Q)^t + v\hat Q v^t.
\end{equation}

\begin{remark} \label{remark: norms}
    In order to prove conditions (1) and (2) we will make use of the following
    norms. If $A$ is a $p\times p$
matrix and $x$ a row vector of size $p$, we define
    \begin{align*}
        & \|A\|_{1\to\infty} := \max_{i,j}|A_{ij}|, \qquad
            \|A\|_{\infty\to\infty} := \max_i\sum_k|A_{ik}|, \qquad
            \|A\|_{\infty\to 1} := \sum_{i,j}|A_{ij}|, \\
        & \|x\|_1 := \sum_i|x_i|,\qquad \|x\|_\infty:=\max_i|x_i|.
    \end{align*}
    Note that $\|\cdot\|_{1\to\infty}$ and
    $\|\cdot\|_{\infty\to\infty}$ are precisely the operator norms
    for linear maps $\ell_1(\mathbb R^p)\to\ell_\infty(\mathbb
    R^p)$ and $\ell_\infty(\mathbb R^p)\to\ell_\infty(\mathbb
    R^p)$, respectively, whereas $\|\cdot\|_{\infty\to 1}$ is
    merely an upper bound on the operator norm for linear maps
    $\ell_\infty(\mathbb R^p)\to\ell_1(\mathbb R^p)$. These
observations imply
    the following inequalities, which are readily verified:
    \begin{align*}
        & \|AB\|_{\infty\to\infty} \leq
                \|B\|_{\infty\to 1}\|A\|_{1\to\infty},
            && \|AB\|_{1\to\infty} \leq
                    \|B\|_{1\to \infty}\|A\|_{\infty\to\infty}, \\[1ex]
        &  |xAx^t| \leq \|A\|_{1\to\infty}\|x\|_1^2,
            && \|AB\|_{\infty\to\infty} \leq
                    \|A\|_{\infty\to \infty}\|B\|_{\infty\to\infty}.
    \end{align*}
\end{remark}

The following two Lemmas are the key ingredients in proving that conditions (1)
and (2) are satisfied.

\begin{lemma} \label{lemma1: invertibility}
    The functions $I$ and $\widehat I$ defined by {\rm (\ref{definition: I})} and
    {\rm (\ref{definition: Ihat})}, respectively, satisfy for all $m\geq 2$ and $q\geq 1$,
    \begin{itemize}
        \item[(a)] $|I(q)-\frac{1}{2}\delta_{1,q}| \leq \frac{D}{2m}$
            with $D=2.22$
        \item[(b)] $|\widehat I(q)-\frac{1}{4}\delta_{1,q}| \leq \frac{C}{2m}$
            with $C=2.18$.
    \end{itemize}
\end{lemma}

\begin{lemma} \label{lemma2: invertibility}
    For all $m\geq 2$,
    \begin{itemize}
        \item[(a)] $\|\hat Q\|_{\infty\to 1} \leq
            m\left(\frac{\pi}{12}+\frac{1}{2}\right)$
        \item[(b)] $\|v\hat Q\|_1\leq 0.3918\sqrt m$
        \item[(c)] $v\hat Q v^t <\frac{1}{\sqrt{2m-1}}$.
    \end{itemize}
\end{lemma}
These Lemmas will be proven in the next two subsections.

\begin{proof}[Proof of conditions (1) and (2).]
    In order to prove condition (1), it follows from (\ref{I-RQ}) that we need
    to show that $I-R_1\hat Q$ is invertible. This is done 
by proving that
    $\|R_1\hat Q\|_{\infty\to\infty}<1$. From the definition of $R_1$ and from
    Lemma \ref{lemma1: invertibility}(b) it follows that
    $\|R_1\|_{1\to\infty} \leq \frac{C}{2m}$. From Remark \ref{remark: norms}
    and Lemma \ref{lemma2: invertibility}(a) we then conclude
    \begin{equation}
        \|R_1\hat Q\|_{\infty\to\infty}
            \leq \|\hat Q\|_{\infty\to 1} \|R_1\|_{1\to\infty}
            \leq \frac{C}{2}\left(\frac{\pi}{12}+\frac{1}{2}\right)\leq 0.381C
            < 1.
    \end{equation}
    This proves that condition (1) is satisfied. Moreover we obtain the bound
    \begin{equation} \label{proof: conditions (1) and (2): eq1}
        \|(I-R_1\hat Q)^{-1}\|_{\infty\to\infty} \leq \frac{1}{1-0.381C}.
    \end{equation}
    It remains to prove condition (2). From equation (\ref{vQ(I-RQ)...}) and
    Lemma \ref{lemma2: invertibility}(c) it suffices to show that
    \[
        |(v\hat Q) [(I-R_1\hat Q)^{-1}R_1](v\hat Q)^t| \leq 1.
    \]
    Using Remark \ref{remark: norms}, equation
    (\ref{proof: conditions (1) and (2): eq1}) and Lemma
    \ref{lemma2: invertibility}(b) we obtain
    \begin{align}
        \nonumber
        |(v\hat Q) [(I-R_1\hat Q)^{-1}R_1](v\hat Q)^t|
            & \leq \|(I-R_1\hat Q)^{-1}R_1\|_{1\to\infty} \|v\hat Q\|_1^2
        \\[2ex]
        \nonumber
            & \leq \|R_1\|_{1\to\infty} \|(I-R_1\hat Q)^{-1}\|_{\infty\to\infty}
                \|v\hat Q\|_1^2
        \\[2ex]
            & \leq \frac{C}{2}\frac{1}{1-0.381C}0.3918^2 < 1.
    \end{align}
    Hence condition (2) is satisfied as well.
\end{proof}
Thus the invertibility of $T_m$ follows from Lemmas \ref{lemma1:
invertibility} and \ref{lemma2: invertibility}. In the remainder of
this Section we will prove that these two Lemmas are true.

\subsection{Proof of Lemma \ref{lemma1: invertibility}}
\label{sec3.2}

Our proof follows the corresponding parts of
\cite[Section 6]{DeiftGioev} and its improved version
in \cite{CostinDeiftGioev}.
Define for $x\in[0,1]$ the auxiliary function $u$ as,
\begin{equation} \label{definition: u}
    u(x)=\frac{1}{h(x^2)}-\frac{1-x^2}{2}+\frac{1}{4m},
\end{equation}
where $h(x)=\frac{4m}{2m-1}\ _2F_1(1,-m+1;-m+3/2;x)$.  Note that
this function $u$ coincides with the function $u$ defined in
\cite[(16)]{CostinDeiftGioev}. We will  use
the following result.
\begin{proposition} {\rm (\cite[Lemma 3]{CostinDeiftGioev})}
 \label{proposition: u}
    For all $m\geq 2$ the following holds.
    \begin{itemize}
        \item[(a)] There exists $x_m\in(0,1)$ such that $u'<0$ on $[0,x_m)$
            and $u'>0$ on $(x_m,1]$.
        \item[(b)] $u(0)=0$, $u(1)=\frac{1}{2m}$ and $u(x_m)>-\frac{1}{4m}$.
    \end{itemize}
\end{proposition}

\subsubsection{Part (a) of Lemma \ref{lemma1: invertibility}}

    In order to analyze $I(q)$ defined by (\ref{definition: I}),
    we apply the substitution
    $\theta=\frac{1}{2}\arccos(2x-1)$ and use (\ref{definition: u}) to arrive at
    \[
        I(q)= \frac{4}{\pi} \int_0^{\frac{\pi}{2}}V_q(\theta)\frac{1}{h(\cos^
        2\theta)} = \frac{4}{\pi}\int_0^{\frac{\pi}{2}}V_q(\theta)\left(u(\cos\theta)
                    +\frac{\sin^2\theta}{2}-\frac{1}{4m}\right)d\theta,
    \]
    where $V_q$ is the function,
    \begin{equation}\label{definition: Vq}
        V_q(\theta)\equiv\frac{\sin(2q-1)\theta}{\sin\theta}=1+2\sum_{k=1}^{q-1}\cos(2k\theta).
    \end{equation}
    Using the elementary facts,
        \[
            \int_0^{\frac{\pi}{2}}V_q(\theta)\sin^ 2\theta d\theta=\frac{\pi}{4}\delta_{1,q},
            \qquad\mbox{and}\qquad
            \int_0^{\frac{\pi}{2}}V_q(\theta)d\theta
            =\frac{\pi}{2},
        \]
    integrating by parts, using the fact that $u(0)=0$ 
(see Proposition \ref{proposition: u}) we obtain,
    \begin{align}
        \nonumber
            I(q)-\frac{1}{2}\delta_{1,q}
                & =\frac{4}{\pi}\int_0^{\pi/2}V_q(\theta)u(\cos\theta)d\theta-\frac{1}{2m}, \\[1ex]
        \label{proposition: Iq: eq1}
                & =\int_0^{\pi/2}W_q(\theta)u'(\cos\theta)\sin\theta d\theta-\frac{1}{2m},\qquad
                \mbox{for all $q\geq 1$,}
    \end{align}
    with $W_q$ the auxiliary function,
    \begin{equation}\label{definition: Wq}
        W_q(\theta)=\frac{4}{\pi}\int_0^{\theta}V_q(s)ds
            =\frac{4}{\pi}\left(\theta+\sum_{k=1}^{q-1}\frac{\sin(2k\theta)}{k}\right),
                \qquad \theta\in[0,\infty).
    \end{equation}
    Here, the expression of $W_q$ as a sum
follows from (\ref{definition: Vq}).
    In order to prove Lemma \ref{lemma1: invertibility}(a) we will make use of
    equation (\ref{proposition: Iq: eq1}), together with Proposition \ref{proposition: u}
    and the following result.

    \begin{proposition} {\rm (cf. \cite[Lemma 4]{CostinDeiftGioev})}
    \label{proposition: wq}
        Let $q\geq 1$. There exists $\theta_q\in(0,\frac{\pi}{2})$ such that the following
        holds.
        \begin{itemize}
        \item[(a)]$W_q$ is increasing on $[0,\theta_q]$ and $0\leq W_q(\theta)\leq W_q(\theta_q)=1.7$
            for $\theta\in[0,\theta_q]$.
        \item[(b)] For $\theta\in[\theta_q,\frac{\pi}{2}]$ we have $1.7\leq W_q(\theta)\leq 2.44$.
        \end{itemize}
    \end{proposition}

    \begin{proof}
        We distinguish three cases. First, in case $q=1$, we have
        $W_1(\theta)=\frac{4}{\pi}\theta$. Then the Proposition is true
        with $\theta_1=\frac{1.7\pi}{4}$. Next, consider the case
        $q=2$. It follows from (\ref{definition: Wq}) that
        $W_2(\theta)=\frac{4}{\pi}(\theta+\sin 2\theta)$ and so
        $W_2$ is increasing on $[0,\pi/3]$ and decreasing on
        $[\pi/3,\pi/2]$. Since $W_2(\pi/3)=4/3+2\sqrt
        3/\pi\in[1.7,2.44]$ and $W_2(\pi/2)=2$ we can define $\theta_2$ to be the
        unique number in $[0,\pi/3]$ such that
        $W_2(\theta_2)=1.7$.

        Finally, we prove that the 
Proposition is satisfied for $q\geq 3$ as well. Define
        a sequence $s_k=k\frac{\pi}{2q-1}$ for integers $k\geq 0$. We first
        prove that
        \begin{equation}\label{estimates: Wq(s1) and Wq(s2)}
            W_q(s_1)\leq 2.44\qquad\mbox{and}\qquad
            W_q(s_2)\geq 1.7,\qquad\mbox{for $q\geq 3$.}
        \end{equation}
        Note that
        \[
            W_q(s_1)=\frac{4}{\pi}\int_0^\pi\frac{\sin
            t}{(2q-1)\sin\left(\frac{t}{2q-1}\right)}dt,
        \]
        and that for every $t\in[0,\pi]$,
        $(2q-1)\sin\left(\frac{t}{2q-1}\right)$ increases in $q$.
        Then $W_q(s_1)$ decreases in $q$, so that for all $q\geq
        3$,
        \[
            W_q(s_1)\leq
            W_2(\pi/3)\leq
            2.44.
        \]
        We now turn to the lower estimate on $W_q(s_2)$ for $q\geq
        3$. We use
        $\sin\left(\frac{t}{2q-1}\right)\leq\frac{t}{2q-1}$ for $t \geq 0$
        and arrive at
        \begin{align*}
            W_q(s_2) &=\frac{4}{\pi}\int_0^\pi\frac{\sin
            t}{(2q-1)\sin\left(\frac{t}{2q-1}\right)}dt+\frac{4}{\pi}\int_\pi^{2\pi}\frac{\sin
            t}{(2q-1)\sin\left(\frac{t}{2q-1}\right)}dt \\[1ex]
            &\geq \frac{4}{\pi}\int_0^\pi\frac{\sin t}{t}+\frac{4}{\pi}\int_\pi^{2\pi}\frac{\sin
            t}{(2q-1)\sin\left(\frac{t}{2q-1}\right)}dt.
        \end{align*}
        Since the last integral is increasing in $q$ we then have
        for $q\geq 3$,
        \begin{align*}
            W_q(s_2)&\geq \frac{4}{\pi}\textrm{Si}(\pi)+\frac{4}{\pi}\int_\pi^{2\pi}\frac{\sin
            t}{5\sin\left(\frac{t}{5}\right)}dt=\frac{4}{\pi}\textrm{Si}(\pi)
            +\frac{4}{\pi}\int_{\pi/5}^{2\pi/5}\frac{\sin
            5t}{\sin t}dt \\[1ex]
            &=\frac{4}{\pi}\textrm{Si}(\pi)+W_3(2\pi/5)-W_3(\pi/5).
        \end{align*}
        The last quantity can be estimated from below using Si$(\pi) \geq 1.851$
        (see e.g.\ \cite{AbramowitzStegun}) and the explicit expression
        (\ref{definition: Wq}) for $W_3$. We then find that $W_q(s_2)\geq 1.7$ for all
        $q\geq 3$.

        Using (\ref{estimates: Wq(s1) and Wq(s2)}) 
we will now complete the proof of the
        Proposition. It is
        immediate that $W_q$ is increasing on $[s_{2k},s_{2k+1}]$
        and decreasing on $[s_{2k+1},s_{2k+2}]$. Furthermore, the
        monotonicity of $1/\sin\theta$ on $[0,\pi/2]$, together
        with (\ref{estimates: Wq(s1) and Wq(s2)}) implies the 
following inequalities for
        the local maxima and minima of $W_q$.
        \begin{align*}
            & 2.44\geq W_q(s_1)\geq W_q(s_3)\geq \ldots \geq
            W_q(s_{2k_1+1}),\qquad\mbox{with
            $k_1=\left[\frac{2q-3}{4}\right]$,} \\[1ex]
            & 1.7\leq W_q(s_2)\leq W_q(s_4)\leq \ldots\leq
            W_q(s_{2k_2}),\qquad\mbox{with $k_2=\left[\frac{2q-1}{4}\right]$.}
        \end{align*}
        Using in addition that $W_q(0)=0$ and that $W_q(\frac{\pi}{2})=2$
        the Proposition now follows by choosing $\theta_q$ to be the
        unique number in the interval $[0,s_1]$ satisfying
        $W_q(\theta_q)=1.7$. Such a number exists since $W_q(0) = 0 < 1.7$ and
        $W_q(s_1) \geq W_q(s_2) \geq 1.7$.
    \end{proof}

    \begin{proof}[Proof of Lemma \ref{lemma1: invertibility}(a).]
        From (\ref{proposition: Iq: eq1}) we have
        \begin{equation}\label{proof: lemma1(a): eq1}
            I(q)-\frac{1}{2}\delta_{1,q} =\int_0
            ^{\theta^*}W_q(\theta)u'(\cos\theta)\sin\theta
            d\theta+\int_{\theta^*}^{\pi/2}W_q(\theta)u'(\cos\theta)\sin\theta
            d\theta-\frac{1}{2m},
        \end{equation}
        where $\theta^*\in[0,\frac{\pi}{2}]$ is defined such that
        $\cos\theta^*=x_m$ (see Proposition
        \ref{proposition: u}).
        With this choice of $\theta^*$ we have from Proposition \ref{proposition: u}(a)
        that
        \begin{equation} \label{proof: lemma1(a): eq2}
            u'(\cos\theta)
            \begin{cases}
                >0, &\mbox{for $\theta\in[0,\theta^*]$,} \\[1ex]
                <0, &\mbox{for
                $\theta\in[\theta^*,\frac{\pi}{2}]$.}
            \end{cases}
        \end{equation}
        Since $0\leq W_q(\theta)\leq 2.44$ for all
        $\theta\in[0,\frac{\pi}{2}]$, we then obtain from (\ref{proof: lemma1(a): eq1})
        and Proposition \ref{proposition: u}(b) that,
        \[
            I(q)-\frac{1}{2}\delta_{1,q}\geq
            \int_{\theta^*}^{\pi/2}W_q(\theta)u'(\cos\theta)\sin\theta
            d\theta-\frac{1}{2m}\geq 2.44 \
            u(\cos\theta^*)-\frac{1}{2m}\geq -\frac{2.22}{2m}.
        \]
        This is the desired lower estimate. 
In order to obtain the upper estimate we distinguish two
        cases. Consider first the case that $q$ is such that $\theta^*\leq
        \theta_q$ (here $\theta^*$ is defined as above and $\theta_q$ is chosen as
        in Proposition \ref{proposition: wq}). Then, since $W_q(\theta)\leq W_q(\theta^*)\leq 1.7$ for
        $\theta\in[0,\theta^*]$ and $W_q(\theta)\geq W_q(\theta^*)$ for
        $\theta\in[\theta^*,\frac{\pi}{2}]$,
        we obtain from (\ref{proof: lemma1(a): eq1}), (\ref{proof: lemma1(a): eq2})
        and Proposition \ref{proposition: u}(b),
        \begin{align*}
            I(q)-\frac{1}{2}\delta_{1,q}
            &\leq
            W_q(\theta^*) \int_0
            ^{\theta^*}u'(\cos\theta)\sin\theta
            d\theta+W_q(\theta^*)
            \int_{\theta^*}^{\pi/2}u'(\cos\theta)\sin\theta
            d\theta-\frac{1}{2m}
            \\[1ex]
            &=-W_q(\theta^*)u(\cos\theta)\Bigl|_0^{\pi/2}-\frac{1}{2m}
            =W_q(\theta^*)\frac{1}{2m}-\frac{1}{2m} \\[1ex]
            & \leq\frac{0.7}{2m} \leq\frac{2.22}{2m}.
        \end{align*}
        Next, consider the case that $q$ is such that $\theta^*\geq\theta_q$.
        Then, since $W_q(\theta)\leq 2.44$ for
        $\theta\in[0,\theta^*]$ and  $W_q(\theta)\geq 1.7$ for
        $\theta\in[\theta^*,\frac{\pi}{2}]$, we obtain
        from (\ref{proof: lemma1(a): eq1}), (\ref{proof: lemma1(a): eq2})
        and Proposition \ref{proposition: u}(b),
        \begin{align*}
            I(q)-\frac{1}{2}\delta_{1,q}
            &\leq
            -2.44 \ u(\cos\theta)\Bigl|_0^{\theta^*}-1.7 \
            u(\cos\theta)\Bigl|_{\theta^*}^{\pi/2}-\frac{1}{2m}
            \\[1ex]
            &=u(x_m)(1.7-2.44)+\frac{2.44}{2m}-\frac{1}{2m}\leq \frac{0.74}{4m}+\frac{1.44}{2m}
            = \frac{1.81}{2m} \leq \frac{2.22}{2m}.
        \end{align*}
        This proves part (a) of Lemma \ref{lemma1: invertibility}.
    \end{proof}

    \subsubsection{Part (b) of Lemma \ref{lemma1: invertibility}}

    The proof of part (b) is analogous to the proof of part (a).
    In this case we introduce the   function
    \begin{equation}
        \hat
        V_q(\theta)\equiv\frac{\sin(2q\theta)\cos\theta}{\sin\theta}
                   = \frac{1}{2} \left(V_{q+1}(\theta) + V_q(\theta) \right).
    \end{equation}
    It is then straightforward to check that
    \begin{align}
        \nonumber
            \widehat I(q)-\frac{1}{4}\delta_{1,q}
                &= \frac{4}{\pi}\int_0^{\pi/2}\hat
                    V_q(\theta)u(\cos\theta)d\theta-\frac{1}{2m} \\[1ex]
        \label{th8}
                &= \int_0^{\pi/2}\hat
                    W_q(\theta)u'(\cos\theta)\sin\theta d\theta-\frac{1}{2m},
    \end{align}
    where $\hat W_q$ is the auxiliary function,
    \[
        \hat W_q(\theta)=\frac{4}{\pi}\int_0^\theta\hat
        V_q(s)ds,\qquad \mbox{$\theta\in[0,\infty)$},
    \]
    which satisfies the following Proposition.

    \begin{proposition}\label{proposition: Wqhat}
        Let $q\geq 1$. There exists $\theta_q\in(0,\frac{\pi}{2})$ such that the following
        holds.
        \begin{itemize}
        \item[(a)]$\hat W_q$ is increasing on $[0,\theta_q]$ and $0\leq \hat W_q(\theta)\leq \hat W_q(\theta_q)=1.7$
            for $\theta\in[0,\theta_q]$.
        \item[(b)] For $\theta\in[\theta_q,\frac{\pi}{2}]$ we have $1.7\leq \hat W_q(\theta)\leq 2.36$.
        \end{itemize}
    \end{proposition}

    \begin{proof}
        The proof is similar to the proof of Proposition \ref{proposition: wq}. Again the case $q=1$ is
        trivial since $\hat W_1$ is monotone increasing on $[0, \frac{\pi}{2}]$ with
        $\hat W_1(0)=0$ and $\hat W_1(\frac{\pi}{2})=2$.

        In order to deal with the case $q \geq 2$ we define $t_k:= k \frac{\pi}{2q}$, $k \geq 0$,
        where $\hat W_q$ attains its local extrema. Using the same arguments as in the proof
        of Proposition \ref{proposition: wq}
 (note that $1/ \tan t$ is decreasing for $t \in
        (0, \frac{\pi}{2})$) it suffices to show that the following estimates hold:
        \begin{itemize}
           \item[(i)] $\hat W_q(t_1)\leq 2.36$
           \item[(ii)] $\hat W_q(t_2) \geq 1.7$.
        \end{itemize}
        In order to prove these two claims we use 
the fact that for every $t \in [0, 2\pi)$ the value of
        $2q \tan \frac{t}{2q}$ decreases in $q$ (for $q \geq 2$) and converges to $t$ as $q$ tends
        to $\infty$. This implies that for all $q \geq 2$ we have
        \[
            \hat W_q(t_1) = \frac{4}{\pi} \int_0^{\pi}
                        \frac{\sin t}{2q \tan \frac{t}{2q}} dt \leq \frac{4}{\pi} \int_0^{\pi}
                        \frac{\sin t}{t} dt = \frac{4}{\pi} \mbox{ Si}(\pi) \leq
                         2.36,
        \]
        and
            \begin{align*}
                       \hat W_q(t_2) &= \frac{4}{\pi} \left( \int_0^{\pi}
                                          \frac{\sin t}{2q \tan \frac{t}{2q}} dt +   \int_{\pi}^{2\pi}
                                           \frac{\sin t}{2q \tan \frac{t}{2q}} dt\right)
                                           \\[1ex]
                                     &\geq \hat W_2(t_1) + \frac{4}{\pi} \int_{\pi}^{2\pi}
                                           \frac{\sin t}{t} dt
                                           = 1 + \frac{4}{\pi}
                                         \left( 1 + \mbox{ Si}(2 \pi) - \mbox{ Si}(\pi) \right) \geq 1.7.
            \end{align*}
    \end{proof}

\begin{proof}[Proof of Lemma \ref{lemma1: invertibility}(b).]
    The proof is completely analogous to the proof of part (a).
    From (\ref{th8}) we have
    \[
            \widehat I(q)-\frac{1}{4}\delta_{1,q} =\int_0
            ^{\theta^*}\hat W_q(\theta)u'(\cos\theta)\sin\theta
            d\theta+\int_{\theta^*}^{\pi/2}\hat W_q(\theta)u'(\cos\theta)\sin\theta
            d\theta-\frac{1}{2m},
    \]
    where again $\theta^*\in[0,\frac{\pi}{2}]$ is defined such that
    $\cos\theta^*=x_m$ (see Proposition \ref{proposition: u}).
    Using Propostion \ref{proposition: Wqhat} with the corresponding
    choice of $\theta_q$ we obtain the lower estimate
        \[
            \widehat I(q)-\frac{1}{4}\delta_{1,q}\geq
            \int_{\theta^*}^{\pi/2}\hat W_q(\theta)u'(\cos\theta)\sin\theta
            d\theta-\frac{1}{2m}\geq 2.36 \
            u(\cos\theta^*)-\frac{1}{2m}\geq -\frac{2.18}{2m}.
        \]
        In order to obtain the upper estimate we distinguish two
        cases. For $\theta^*\leq
        \theta_q$ we have
        \begin{align*}
            \widehat I(q)-\frac{1}{4}\delta_{1,q}
            &\leq
            \hat W_q(\theta^*) \int_0
            ^{\theta^*}u'(\cos\theta)\sin\theta
            d\theta+\hat W_q(\theta^*)
            \int_{\theta^*}^{\pi/2}u'(\cos\theta)\sin\theta
            d\theta-\frac{1}{2m}
            \\[1ex]
            &
            =\hat W_q(\theta^*)\frac{1}{2m}-\frac{1}{2m}
             \leq \frac{0.7}{2m}  \leq \frac{2.18}{2m}.
        \end{align*}
        For $\theta^*\geq \theta_q$ we have

        \begin{align*}
            \widehat I(q)-\frac{1}{4}\delta_{1,q}
            &\leq
            -2.36 \ u(\cos\theta)\Bigl|_0^{\theta^*}-1.7 \
            u(\cos\theta)\Bigl|_{\theta^*}^{\pi/2}-\frac{1}{2m}
            \\[1ex]
            &=u(x_m)(1.7-2.36)+\frac{2.36}{2m}-\frac{1}{2m}\leq \frac{0.66}{4m}+\frac{1.36}{2m}
            = \frac{1.69}{2m} \leq \frac{2.18}{2m}.
        \end{align*}
        This completes the proof of Lemma \ref{lemma1: invertibility}.
    \end{proof}

    \subsection{Proof of Lemma \ref{lemma2: invertibility}}
        \label{subsection: proof invertibility: third part}

    \subsubsection{Part (a) of Lemma \ref{lemma2: invertibility}}

    We start by introducing the convenient notation
    $d_k=\sum_{j=k+1}^{m-1}c_j$ for $k=0,\ldots ,m-1$,
  $d_{m-1}\equiv 0$ (cf. \eqref{definition: cl}).
    We state the following technical Proposition.

    \begin{proposition} \label{proposition: c1 sum dj}
        For all $m\geq 2$,
        \begin{align}
            \label{estimate: c1 gamma}
            & c_1=\frac{2m-2}{2m-1}<1, \qquad\mbox{and}\qquad
                \gamma\equiv1-\frac{c_1}{4}>\frac{3}{4}, \\[1ex]
            \label{sum dj}
            & \sum_{j=0}^{m-1}d_j=\frac{m}{2}c_1, \\[1ex]
            \label{estimate: d0}
            & \frac{1}{2}\sqrt{m\pi}-1 \leq d_0 \leq \frac{1}{2}\sqrt{m\pi}.
        \end{align}
    \end{proposition}

    \begin{proof}
        By definition, we have
        \[
            c_1 = \frac{2^{2-2m}}{A_m}\binom{2m-2}{m-2}
                = 2^{2-2m}\frac{(2m-2)!}{(m-2)!m!} \prod_{j=1}^m\frac{2j}{2j-1}.
        \]
        Now, since
        \[
            \frac{(2m-2)!}{\prod_{j=1}^m (2j-1)} = 2^{m-1}\frac{(m-1)!}{2m-1},
            \qquad \mbox{and} \qquad
            \frac{\prod_{j=1}^m 2j}{m!}=2^m,
        \]
        we obtain
        $ c_1=\frac{2m-2}{2m-1}<1$ and hence
        $\gamma>\frac{3}{4}$. Hence the first part of the Proposition is proved.

        In order to prove the second part, we observe that by
        definition $\sum_{j=0}^{m-1}d_j =\sum_{j=1}^{m-1}jc_j$. This implies
          that,
        \[
            \sum_{j=0}^{m-1}d_j =
                \frac{2^{2-2m}}{A_m}\sum_{k=0}^{m-2}(m-1-k) \binom{2m-2}{k}.
        \]
        Since
        \[
            (k+1)\binom{2m-2}{k+1}-k\binom{2m-2}{k}=2(m-1-k)\binom{2m-2}{k},
        \]
        we arrive at
        \begin{align*}
            \sum_{j=0}^{m-1}d_j
                & = \frac{2^{2-2m}}{2A_m}
                \sum_{k=0}^{m-2}\left[(k+1)\binom{2m-2}{k+1}-k\binom{2m-2}{k}\right]
            \\[2ex]
                & = \frac{2^{2-2m}}{2A_m}(m-1)\binom{2m-2}{m-1}
                = \frac{m}{2}\frac{2^{2-2m}}{A_m}\binom{2m-2}{m-2}=\frac{m}{2}c_1.
        \end{align*}
        This proves the second part of the Proposition.

        It now remains to prove the last part of the Proposition.
        First, we will derive a convenient expression
        for $d_0$. Since $\sum_{j=0}^{2m-2}\binom{2m-2}{j}=2^{2m-2}$ we have
        \begin{align*}
            d_0 &=\frac{2^{2-2m}}{A_m}\sum_{k=0}^{m-2}\binom{2m-2}{k}
                =\frac{2^{2-2m}}{2A_m}\left[2^{2m-2}-\binom{2m-2}{m-1}\right]
            \\[1ex]
            &= \frac{1}{2A_m}-\frac{2^{2-2m}}{2A_m}\binom{2m-2}{m-1}
                     =\frac{1}{2A_m}-\frac{1}{2}\frac{m}{m-1}c_1.
        \end{align*}
        Using the definition of $A_m$ we obtain
        \begin{equation}\label{proof: proposition: lemma2(a) invertibility: eq1}
            d_0=\frac{\sqrt\pi}{2}\frac{\Gamma(m+1)}{\Gamma(m+1/2)}-\frac{m}{2m-1}.
        \end{equation}
        Next, we note the following estimate for   the quotient of Gamma functions 
        \begin{eqnarray}
        \label{equation: stirling}
            0 \leq \ln\Gamma(z) - (z-\frac{1}{2})\ln
            z+z-\frac{1}{2}\ln(2\pi) \leq \frac{1}{12 z},
        \end{eqnarray}
        for $z > 1$ (see e.g. \cite[(6.1.42)]{AbramowitzStegun}). Thus
        \begin{align*}
            \ln\Gamma(m+1) &\geq(m+\frac{1}{2})\ln(m+1)-(m+1)+\frac{1}{2}\ln(2\pi)
            \\[1ex]
            \ln\Gamma(m+\frac{1}{2}) &\leq
            m\ln(m+\frac{1}{2})-(m+\frac{1}{2})+\frac{1}{2}\ln(2\pi)+\frac{1}{12(m+\frac{1}{2})},
        \end{align*}
        so that
        \[
            \ln\Gamma(m+1)-\ln\Gamma(m+\frac{1}{2})\geq\frac{1}{2}\ln
            m+m\ln\left(\frac{m+1}{m+\frac{1}{2}}\right)-\frac{1}{2}-\frac{1}{12
            m}.
        \]
        Further, since
        \[
            \ln\left(\frac{m+1}{m+\frac{1}{2}}\right)\geq
            \frac{1}{2m+1}-\frac{1}{2(2m+1)^2},
        \]
        we arrive at
        \begin{align*}
            \ln\Gamma(m+1)-\ln\Gamma(m+\frac{1}{2})&\geq\frac{1}{2}\ln
            m+\frac{m}{2m+1}-\frac{1}{2}-\frac{m}{2(2m+1)^2}-\frac{1}{12
            m}\\[1ex]
            &\geq \frac{1}{2}\ln m-\frac{1}{2m}.
        \end{align*}
        Therefore,
        \[
            \frac{\Gamma(m+1)}{\Gamma(m+\frac{1}{2})}\geq\sqrt m
            e^{-\frac{1}{2m}}\geq\sqrt m (1-\frac{1}{2m})=\sqrt
            m-\frac{1}{2\sqrt m}.
        \]
        Inserting this inequality
 into (\ref{proof: proposition: lemma2(a) invertibility: eq1}) we then have
        \[
            d_0\geq\frac{\sqrt{\pi m}}{2}-\frac{\sqrt\pi}{4\sqrt
            m}-\frac{m}{2m-1}\geq\frac{\sqrt{\pi
            m}}{2}-1,\qquad\mbox{for $m\geq 2$.}
        \]
        In order to prove the upper
        bound we deduce from (\ref{equation: stirling}) that
          \begin{align*}
            \ln\Gamma(m+1)-\ln\Gamma(m+\frac{1}{2})&\leq\frac{1}{2}\ln
            (m+1)+m\ln\left(\frac{m+1}{m+\frac{1}{2}}\right)-\frac{1}{2}+\frac{1}{12
            m}.
        \end{align*}
      Using
      \[
         \ln\left(\frac{m+1}{m+\frac{1}{2}}\right) \leq \frac{1}{2m+1} \quad \mbox{ and }
         \quad \ln(m+1) \leq \ln m + \frac{1}{m}
      \]
      we obtain for $m \geq 2$ 
         \[
              \frac{\Gamma(m+1)}{\Gamma(m+\frac{1}{2})}\leq\sqrt m
              e^{\frac{2}{5m}}.
          \]
      The claim then follows from (\ref{proof: proposition: lemma2(a) invertibility: eq1}) and from the inequalities
      \[
          \frac{\sqrt{\pi m}}{2}\left(e^{\frac{2}{5m}} -1\right) - \frac{m}{2m-1}
          \leq \frac{1}{2}\left(\sqrt{\pi m} e^{\frac{1}{5}} \frac{2}{5m} -1 \right)
          \leq \frac{1}{2}\left(\sqrt{\frac{\pi}{2}} e^{\frac{1}{5}} \frac{2}{5} -1 \right) < 0
      \]
      for $m \geq 2$.

    \end{proof}

    The next result will be used in the 
proofs of all parts of Lemma \ref{lemma2: invertibility}.

    \begin{proposition}\label{proposition: normQhat}
        The following exact relation holds,
        \begin{equation}\label{proposition: normQhat: eq1}
            \|\hat Q\|_{\infty\to 1}=\frac{d_0^2}{4\gamma}+\frac{m}{2}c_1.
        \end{equation}
    \end{proposition}

    \begin{proof}
        A straightforward calculation, using (\ref{sum dj}), shows that
        \[
            \|\hat Q\|_{\infty\to 1} = \frac{1}{\gamma}\left(d_0+d_1+\frac{1}{4}d_1^2
            +\gamma\sum_{j=2}^{m-1}d_j\right) =\frac{1}{\gamma}\left((1-\gamma)(d_0+d_1)+\frac{1}{4}d_1^2+
            \gamma\frac{m}{2}c_1\right).
        \]
        The result then follows from the facts that $1-\gamma=\frac{c_1}{4}$ and $d_1=d_0-c_1$.
    \end{proof}

    \begin{proof}[Proof of Lemma \ref{lemma2: invertibility}(a).]
        The first part of the Lemma follows easily from (\ref{proposition: normQhat: eq1}),
        (\ref{estimate: c1 gamma}) and (\ref{estimate: d0}).
    \end{proof}

    \subsubsection{Parts (b) and (c) of Lemma \ref{lemma2: invertibility}}

    For convenience, we will write the $(m-1)$-vector $v$ as a sum of
    two vectors $v=v^0+v^1$ with $v^0$ and $v^1$ given by,
    \begin{equation}\label{definition: v0}
        v^0 =
            \left[
                \frac{1}{2}\sqrt{\frac{m}{2m-1}}-\frac{1}{2\sqrt m},
                -\frac{1}{2\sqrt m},\ldots ,-\frac{1}{2\sqrt m}
            \right],
    \end{equation}
    and
    \begin{equation}\label{definition: v1}
        v^1 = \sqrt{\frac{m}{2m-1}}
            \left[
                I(1)-\frac{1}{2}, I(2),\ldots ,I(m-1)
        \right].
    \end{equation}
    The main feature of this splitting is that the entries of $v^0$ do not depend
    on the $I$-functions and that, by Lemma \ref{lemma1:
    invertibility}, the entries of $v^1$ can be estimated by
    \begin{equation}
        |v^1_j|\leq\frac{D}{2m} \sqrt{\frac{m}{2m-1}},\qquad
        \mbox{for all $j=1,\ldots ,m-1$}.
    \end{equation}
    Recalling that $\hat Q$ is symmetric, it is straightforward to check that we have
    the following estimates on $\|v\hat Q\|_1$ and $v\hat Q v^t$:
    \begin{align}
        \label{estimate: norm_vhatQ: eq1}
        & \|v\hat Q\|_1 \leq
            \|v^0\hat Q\|_1 + \frac{D}{2m}\sqrt{\frac{m}{2m-1}}\|\hat Q\|_{\infty\to
            1}, \\[2ex]
        \label{estimate: norm_vhatQ: eq2}
        & v\hat Q v^t \leq v^0 \hat Q (v^0)^t
        +\frac{D}{m}\sqrt{\frac{m}{2m-1}}\|v^0\hat Q\|_1 +\frac{D^2}{4m(2m-1)}\|\hat Q\|_{\infty\to
        1}.
    \end{align}
    It will turn out that we need to prove parts
 (b) and (c) of Lemma \ref{lemma2: invertibility}
    in two steps. First, we consider the case $2\leq m \leq 32$
    and we let Maple explicitly calculate the right hand sides of
    the above estimates. We then need explicit expressions for $\|v^0\hat
    Q\|_1$ and $v^0\hat Q (v^0)^t$ (recall that we already have
    an explicit expression for $\|\hat Q\|_{\infty\to 1}$).
    For the proof in the case $m\geq 33$ we will determine estimates for the right hand sides of
    (\ref{estimate: norm_vhatQ: eq1}) and (\ref{estimate: norm_vhatQ:
    eq2}). In particular we need to determine estimates on $\|v^0\hat
    Q\|_1$ and $v^0 \hat Q (v^0)^t$. In order to get a good
    estimate on $\|v^0\hat
    Q\|_1$ we will use the following Proposition.

    \begin{proposition}\label{proposition: dj over cj}
    For $j=1,\ldots ,m-1$,
    \[
        0\leq \frac{d_j}{c_j}\leq \frac{d_1}{c_1}.
    \]
    \end{proposition}

\begin{proof}
    Define $a_j=\frac{d_{m-j}}{c_{m-j}}$, for $j=1,\ldots ,m-1$.
    Since $c_{m-j}=c_{m-j+1}\frac{2m-j}{j-1}$ for $j \geq 2$ we have the
    recursion relation,
    \[
        a_j=\frac{j-1}{2m-j}(a_{j-1}+1),\quad \mbox{ for } 2 \leq j \leq m-1; \qquad \qquad a_1=0.
    \]
    We now prove that $a_j$ is increasing, which proves the Proposition.
    We prove by induction that $a_j\leq a_{j+1}$. For $j=1$ this is
    obvious. Next, suppose that it is true for $j$. Then
    \begin{align*}
        a_{j+1} &= (a_j+1)\frac{j}{2m-j-1}\leq
        (a_{j+1}+1)\frac{j}{2m-j-1} \\[1ex]
        &=a_{j+2}\frac{2m-j-2}{2m-j-1}\frac{j}{j+1}\leq a_{j+2}
    \end{align*}
    which completes the proof.
\end{proof}

    \begin{proposition}\label{proposition: lemma2(ii): invertibility}
        For $m\geq 2$,
        \begin{align}
            \label{proposition: lemma2(ii): invertibility: eq1}
            &\|v^0\hat Q\|_1=\sum_{j=1}^{m-1}\frac{c_j}{2\gamma}
                \left|\frac{1}{\sqrt m}\left(1+\frac{d_1}{4}
                +\gamma
                \frac{d_j}{c_j}\right)-\sqrt{\frac{m}{2m-1}}\right|,
                \\[2ex]
            \label{proposition: lemma2(ii): invertibility: eq2}
            &\|v^0\hat Q\|_1\leq 0.2869\sqrt m.
        \end{align}
    \end{proposition}

    \begin{proof}
        A straightforward calculation using the fact that
        $\frac{d_0}{c_1}=1+\frac{d_1}{4}+\gamma\frac{d_1}{c_1}$
        shows that the $j$-th entry of $v^0\hat Q$ is given by
        \begin{equation}\label{j-th entry of v0Qhat}
            (v^0\hat Q)_j=
            \frac{c_j}{2\gamma}\left(\sqrt{\frac{m}{2m-1}}-\frac{1}{\sqrt m}(1+\frac{d_1}{4}
            +\gamma \frac{d_j}{c_j})\right).
        \end{equation}
        This proves the first part of the Proposition.
 In order to prove the second part we
        obtain an estimate for the absolute value term in
        (\ref{proposition: lemma2(ii): invertibility: eq1}). For all $m\geq 2$ we
        have by (\ref{estimate: d0}) and Proposition \ref{proposition: dj over cj} that
        \begin{align*}
            \frac{1}{\sqrt m}\left(1+\frac{d_1}{4}
            +\gamma \frac{d_j}{c_j}\right)-\sqrt{\frac{m}{2m-1}} &\leq \frac{1}{\sqrt m}\left(1+\frac{d_1}{4}
        +\gamma
        \frac{d_1}{c_1}\right)-\sqrt{\frac{m}{2m-1}} \\[1ex]
        &= \frac{1}{\sqrt m}\frac{d_0}{c_1}-\sqrt{\frac{m}{2m-1}}
        \\[1ex]
        &\leq \begin{cases}
                {\displaystyle \frac{\sqrt \pi}{2}\frac{2m-1}{2m-2}-\frac{1}{\sqrt
                2}\leq 0.41}, & \mbox{for $m\geq 3$.} \\[2ex]
                0, & \mbox{for $m=2$,}
                \end{cases}
        \end{align*}
and
\begin{align*}
    \frac{1}{\sqrt m}\left(1+\frac{d_1}{4}
        +\gamma
        \frac{d_j}{c_j}\right)-\sqrt{\frac{m}{2m-1}} &\geq
            \frac{1}{\sqrt m}\left(1+\frac{d_1}{4}\right)
            -\sqrt{\frac{m}{2m-1}}\\[1ex]
            &=\frac{1}{\sqrt
            m}\left(\gamma+\frac{d_0}{4}\right)-\sqrt{\frac{m}{2m-1}}\\[1ex]
            &\geq\frac{\sqrt
            \pi}{8}-\sqrt{\frac{m}{2m-1}}+\frac{1}{2\sqrt m}
            \geq \frac{\sqrt \pi }{8}-\frac{1}{\sqrt 2}.
\end{align*}
This then implies by (\ref{proposition: lemma2(ii): invertibility:
eq1}) that,
\begin{equation}
    \|v^0\hat Q\|_1 \leq \frac{d_0}{2\gamma}
    \left(\frac{1}{\sqrt
        2}-\frac{\sqrt \pi }{8}\right)\leq \frac{\sqrt \pi}{3}\left(\frac{1}{\sqrt
        2}-\frac{\sqrt \pi }{8}\right)\sqrt m \leq 0.2869\sqrt m,\qquad\mbox{for $m\geq 2$,}
\end{equation}
and the Proposition is proved.
    \end{proof}
    \begin{proposition}\label{proposition: lemma2(iii): invertibility}
        For $m\geq 2$,
        \begin{equation}\label{proposition: lemma2(iii): invertibility: eq1}
            v^0\hat Q (v^0)^t=-\frac{1}{2\gamma}\frac{d_0}{\sqrt
            m}\sqrt{\frac{m}{2m-1}}+\frac{1}{2\gamma}\frac{m(m-1)}{(2m-1)^2}+\frac{c_1}{8}+\frac{d_0^2}{16\gamma
            m}.
        \end{equation}
        Further,
        \begin{equation}\label{proposition: lemma2(iii): invertibility: eq2}
            v^0\hat Q (v^0)^t\leq
            \frac{0.246}{\sqrt{2m-1}},\qquad\mbox{for $m\geq 33$.}
        \end{equation}
    \end{proposition}

    \begin{proof}
        From (\ref{j-th entry of v0Qhat}), (\ref{definition: v0}) and the fact that
        $\frac{d_0}{c_1}=1+\frac{d_1}{4}+\gamma\frac{d_1}{c_1}$ it follows that
        \begin{align*}
            v^0\hat Q (v^0)^t
                &=\sum_{j=1}^{m-1} \frac{c_j}{2\gamma}\left(\sqrt{\frac{m}{2m-1}}-\frac{1}{\sqrt m}(1+\frac{d_1}{4}
                    +\gamma \frac{d_j}{c_j})\right)v_j^0 \\[2ex]
                &= \frac{c_1}{2\gamma}\left(\sqrt{\frac{m}{2m-1}}-\frac{1}{\sqrt m}\frac{d_0}{c_1}\right)
                    \frac{1}{2}\sqrt{\frac{m}{2m-1}}\\[2ex]
                &\qquad\qquad-\frac{1}{4\gamma}\frac{1}{\sqrt
                     m}\sum_{j=1}^{m-1} \left(c_j\sqrt{\frac{m}{2m-1}}-\frac{1}{\sqrt m}(c_j+c_j\frac{d_1}{4}
                    +\gamma d_j)\right).
        \end{align*}
        Now, from (\ref{sum dj}) and from the fact that
        $1-\gamma=\frac{c_1}{4}$ we have
        \begin{align*}
            \sum_{j=1}^{m-1}\left(c_j+c_j\frac{d_1}{4}
                    +\gamma d_j\right) &=
                    d_0+\frac{1}{4}d_0d_1+\gamma\left(\frac{m}{2}c_1-d_0\right)
                    \\[1ex]
                    &= \gamma\frac{m}{2}c_1+d_0\left(1+\frac{d_1}{4}-\gamma\right)
                    =\gamma\frac{m}{2}c_1+\frac{d_0^2}{4}.
        \end{align*}
        We obtain
        \[
            v^0\hat Q (v^0)^t = \frac{c_1}{4\gamma}\frac{m}{2m-1}-\frac{1}{2\gamma}\frac{d_0}{\sqrt
        m}\sqrt{\frac{m}{2m-1}}+\frac{1}{4\gamma
        m}\left(\gamma\frac{m}{2}c_1+\frac{d_0^2}{4}\right).
        \]
        The first part of the Proposition then follows from (\ref{estimate: c1
        gamma}). Next, from (\ref{proposition: lemma2(iii): invertibility: eq1}),
        (\ref{estimate: c1 gamma}), (\ref{estimate: d0}) and from the fact that
        $\frac{m(m-1)}{(2m-1)^2}<\frac{1}{4}$ we have
\begin{align*}
            v^0\hat Q (v^0)^t-\frac{0.246}{\sqrt{2m-1}} & \leq
            -\frac{\sqrt\pi}{4\gamma}\sqrt{\frac{m}{2m-1}}+\frac{1}{2\gamma\sqrt{2m-1}}
            +\frac{1}{6}+\frac{1}{8}+\frac{\pi}{64\gamma}-\frac{0.246}{\sqrt{2m-1}} \\[1ex]
        &\leq -\frac{\sqrt\pi}{4\sqrt 2\gamma}
        +\frac{7}{24}+\frac{\pi}{48}+\frac{2/3-0.246}{\sqrt{2m-1}} \\[1ex]
        & < 0,\qquad \mbox{for $m\geq 33$.}
    \end{align*}
    In the last inequality we have used the fact that $\gamma\leq 0.754$ for $m\geq 33$.
    \end{proof}

    \begin{proof}[Proof of Lemma \ref{lemma2: invertibility} (b) and (c).]
        First, consider the case $2\leq m\leq 32$. From (\ref{estimate: norm_vhatQ:
        eq1}), (\ref{proposition: lemma2(ii): invertibility: eq1}) and
        (\ref{proposition: normQhat: eq1}) we obtain,
        \[
            \|v\hat Q\|_1\leq \sum_{j=1}^{m-1}\frac{c_j}{2\gamma}
                \left|\frac{1}{\sqrt m}\left(1+\frac{d_1}{4}
                +\gamma
                \frac{d_j}{c_j}\right)-\sqrt{\frac{m}{2m-1}}\right|+ \frac{D}{2m}\sqrt{\frac{m}{2m-1}}
                \left(\frac{d_0^2}{4\gamma}+\frac{m}{2}c_1\right),
        \]
        and from (\ref{estimate: norm_vhatQ: eq2}), (\ref{proposition: lemma2(iii): invertibility: eq1}),
        (\ref{proposition: lemma2(ii): invertibility: eq2}) and (\ref{proposition: normQhat: eq1})
        we obtain
        \begin{multline*}
            v\hat Qv^t -\frac{1}{\sqrt{2m-1}}\leq -\frac{d_0}{2\gamma}\frac{1}{\sqrt
            {2m-1}}+\frac{1}{2\gamma}\frac{m(m-1)}{(2m-1)^2}+\frac{c_1}{8}+\frac{d_0^2}{16\gamma
            m} \\[2ex] +(0.2869 D-1)\frac{1}{\sqrt{2m-1}}
            +\frac{D^2}{4m(2m-1)}\left(\frac{d_0^2}{4\gamma}+\frac{m}{2}c_1\right).
        \end{multline*}
        We now let Maple calculate explicitly the right hand sides of
        these estimates for $2\leq m \leq 32$, and we see that the Lemma is
        indeed satisfied in this case.

        Next, we consider the case $m\geq 33$.
        From equations (\ref{estimate: norm_vhatQ: eq1}) and
        (\ref{proposition: lemma2(ii): invertibility: eq2})
        and from Lemma \ref{lemma2: invertibility}(a) we have
        \[
            \|v\hat Q\|_1\leq
            \left[0.2869+\frac{D}{2\sqrt{2m-1}}\left(\frac{\pi}{12}+\frac{1}{2}\right)\right]\sqrt
            m\leq 0.3918\sqrt m,\qquad\mbox{for $m\geq 33$.}
        \]
        Further, from (\ref{estimate: norm_vhatQ: eq2}), (\ref{proposition: lemma2(ii): invertibility:
        eq2}), (\ref{proposition: lemma2(iii): invertibility: eq2})
        and Lemma \ref{lemma2: invertibility}(a) it follows that
        \begin{align*}
            v\hat Q v^t &\leq \left[0.246+0.2869 D+
            \frac{D^2}{4\sqrt{2m-1}}\left(\frac{\pi}{12}+\frac{1}{2}\right)\right]\frac{1}{\sqrt{2m-1}}
            \\[2ex]
            &< \frac{1}{\sqrt{2m-1}},\qquad\mbox{for $m\geq 33$}.
        \end{align*}
       This concludes the proof of Lemma \ref{lemma2: invertibility}.
    \end{proof}

\section{Asymptotics of $\phi_n$, $\psi_1$ and $\psi_2$ on the
positive real line}
    \label{section: asymptotics polynomials}

The goal of this section is to derive the leading order behavior and error
bounds for the functions $\phi_n$, $\psi_1$ and $\psi_2$ which appear in the
basis of $\mathfrak g_1$, $\mathfrak g_2$ (see Lemma \ref{lemma: basis g}).
These results are stated in Lemmas \ref{lemma: asymptotics
phin}--\ref{lemma: asymptotics psi: exponential} below. They will be used in
the subsequent Section \ref{section: asymptotics B12 matrix} to determine the
asymptotic behavior of the matrix $B$ defined by (\ref{definition:
B12-matrix}). We present our results for the rescaled functions
\begin{equation}\label{definition: phinhat psirhat}
    \hat\phi_n(x)=\sqrt{\beta_n}\phi_n(\beta_n x),\qquad
    \hat\psi_r(x)=\sqrt{\beta_n}\psi_r(\beta_n x),\qquad r=1,2,
\end{equation}
where $\beta_n$ denotes the Mhaskar--Rakhmanov--Saff number (see Subsection
\ref{subsection: relevant result from v6} below). In this rescaling all zeros
of $\hat\phi_n$ lie in the interval $[0,1]$.

As is well-known in the theory of classical orthogonal polynomials, there are
different asymptotic descriptions of the orthogonal polynomials in different
parts of the complex plane. For our purposes it will suffice to consider
$\hat\phi_n$, $\hat\psi_1$, $\hat\psi_2$ on $\mathbb R_+$. We find it most
convenient for the analysis of Section \ref{section: asymptotics B12 matrix} to
split $(0,\infty)$ into four regions $(0,n^{-1}]$,
$[n^{-1},1-n^{\kappa-\frac{2}{3}}]$,
$[1-n^{\kappa-\frac{2}{3}},1+n^{\kappa-\frac{2}{3}}]$ and
$[1+n^{\kappa-\frac{2}{3}},\infty)$, which are called the Bessel-, bulk-, Airy-
and exponential regions, respectively. Here, $\kappa$ could be any sufficiently
small positive constant. To be definite we choose once and for all,
\begin{equation}\label{kappa_defn}
    \kappa=\frac{1}{12}.
\end{equation}

The results of this section are corollaries of \cite{v6}, where the asymptotic
behavior of orthogonal polynomials of Laguerre type has been derived. For the
convenience of the reader we summarize the relevant results 
from \cite{v6} in Subsection
\ref{subsection: relevant result from v6}. After some auxiliary considerations
in Subsection \ref{subsection: auxiliary results} we then derive the asymptotic
description for $\hat\phi_n$ in Subsection \ref{subsection: asymptotics Phin}
(Lemma \ref{lemma: asymptotics phin}) and for $\hat\psi_r$ ($r=1,2$) in
Subsection \ref{subsection: asymptotics Psir} (Lemmas \ref{lemma: asymptotics
psi: bessel}--\ref{lemma: asymptotics psi: exponential}).

\subsection{Relevant results from \cite{v6}}
    \label{subsection: relevant result from v6}

In order to describe the asymptotics of the functions $\hat\phi_n$ and
$\hat\psi_r$ ($r=1,2$) on $\mathbb R_+$ we first introduce the sequence of
Mhaskar--Rakhmanov--Saff numbers, which we denote by $\beta_n$.
For $V$ as in \eqref{def:V}, these numbers are
uniquely determined for $n$ sufficiently large by the equation,
cf.~\cite[(2.1)]{v6}
\begin{equation}
\label{defrel: MRS}
    \frac{1}{2\pi}\int_0^{\beta_n}V'(x)\sqrt{\frac{x}{\beta_n-x}}dx=n,
\end{equation}
and they have a convergent power series expansion
of the form, cf.~\cite[Proposition
3.4]{v6}
\begin{equation}\label{definition: betan auxiliary results}
    \beta_n= n^{1/m}\sum_{k=0}^\infty\beta^{(k)}n^{-k/m},\qquad \beta^{(0)}=(\frac{1}{2}mq_m
    A_m)^{-1/m},\qquad A_m=\prod_{j=1}^m\frac{2j-1}{2j}.
\end{equation}
Next, we introduce the equilibrium measure $\mu_n$ on $[0,\infty)$ in the
presence of the rescaled external
field $V_n(x)=\frac{1}{n}V(\beta_n x)$. This measure 
is absolutely continuous with respect to Lebesgue measure and its density 
$\omega_n$ is given by, cf.~\cite[Proposition 3.12]{v6}
\begin{equation} \label{def: equmeas}
    \omega_n(x) = \frac{d\mu_n}{dx}(x) =\frac{1}{2\pi}\sqrt{\frac{1-x}{x}}h_n(x)\chi_{(0,1]},
\end{equation}
where $h_n(x)=\sum_{k=0}^{m-1}h_{n,k}x^k$ is a real polynomial of degree $m-1$,
and satisfies
\begin{equation}\label{property: equilibrium measure}
    \int_0^1\sqrt\frac{1-s}{s}h_n(s)ds=2\pi.
\end{equation}
The coefficients $h_{n,k}$ can be expanded to any order in powers of
$n^{-1/m}$. In particular, to any order $q=1,2,\ldots$, as $n\to\infty$, we
have uniformly for $x$ in compact sets
\begin{equation}\label{property: hn: auxiliary results}
    h_n(x)=h(x)+\sum_{k=1}^q h_{(k)}(x) n^{-k/m}+\bigO(n^{-(q+1)/m}),
\end{equation}
where $h$ is given by (\ref{definition: h}),  cf.~\cite[Proposition 3.9 and
Remark 3.10]{v6}. Furthermore, there exists a constant $h_0>0$ such that
$h_n(x)\leq h_0$ for all $n$ sufficiently large and $x\in[0,\infty)$,
cf.~\cite[Proposition 3.9]{v6}.

Let $f_n$ and $\tilde f_n$ be the biholomorpic maps (near 1 and 0, resp.)
 as defined
in \cite[Remark 3.20]{v6} and \cite[Remark 3.26]{v6}, respectively. These maps
are of the form
\begin{equation}\label{definition: fn fntilde}
    f_n(x) = c_n n^{2/3}(x-1) \hat f_n(x),\qquad \mbox{and}\qquad
    \tilde f_n(x) = -\tilde c_n n^2 x \hat{\tilde f}_n(x),
\end{equation}
where $\hat f_n$ and $\hat{\tilde f}_n$ are real analytic near 1 and 0,
respectively, satisfying for $n$ sufficiently large, cf.~\cite[Remarks 3.20 and
3.26]{v6}
\begin{align}
    & |\hat f_n(z)-1|\leq C|z-1|,\qquad \mbox{for $|z-1|$ small,}\\
    & |\hat{\tilde f}_n(z)-1|\leq C|z|,\qquad\mbox{for $|z|$ small,}
\end{align}
for some constant $C>0$. The numbers $c_n$ and $\tilde c_n$ are
given by,
cf.~\cite[Remarks 3.20, 3.26 and 2.2]{v6}
\begin{align}\label{definition: cn auxiliary results}
    c_n &= \bigl(\frac{1}{2}h_n(1)\bigr)^{2/3}=\sum_{k=0}^\infty c^{(k)}n^{-k/m},\qquad c^{(0)}=(2m)^{2/3},\\[1ex]
    \label{definition: cntilde auxiliary results}
    \tilde c_n &= \bigl(\frac{1}{2}h_n(0)\bigr)^2=\sum_{k=0}^\infty \tilde c^{(k)}n^{-k/m},\qquad
    \tilde c^{(0)}=\left(\frac{2m}{2m-1}\right)^2.
\end{align}

Further, we will need the conformal map $\varphi$ from $\mathbb
C\setminus[0,1]$ onto the exterior of the unit circle, cf.~\cite[(2.11)]{v6}
\[
    \varphi(z)=2(z-1/2)+2z^{1/2}(z-1)^{1/2},\qquad\mbox{for $z\in\mathbb
    C\setminus[0,1]$}.
\]

For notational convenience, we also introduce for
$z\in\mathbb{C}\setminus((-\infty,0]\cup[1,\infty))$ and $j=1,2$, the scalar
functions, cf.~\cite[(5.3) and (5.13)]{v6}
\begin{align}
    & \eta_j(z)=\frac{1}{2}(\alpha\pm 1)\arccos(2z-1),
    \\[1ex]
    &\zeta_j(z)=\eta_j(z)-\frac{\pi\alpha}{2}.
\end{align}
Here and below, the $+$ sign in $\pm$ holds
 for $\eta_1$ whereas the $-$ sign holds for
$\eta_2$. The function $\arccos z$ is defined as the inverse function of $\cos
z: \{0<\Re z<\pi\}\to
\mathbb{C}\setminus\left((-\infty,-1]\cup[1,\infty)\right)$. Further, introduce
for $j=1,2$,
\begin{equation}\label{definition: Fnj}
    F_{n,j}(x) =-\frac{n}{2}\int_1^x\sqrt{\frac{1-s}{s}}h_n(s)ds
        +\eta_j(x)-\frac{\pi}{4},\qquad \mbox{for $x\in[0,1]$.}
\end{equation}
Throughout the rest of this paper we denote $F_{n,1}$ 
by $F_n$  for brevity.

\begin{theorem} {\rm (\cite[Theorem 2.4]{v6})}
    The functions $\hat\phi_n(x)=\sqrt{\beta_n}\phi_n(\beta_n x)$ have the
    following asymptotic behavior on the positive real line as $n\to\infty$.
    There exists $\delta>0$ (sufficiently small) such that:
    \begin{itemize}
        \item[(i)] Uniformly for $x\in(0,\delta]$,
            \begin{multline}\label{theorem recall: asymptotics phin: bessel}
                \hat\phi_n(x) =
                        (-1)^n \frac{\sqrt 2(-\tilde f_n(x))^{1/4}}{x^{1/4}(1-x)^{1/4}}
                        \left[ \sin\zeta_1(x) J_\alpha\left(2(-\tilde f_n(x))^{1/2}\right)
                        (1+\bigO(1/n)) \right.
                \\
                        \left. +\, \cos\zeta_1(x) J_\alpha'\left(2(-\tilde f_n(x))^{1/2}\right)
                        (1+\bigO(1/n)) \right].
        \end{multline}
        \item[(ii)] Uniformly for $x\in[\delta,1-\delta]$,
            \begin{equation}\label{theorem recall: asymptotics phin: bulk}
                \hat\phi_n(x) =
                        \sqrt{\frac{2}{\pi}}\frac{\cos F_n(x)}{x^{1/4}(1-x)^{1/4}}
                        + \bigO\left(\frac{1}{nx^{1/4}(1-x)^{1/4}}\right),
            \end{equation}
            where $F_n=F_{n,1}$ is defined by {\rm (\ref{definition: Fnj})}.
        \item[(iii)] Uniformly for $x\in[1-\delta,1+\delta]$,
            \begin{multline}\label{theorem recall: asymptotics phin: airy}
                \hat\phi_n(x) =
                        \frac{\sqrt{2}}{x^{1/4}}\left[\cos\eta_{1}(x)
                        \left|\frac{f_n(x)}{x-1}\right|^{1/4} \Ai(f_n(x))
                        (1+\bigO(1/n)) \right.
                \\
                        -\,\left. \frac{\sin\eta_1(x)}{(1-x)^{1/2}}
                        \left|\frac{f_n(x)}{x-1}\right|^{-1/4} \Ai'(f_n(x))
                        (1+\bigO(1/n)) \right].
        \end{multline}
        \item[(iv)] Uniformly for $x\in[1+\delta,\infty]$,
            \begin{equation}\label{theorem recall: asymptotics phin: exponential}
                \hat\phi_n(x) =
                        \frac{1}{\sqrt{2\pi}}
                        \frac{\varphi(x)^{\frac{1}{2}(\alpha+1)}}{x^{1/4}(x-1)^{1/4}}
                        \exp\left[-\frac{n}{2}\int_1^x\sqrt{\frac{s-1}{s}}h_n(s)ds\right](1+\bigO(1/n))
        \end{equation}
    \end{itemize}
\end{theorem}

\begin{remark}\label{remark: eta functions}
    Note that the functions $\eta_j$ are only analytic in
    $\mathbb{C}\setminus((-\infty,0]\cup[1,\infty))$. However, since $\eta_{j,+}=-\eta_{j,-}$
    on $(1,\infty)$ the functions $\cos\eta_j(z)$ and $\frac{\sin\eta_j(z)}{(1-z)^{1/2}}$
    are analytic near 1. Furthermore, the reader can verify that these functions
    have the following behavior near 1,
    \begin{equation}\label{remark: eta functions: eq1}
        \cos\eta_{1,2}(x)=1+\bigO(x-1),\qquad \frac{\sin\eta_{1,2}(x)}{(1-x)^{1/2}}=(\alpha\pm 1)+\bigO(x-1),
        \qquad\mbox{as $x\to 1$,}
    \end{equation}
    and using the fact that $\varphi(z)=e^{i\arccos(2z-1)}$ for
    $z\in\mathbb{C}_+$ one can verify that for $x>1$,
    \begin{align}
        \label{remark: eta functions: eq2}
        & \cos\eta_{1,2}(x)=\frac{1}{2}\left(\varphi(x)^{\frac{1}{2}(\alpha\pm 1)}
            +\varphi(x)^{-\frac{1}{2}(\alpha\pm 1)}\right),
        \\[1ex]
        \label{remark: eta functions: eq3}
        & \frac{\sin\eta_{1,2}(x)}{(1-x)^{1/2}}= \frac{1}{2\sqrt{x-1}}\left(\varphi(x)^{\frac{1}{2}(\alpha\pm 1)}
            -\varphi(x)^{-\frac{1}{2}(\alpha\pm 1)}\right).
    \end{align}
     For later reference we observe  that
    \begin{equation}\label{remark: eta functions: eq4}
        \zeta_{1,2}(z)=\pm\frac{\pi}{2}-(\alpha\pm
        1)z^{1/2}(1+\bigO(z)),\qquad\mbox{as $z\to 0$.}
    \end{equation}
\end{remark}

In order to obtain the asymptotics of the functions $\hat\psi_r$ ($r=1,2$), see
(\ref{definition: phinhat psirhat}), we write them in terms of the RH problem
for orthogonal polynomials
due to Fokas, Its and Kitaev \cite{FokasItsKitaev}.
Let $Y$ be the solution of the RH problem for orthogonal polynomials associated
to the weight $x^\alpha e^{-V(x)}$ on $[0,\infty)$,
\[
    Y(z)=\begin{pmatrix}
        \frac{1}{\gamma_n}p_n(z) & \frac{1}{\gamma_n}C(p_n w)(z) \\
        -2\pi i\gamma_{n-1}p_{n-1}(z) & -2\pi i\gamma_{n-1}C(p_{n-1}w)(z)
    \end{pmatrix},\qquad\mbox{for $z\in\mathbb{C}\setminus[0,\infty)$,}
\]
where $\gamma_n>0$ is the leading coefficient of $p_n(z)$,
$p_n(z)=\gamma_n z^n+\cdots$.
Define a $2\times 2$ matrix valued function $U$ by
\[
    U(z)=\beta_n^{-(n+\frac{\alpha}{2})\sigma_3} Y(\beta_n z)
    \beta_n^{\frac{1}{2}\alpha\sigma_3},\qquad \mbox{for
    $z\in\mathbb{C}\setminus[0,\infty)$,}
\]
where $\sigma_3=\left(\begin{smallmatrix}1 & 0 \\ 0 &
-1\end{smallmatrix}\right)$ is the third Pauli matrix, cf.~\cite[(3.14)]{v6}. Using equations
(\ref{th0}), (\ref{definition: psi1tilde}), (\ref{definition: psi2tilde}), together with the 
defining relation for the rescaled
external field $V_n(x) = \frac{1}{n} V(\beta_n x)$,
it is straightforward to verify that
\begin{align*}
    \begin{pmatrix}
        \hat\psi_2(x) \\
        \hat\psi_1(x)
    \end{pmatrix}
        &=
            \frac{n^{-1/2}}{x}
            \begin{pmatrix}
                1 & 0 \\
                0 & \frac{i\alpha}{2\pi}
            \end{pmatrix}
            (-1/d_n)^{\sigma_3}
            Y(0)^{-1}Y(\beta_n x)
            \begin{pmatrix}
                1 \\
                0
            \end{pmatrix}
            (\beta_n x)^{\frac{\alpha}{2}}e^{-\frac{1}{2}V(\beta_n x)}
    \\[2ex]
        &=
            \frac{n^{-1/2}}{x \sqrt\pi}
            \begin{pmatrix}
                1 & 0 \\
                0 & \frac{i\alpha}{2}
            \end{pmatrix}
            \left(-\frac{\sqrt\pi}{d_n}\beta_n^{\frac{1}{2}\alpha}\right)^{\sigma_3}
            U(0)^{-1}U(x)
            \begin{pmatrix}
                1 \\
                0
            \end{pmatrix}
            x^{\frac{\alpha}{2}}e^{-\frac{1}{2}nV_n(x)}.
\end{align*}
The constant matrix $U(0)^{-1}$ has been determined in \cite[Remark 5.5]{v6}.
Inserting this information and the defining relation
\begin{equation}\label{definition: dn}
    - \frac{1}{d_n}  \equiv
            \frac{\tilde c_n^{\frac{\alpha}{2}} n^\alpha e^{\frac{1}{2}V(0)}}{\Gamma(\alpha)}
            \beta_n^{-\frac{1}{2}\alpha},
\end{equation}
 into the previous equation, we obtain
\begin{multline}\label{psi in U}
    \begin{pmatrix}
        \hat\psi_2(x) \\
        \hat\psi_1(x)
    \end{pmatrix}=(-1)^n\frac{n^{-1/2}}{x\sqrt\pi}
        \begin{pmatrix}
            \frac{\alpha}{4} & \frac{1}{2} \\[1ex]
            -\frac{\alpha}{4} & \frac{1}{2}
        \end{pmatrix}
        (\tilde c_n n^2)^{-\frac{1}{4}\sigma_3}
        \begin{pmatrix}
            1-\alpha & -i(\alpha+1) \\
            1 & i
        \end{pmatrix}\\[1ex]
        \times 2^{\alpha\sigma_3}R(0)^{-1}e^{-\frac{1}{2}n\ell_n\sigma_3}U(x)\begin{pmatrix}
    1 \\ 0 \end{pmatrix}
    x^{\frac{\alpha}{2}}e^{-\frac{1}{2}nV_n(x)},
\end{multline}
where $R$ is the result of the series of transformations $Y\mapsto
U\mapsto T\mapsto S\mapsto R$ in the Deift--Zhou steepest-descent analysis of
the RH problem for $Y$, see \cite[Section 3]{v6}, and where $\ell_n$ is 
the Lagrange multiplier given in \cite[Proposition 3.12]{v6}. The first column
of $U$ has been determined in \cite[Section 5]{v6}, and in the next theorem we
summarize its description on $\mathbb{R}_+$.

\begin{theorem}
    The first column of $U$ has the following description on $\mathbb R_+$.
    \begin{itemize}
        \item[(i)] {\rm \cite[(5.14)]{v6}} For $x\in(0,\delta]$,
            \begin{multline} \label{theorem: first column U: bessel}
                U(x)
                \begin{pmatrix}
                    1 \\
                    0
                \end{pmatrix} =
                        x^{-\frac{\alpha}{2}}e^{\frac{1}{2}nV_n(x)}e^{\frac{1}{2}n\ell_n\sigma_3}
                        (-1)^n \frac{\sqrt\pi(-\tilde f_n(x))^{1/4}}{x^{1/4}(1-x)^{1/4}}
                        \\
                        \times\, R(x) 2^{-\alpha\sigma_3}
                        \begin{pmatrix}
                            \sin\zeta_1(x) &  \cos\zeta_1(x) \\
                            -i\sin\zeta_2(x) & -i\cos\zeta_2(x)
                        \end{pmatrix}
                        \begin{pmatrix}
                            J_\alpha(2(-\tilde f_n(x))^{1/2}) \\[1ex]
                            J_\alpha'(2(-\tilde f_n(x))^{1/2})
                        \end{pmatrix}.
            \end{multline}
        \item[(ii)] {\rm \cite[(5.6)]{v6}} For $x\in[\delta,1-\delta]$,
            \begin{equation} \label{theorem: first column U: bulk}
                U(x)
                \begin{pmatrix}
                    1 \\
                    0
                \end{pmatrix} =
                        x^{-\frac{\alpha}{2}}e^{\frac{1}{2}nV_n(x)}e^{\frac{1}{2}n\ell_n\sigma_3}
                        \frac{1}{x^{1/4}(1-x)^{1/4}} R(x)2^{-\alpha\sigma_3}
                        \begin{pmatrix}
                            \cos F_{n,1}(x) \\
                            -i\cos F_{n,2}(x)
                        \end{pmatrix},
            \end{equation}
            where $F_{n,j}$ is defined by {\rm (\ref{definition: Fnj})}.
        \item[(iii)] {\rm \cite[(5.9)]{v6}} For $x\in[1-\delta,1+\delta]$,
            \begin{multline} \label{theorem: first column U: airy}
                U(x)
                \begin{pmatrix}
                    1 \\
                    0
                \end{pmatrix} =
                        x^{-\frac{\alpha}{2}}e^{\frac{1}{2}nV_n(x)}e^{\frac{1}{2}n\ell_n\sigma_3}
                        \frac{\sqrt\pi}{x^{1/4}}R(x)2^{-\alpha\sigma_3}
                        \\
                        \times\,
                        \begin{pmatrix}
                            \cos\eta_1(x) & -\frac{\sin\eta_1(x)}{(1-x)^{1/2}}
                            \\[2ex]
                            -i\cos\eta_2(x) & i\frac{\sin\eta_2(x)}{(1-x)^{1/2}}
                        \end{pmatrix}
                        \left|\frac{f_n(x)}{x-1}\right|^{\sigma_3/4}
                        \begin{pmatrix}
                            \Ai(f_n(x)) \\
                            \Ai'(f_n(x))
                        \end{pmatrix}.
            \end{multline}
        \item[(iv)] {\rm \cite[(5.4), see also (3.41) and (2.8)]{v6}} For $x\in[1+\delta,\infty]$,
            \begin{multline} \label{theorem: first column U: exponential}
                U(x)
                \begin{pmatrix}
                    1 \\
                    0
                \end{pmatrix} =
                        x^{-\frac{\alpha}{2}}e^{\frac{1}{2}nV_n(x)}e^{\frac{1}{2}n\ell_n\sigma_3}
                        \frac{1}{2x^{1/4}(x-1)^{1/4}}
                        \\[1ex]
                        \times\, R(x)2^{-\alpha\sigma_3}
                        \begin{pmatrix}
                            \varphi(x)^{\frac{1}{2}(\alpha+1)} \\
                            -i\varphi(x)^{\frac{1}{2}(\alpha-1)}
                        \end{pmatrix}
                        \exp\left(-\frac{n}{2}\int_1^x\sqrt{\frac{s-1}{s}}h_n(s)ds\right).
            \end{multline}
    \end{itemize}
\end{theorem}

\subsection{Auxiliary results}
    \label{subsection: auxiliary results}

In order to determine the asymptotics of the functions $\hat\phi_n$,
$\hat\psi_1$ and $\hat\psi_2$ on the positive real line we will make use of the
following auxiliary results.

\begin{proposition}\label{proposition: matching formulas}
    Let $j=1,2$. The following matching formulae hold.
    \begin{itemize}
        \item[(i)]
        Uniformly for $x\in[\frac{1}{2}n^{-1},\delta]$, as $n\to\infty$,
        \begin{multline}\label{proposition: matching formulas: eq1}
            \left(-\tilde f_n(x)\right)^{1/4}\left[\sin\zeta_j(x)J_\alpha\left(2(-\tilde f_n(x))^{1/2}\right)
            +\cos\zeta_j(x)J_\alpha'\left(2(-\tilde f_n(x))^{1/2}\right)\right]\\[1ex]
            = \frac{(-1)^n}{\sqrt\pi}\Bigl(\cos F_{n,j}(x)+\tau_n(x)\sin F_{n,j}(x)\Bigr)
                + \bigO(1/n),
        \end{multline}
        with $\tau_n(x)=\frac{4\alpha^2-1}{16(-\tilde f_n(x))^{1/2}}$, and with
        $F_{n,j}$ given by {\rm (\ref{definition: Fnj})}.
        \item[(ii)] 
       Uniformly for
               $x\in[1-\delta,1-\frac{1}{2}n^{\kappa-\frac{2}{3}}]$, as $n\to\infty$,
    \begin{multline}\label{proposition: matching formulas: eq2}
        \cos\eta_j(x)|f_n(x)|^{1/4}\Ai(f_n(x))-\sin\eta_j(x)|f_n(x)|^{-1/4}\Ai'(f_n(x))
        \\[1ex]
        =\frac{1}{\sqrt\pi}\cos
        F_{n,j}(x)+\bigO\left(\frac{1}{n(1-x)^{3/2}}\right).
    \end{multline}
    \item[(iii)]
        Uniformly for
        $x\in[1+\frac{1}{2}n^{\kappa-\frac{2}{3}},1+\delta]$, as $n\to\infty$,
        \begin{equation}
            f_n(x)^{1/4}
                \Ai(f_n(x))=\frac{1}{2\sqrt\pi}\exp\left(-\frac{n}{2}\int_1^x\sqrt\frac{s-1}{s}h_n(s)ds\right)
                \left(1+\bigO(n^{-\frac{3}{2}\kappa})\right),
        \end{equation}
        and
        \begin{align}
            & f_n(x)^{-1/4}
                \Ai'(f_n(x))=-\frac{1}{2\sqrt\pi}\exp\left(-\frac{n}{2}\int_1^x\sqrt\frac{s-1}{s}h_n(s)ds\right)
                \left(1+\bigO(n^{-\frac{3}{2}\kappa})\right).
        \end{align}
    \end{itemize}
\end{proposition}

\begin{proof}
    (i) From (\ref{definition: fn fntilde}) and the fact that $\tilde c_n$ and
    $\hat{\tilde f}_n$ are positive, we have
    \[
        2(-\tilde f_n(x))^{1/2}=i\lim_{z\to x+i0} 2\tilde
        f_n(z)^{1/2},\qquad\mbox{for $x\in(0,\delta]$.}
    \]
    Using in addition \cite[(2.10) and (2.8)]{v6}, (\ref{property: equilibrium measure}),
    (\ref{definition: Fnj}) and the fact
    that $\eta_j=\zeta_j+\frac{\pi\alpha}{2}$ we arrive at,
    \begin{align*}
        2(-\tilde f_n(x))^{1/2} &=
        \frac{n}{2}\int_0^x\sqrt\frac{1-s}{s}h_n(s)ds =
            \frac{n}{2}\int_1^x\sqrt{\frac{1-s}{s}}h_n(s)ds+\pi n \\[1ex]
        & =
            -F_{n,j}(x)+\zeta_j(x)+\frac{\pi\alpha}{2}-\frac{\pi}{4}+\pi n,
            \qquad \mbox{for $x\in(0,\delta]$.}
    \end{align*}
    By \cite[(9.2.5), (9.2.9) and (9.2.10)]{AbramowitzStegun} this 
implies, uniformly
    for $x\in[\frac{1}{2}n^{-1},\delta]$, as $n\to\infty$,
    \begin{align}
        \nonumber
        & \sqrt\pi(-\tilde f_n(x))^{1/4}J_\alpha\left(2(-\tilde f_n(x))^{1/2}\right) \\[1ex]
        \nonumber
        & \qquad =\cos\left(2(-\tilde f_n(x))^{1/2}-\frac{\pi\alpha}{2}-\frac{\pi}{4}\right)
            -\tau_n(x)\sin\left(2(-\tilde f_n(x))^{1/2}-\frac{\pi\alpha}{2}-\frac{\pi}{4}\right)
            +\bigO\left(\frac{1}{n^2 x}\right)\\[1ex]
        \label{proof: proposition: matching formulas: eq1}
        &\qquad =(-1)^n\left[-\sin\Bigl(F_{n,j}(x)-\zeta_j(x)\Bigr)+\tau_n(x)\cos\Bigl(F_{n,j}(x)-\zeta_j(x)\Bigr)
        +\bigO\left(\frac{1}{n}\right)\right],
    \end{align}
    and similarly by \cite[(9.2.11), (9.2.15) and (9.2.16)]{AbramowitzStegun},
    \begin{align}
        \nonumber
        & \sqrt\pi(-\tilde f_n(x))^{1/4}J_\alpha'\left(2(-\tilde f_n(x))^{1/2}\right)\\[1ex]
        \nonumber
        &\qquad=(-1)^n
        \left[\cos\Bigl(F_{n,j}(x)-\zeta_j(x)\Bigr)+\frac{4\alpha^2+3}{16(-\tilde f_n(x))^{1/2}}
        \sin\Bigl(F_{n,j}(x)-\zeta_j(x)\Bigr)
        +\bigO\left(\frac{1}{n^2 x}\right)\right] \\[1ex]
        \label{proof: proposition: matching formulas: eq2}
        &\qquad =(-1)^n
        \left[\cos\Bigl(F_{n,j}(x)-\zeta_j(x)\Bigr)+\tau_n(x)
        \sin\Bigl(F_{n,j}(x)-\zeta_j(x)\Bigr)
        +\bigO\left(\frac{1}{n \sqrt x}\right)\right].
    \end{align}
    Together with the fact that $\cos\zeta_j(x)=\bigO(\sqrt x)$ as $x\to 0$,
which follows
    from (\ref{remark: eta functions: eq4}),
 this yields (\ref{proposition: matching formulas: eq1}).

    (ii) From (\ref{definition: fn fntilde}) and the fact that $c_n$ and $\hat f_n$ are positive, we have
    \[
        \frac{2}{3}(-f_n(x))^{2/3}=i\lim_{z\to
        x+i0}\frac{2}{3}f_n(z)^{3/2},\qquad\mbox{for $x\in[1-\delta,1)$.}
    \]
    From \cite[(2.9) and (2.8)]{v6} and (\ref{definition: Fnj}) we
    then obtain,
    \[
        \frac{2}{3}(-f_n(x))^{3/2}=
            -\frac{n}{2}\int_1^x\sqrt{\frac{1-s}{s}}h_n(s)ds=F_{n,j}(x)-\eta_j(x)+\frac{\pi}{4},
            \qquad \mbox{for $x\in[1-\delta,1)$.}
    \]
    This implies by \cite[(10.4.60)]{AbramowitzStegun}, uniformly for
    $x\in[1-\delta,1-\frac{1}{2}n^{\kappa-\frac{2}{3}}]$, as $n\to\infty$,
    \begin{align}
        \nonumber
        |f_n(x)|^{1/4}\Ai(f_n(x)) &=
            \frac{1}{\sqrt\pi}\sin\left(\frac{2}{3}(-f_n(x))^{3/2}+\frac{\pi}{4}\right)
            + \bigO\left(\frac{1}{n(1-x)^{3/2}}\right) \\[2ex]
        \label{proof: proposition: matching formulas: eq3}
        &=
            \frac{1}{\sqrt\pi}
            \cos\Bigl(F_{n,j}(x)-\eta_j(x)\Bigr)+\bigO\left(\frac{1}{n(1-x)^{3/2}}\right),
    \end{align}
    and similarly by \cite[(10.4.62)]{AbramowitzStegun},
    \begin{equation} \label{proof: proposition: matching formulas: eq4}
        |f_n(x)|^{-1/4}\Ai'(f_n(x))=\frac{1}{\sqrt\pi}
            \sin\Bigl(F_{n,j}(x)-\eta_j(x)\Bigr)+\bigO\left(\frac{1}{n(1-x)^{3/2}}\right).
    \end{equation}
    After a straightforward calculation
 we obtain   (\ref{proposition: matching formulas: eq2}).

    (iii) From \cite[(2.9) and (2.8)]{v6} we have
    \[
        \frac{2}{3}f_n(x)^{3/2}=-\frac{n}{2}\int_x^1
        \sqrt{\frac{s-1}{s}}h_n(s)ds,\qquad\mbox{for $x\in(1,1+\delta]$.}
    \]
    Using in addition \cite[(10.4.59) and (10.4.61)]{AbramowitzStegun} it
    is simple  to check that the last part of the Proposition is also
    satisfied.
\end{proof}

\begin{proposition}\label{proposition: bessel functions}
    For every $L>0$ we have as $n\to\infty$,
    \begin{multline}\label{proposition: bessel functions: eq1}
        J_\alpha(2(-\tilde f_n(x))^{1/2})=J_\alpha(2\tilde c_n^{1/2}n\sqrt x) \\
        + \begin{cases}
                \bigO(n^\alpha x^{\frac{\alpha}{2}+1}), &\mbox{uniformly for $x\in(0,Ln^{-2}]$,} \\
                \bigO(n^{1/2}x^{5/4}), & \mbox{uniformly for $x\in[n^{-2},2n^{-1}]$,}
            \end{cases}
    \end{multline}
    \begin{multline}\label{proposition: bessel functions: eq2}
        J_\alpha'(2(-\tilde f_n(x))^{1/2})=J_\alpha'(2\tilde c_n^{1/2}n\sqrt x) \\[1ex]
        + \begin{cases}
                \bigO(n^{\alpha-1} x^{\frac{\alpha}{2}+\frac{1}{2}}), &\mbox{uniformly for $x\in(0,Ln^{-2}]$,} \\
                \bigO(n^{1/2}x^{5/4}), & \mbox{uniformly for $x\in[n^{-2},2n^{-1}]$.}
            \end{cases}
    \end{multline}
\end{proposition}

\begin{proof}
    Note that
    \[
        (-\tilde f_n(x))^{1/2}=\tilde c_n^{1/2}n\sqrt x (1+\bigO(x)),
        \qquad\mbox{uniformly for $x\in(0,2n^{-1}]$, as $n\to\infty$.}
    \]
    Since $\sup_{y\in[0,C]}|y^{-(\alpha-1)}J_\alpha'(y)|<\infty$ 
for any $C>0$, it is
    then simple to check that,
    \begin{align*}
        \nonumber
        & J_\alpha\left(2(-\tilde f_n(x))^{1/2}\right)-J_\alpha\left(2\tilde c_n^{1/2}n\sqrt x\right)
        \\[1ex]
        \nonumber
        & \qquad\qquad = \left(2(-\tilde f_n(x))^{1/2}-2\tilde c_n^{1/2}n\sqrt x\right)
                            \int_0^1 J_\alpha'\left((1-t)2(-\tilde f_n(x))^{1/2}+2t\tilde c_n^{1/2}n\sqrt
                            x\right) dt
        \\[1ex]
        & \qquad\qquad = \bigO(n^\alpha x^{\frac{\alpha}{2}+1}),
    \end{align*}
    uniformly for $x\in(0,Ln^{-2}]$, as $n\to\infty$. The determination of the
    error term in $[n^{-2},2n^{-1}]$ is analogous using
    $\sup_{y\in[C,\infty)}|\sqrt y J_\alpha'(y)|<\infty$ for any $C>0$.

    Similarly, using the facts $\sup_{y\in[0,C]}|y^{-(\alpha-2)}J_\alpha''(y)|<\infty$
    and
    $\sup_{y\in[C,\infty)}|\sqrt y J_\alpha''(y)|<\infty$ for any $C>0$, 
one proves  (\ref{proposition: bessel functions: eq2}).
\end{proof}

\begin{corollary}\label{corollary: bessel functions}
    For every $L>0$ we have as $n\to\infty$,
    \begin{equation}\label{lemma: bessel functions 1: eq1}
        J_\alpha(2(-\tilde f_n(x))^{1/2})=
            \begin{cases}
                \bigO(n^\alpha x^{\frac{\alpha}{2}}), &\mbox{uniformly for $x\in(0,Ln^{-2}]$,} \\
                \bigO(n^{-1/2}x^{-1/4}), & \mbox{uniformly for $x\in[n^{-2},2n^{-1}]$.}
            \end{cases}
    \end{equation}
    \begin{equation}\label{lemma: bessel functions 1: eq2}
        J_\alpha'(2(-\tilde f_n(x))^{1/2})=
            \begin{cases}
                \bigO(n^{\alpha-1} x^{\frac{\alpha}{2}-\frac{1}{2}}), &\mbox{uniformly for $x\in(0,Ln^{-2}]$,} \\
                \bigO(n^{-1/2}x^{-1/4}), & \mbox{uniformly for $x\in[n^{-2},2n^{-1}]$.}
            \end{cases}
    \end{equation}
\end{corollary}

\begin{proof}
    This follows from the facts 
$$
\begin{aligned}
   \null&\sup_{y\in[0,C]}|y^{-\alpha}J_\alpha(y)|<\infty,
  \qquad  \sup_{y\in[C,\infty)}|\sqrt y J_\alpha(y)|<\infty,\\
  \null&\sup_{y\in[0,C]}|y^{-(\alpha-1)}J_\alpha'(y)|<\infty,
  \qquad \sup_{y\in[C,\infty)}|\sqrt y J_\alpha'(y)|<\infty
\end{aligned}
$$ 
for any $C>0$.
\end{proof}

\begin{proposition}\label{proposition: airy functions}
    Uniformly for
    $x\in[1-2n^{\kappa-\frac{2}{3}}, 1+2n^{\kappa-\frac{2}{3}}]$,
as $n\to\infty$,
    \begin{align} \label{proposition: airy functions: eq1}
        & \left|\frac{f_n(x)}{x-1}\right|^{1/4} \Ai(f_n(x))
            = c_n^{1/4} n^{1/6} \Ai\left(c_n n^{2/3}(x-1)\right)
            + \bigO(n^{-1/2+\frac{9}{4}\kappa}),
        \\[2ex]
        \label{proposition: airy functions: eq2}
        & \left|\frac{f_n(x)}{x-1}\right|^{-1/4} \Ai'(f_n(x))
            = \bigO(n^{-1/6+\frac{1}{4}\kappa}).
    \end{align}
\end{proposition}

\begin{proof}
    Note that as $n\to\infty$
    \begin{equation}\label{proof: proposition: airy functions: eq1}
        f_n(x) = c_n n^{2/3} (x-1) (1+\bigO(n^{\kappa-\frac{2}{3}})),
    \end{equation}
    uniformly for
    $x\in[1-2n^{\kappa-\frac{2}{3}},1+2n^{\kappa-\frac{2}{3}}]$. Together
    with $|\Ai'(\xi)|\leq C(1+|\xi|)^{1/4}$ for $\xi\in\mathbb{R}$ and
    $C$ some positive    constant, one can then verify that
    \begin{align}
        \nonumber
        & \Ai(f_n(x)) - \Ai(c_n n^{2/3}(x-1))
        \\[1ex]
        \nonumber
        &\qquad\qquad = \left(f_n(x)-c_n n^{2/3}(x-1)\right)
                        \int_0^1 \Ai'\left((1-t)f_n(x)+tc_n n^{2/3}(x-1)\right)dt
        \\[1ex]
        &\qquad\qquad = \bigO(n^{-2/3+\frac{9}{4}\kappa}).
    \end{align}
    Equation (\ref{proposition: airy functions: eq1}) now follows from this equation
    together with (\ref{proof: proposition: airy functions: eq1}) and the
    fact that the Airy function is bounded on the real line.

    From (\ref{proof: proposition: airy functions: eq1}) and from the fact that
    $|\Ai'(\xi)|\leq C(1+|\xi|)^{1/4}$ we have $\Ai'(f_n(x))=\bigO(n^{\frac{1}{4}\kappa})$. Together
    with (\ref{proof: proposition: airy functions: eq1}) this proves equation
    (\ref{proposition: airy functions: eq2}).
\end{proof}

\subsection{Asymptotic behavior of $\hat\phi_n$}
    \label{subsection: asymptotics Phin}

The asymptotic behavior of $\hat\phi_n$ on the positive real line is now given
by the following Lemma.

\begin{lemma}\label{lemma: asymptotics phin}
    The functions $\hat\phi_n(x)=\sqrt{\beta_n}\phi_n(\beta_n x)$ have the
    following asymptotic behavior on the positive real line, as $n\to\infty$.
    \begin{enumerate}
        \item[(i)] Bessel region: For every $L>0$,
            \begin{equation}\label{lemma: asymptotics phin: bessel}
                \hat\phi_n(x) =
                \begin{cases}
                    \bigO(n^{\alpha+\frac{1}{2}}x^{\frac{\alpha}{2}}),
                        & \mbox{uniformly for $x\in(0,Ln^{-2}]$,} \\
                    \bigO(x^{-1/4}),
                        & \mbox{uniformly for $x\in[n^{-2},2n^{-1}]$.}
                \end{cases}
            \end{equation}
        \item[(ii)] Bulk region:
            \begin{equation}\label{lemma: asymptotics phin: bulk}
                \hat\phi_n(x) =
                    \sqrt{\frac{2}{\pi}}\frac{\cos F_n(x)}{x^{1/4}(1-x)^{1/4}}
                    + \bigO\left(\frac{1}{nx^{3/4}(1-x)^{7/4}}\right),
            \end{equation}
            uniformly for $x\in
            [\frac{1}{2}n^{-1},1-\frac{1}{2}n^{\kappa-\frac{2}{3}}]$.
        \item[(iii)] Airy region:
            \begin{equation}\label{lemma: asymptotics phin: airy}
                \hat\phi_n(x)=
                    \sqrt 2 c_n^{1/4}n^{1/6} \Ai\left(c_n n^{2/3}(x-1)\right)
                    + \bigO(n^{-1/6+\frac{1}{4}\kappa}),
            \end{equation}
            uniformly for
            $x\in[1-2n^{\kappa-\frac{2}{3}},1+2n^{\kappa-\frac{2}{3}}]$.
        \item[(iv)] Exponential region: there exists a constant $c>0$ such
            that,
            \begin{equation}\label{lemma: asymptotics phin: exponential region}
                \hat\phi_n(x)=\bigO\left(e^{-c(x-1)n^{2/3}}\right), \qquad
                \mbox{uniformly for $x\in[1+\frac{1}{2}n^{\kappa-\frac{2}{3}},\infty)$.}
            \end{equation}
    \end{enumerate}
\end{lemma}

\begin{proof}
    (i) Using equation (\ref{theorem recall: asymptotics phin: bessel}),
    Corollary \ref{corollary: bessel functions} and
    the facts that $(-\tilde f_n(x))^{1/4}=\bigO(n^{1/2}x^{1/4})$,
    as $n\to\infty$, and $\cos\zeta_1(x)=\bigO(x^{1/2})$ as $x\to 0$, we obtain
    (\ref{lemma: asymptotics phin: bessel}).

    (ii) By  (\ref{proof: proposition: matching formulas: eq1}) and
    (\ref{proof: proposition: matching formulas: eq2}),
    \[
        (-\tilde f_n(x))^{1/4}J_\alpha(2(-\tilde f_n(x))^{1/4})=\bigO(1),
        \qquad
        (-\tilde f_n(x))^{1/4}J_\alpha'(2(-\tilde f_n(x))^{1/4})=\bigO(1),
    \]
    as $n\to\infty$, uniformly for $x\in[\frac{1}{2}n^{-1},\delta]$. From
    (\ref{theorem recall: asymptotics phin: bessel}), 
(\ref{proposition: matching formulas:
    eq1}), and the estimate
 $\tau_n(x)=\bigO\left(\frac{1}{n\sqrt x}\right)$, we then
    obtain,
    \begin{align}
        \nonumber
        \hat\phi_n(x)
            &=  \frac{(-1)^n\sqrt 2}{x^{1/4}(1-x)^{1/4}}\left[(-\tilde f_n(x))^{1/4}\sin\zeta_1(x)
                J_\alpha(2(-\tilde f_n(x))^{1/2})\right.
        \\
        \nonumber
            & \qquad\qquad\qquad\qquad\qquad\qquad
                \left.+(-\tilde f_n(x))^{1/4}\cos\zeta_1(x)
                J_\alpha'(2(-\tilde f_n(x))^{1/2})+\bigO(1/n)\right]
        \\[1ex]
        \label{proof: lemma: asymptotics phin: eq1}
            &= \sqrt{\frac{2}{\pi}}\frac{1}{x^{1/4}(1-x)^{1/4}}
            \left[\cos F_n(x)+\bigO\left(\frac{1}{n\sqrt x}\right)\right],
    \end{align}
    as $n\to\infty$, uniformly for $x\in[\frac{1}{2}n^{-1},\delta]$, where we
    recall that that $F_{n,1}\equiv F_n$.
 Further, from (\ref{proof: proposition: matching formulas: eq3})
    and (\ref{proof: proposition: matching formulas: eq4}), we have
    \[
        |f_n(x)|^{1/4}\Ai(f_n(x))=\bigO(1),\qquad
        |f_n(x)|^{-1/4}\Ai'(f_n(x))=\bigO(1),
    \]
    as $n\to\infty$, uniformly for $x\in[1-\delta,1-n^{\kappa-\frac{2}{3}}]$. By
    (\ref{theorem recall: asymptotics phin: airy})
 and (\ref{proposition: matching formulas:
    eq2}) we then obtain
    \begin{align}
        \nonumber
        \hat\phi_n(x)
            &= \frac{\sqrt
            2}{x^{1/4}(1-x)^{1/4}}\left[\cos\eta_1(x)|f_n(x)|^{1/4}\Ai(f_n(x))\right.
        \\
        \nonumber
            & \qquad\qquad\qquad\qquad\qquad\qquad
                \left. -
                \sin\eta_1(x)|f_n(x)|^{-1/4}\Ai'(f_n(x))+\bigO\left(\frac{1}{n(1-x)^{3/2}}\right)\right]
        \\[1ex]
        \label{proof: lemma: asymptotics phin: eq2}
            &= \sqrt{\frac{2}{\pi}}\frac{1}{x^{1/4}(1-x)^{1/4}}
            \left[\cos F_n(x)+\bigO\left(\frac{1}{n(1-x)^{3/2}}\right)\right],
    \end{align}
    as $n\to\infty$, uniformly for $x\in[1-\delta,1-n^{\kappa-\frac{2}{3}}]$.
    Equations (\ref{theorem recall: asymptotics phin: bulk}),
    (\ref{proof: lemma: asymptotics phin: eq1}) and
    (\ref{proof: lemma: asymptotics phin: eq2}) then yield (\ref{lemma: asymptotics phin: bulk}).

    (iii) Now, we prove the third part of the Proposition. From
    equations (\ref{theorem recall: asymptotics phin: airy}) and
    (\ref{remark: eta functions: eq1}) it follows readily that,
    \begin{multline}
        \hat\phi_n(x) =
                \sqrt 2\left|\frac{f_n(x)}{x-1}\right|^{1/4}\Ai(f_n(x))
                \left(1+\bigO(n^{\kappa-\frac{2}{3}})\right)
        \\-\sqrt
                2(\alpha+1)\left|\frac{f_n(x)}{x-1}\right|^{-1/4}\Ai'(f_n(x))
                \left(1+\bigO(n^{\kappa-\frac{2}{3}})\right),
    \end{multline}
    as $n\to\infty$, uniformly for $x\in[1-2n^{\kappa-\frac{2}{3}},1+2n^{\kappa-\frac{2}{3}}]$.
    Using Proposition \ref{proposition: airy functions} and the fact that the Airy
    function is bounded on the real line, we then arrive at equation
    (\ref{lemma: asymptotics phin: airy}).

    (iv) Finally, (\ref{theorem recall: asymptotics phin: airy}),
 (\ref{theorem recall: asymptotics phin:
    exponential}), (\ref{remark: eta functions: eq2}),
 (\ref{remark: eta functions: eq3})
    and Proposition \ref{proposition: matching formulas}(iii) lead to,
    \begin{align*}
        \hat\phi_n(x)
        &=
                \frac{1}{\sqrt{2\pi}}\frac{\varphi(x)^{\frac{1}{2}(\alpha+1)}}{x^{1/4}(x-1)^{1/4}}
                \exp\left[-\frac{n}{2}\int_1^x\sqrt{\frac{s-1}{s}}h_n(s)ds\right]
                \left(1+\bigO(n^{-\frac{3}{2}\kappa})\right),
        \\[1ex]
        &=
                \exp\left[-\frac{n}{2}\int_1^x\sqrt{\frac{s-1}{s}}h_n(s)ds\right]
                \bigO\left(x^{\frac{1}{2}\alpha}n^{\frac{1}{6}-\frac{1}{4}\kappa}\right),
    \end{align*}
    as $n\to\infty$, uniformly for
    $x\in[1+\frac{1}{2}n^{\kappa-\frac{2}{3}},\infty)$. Since there exists $h_0>0$ such that $h_n(s)\geq h_0>0$
    for $n$ sufficiently large, and
 as $\frac{1}{\sqrt s}\leq \frac{1}{\sqrt x}$ for $s\in[1,x]$,
    one then proves that 
    \begin{align}\label{proof: asymptotics phi: exponential estimate}
        & \exp\left[-\frac{n}{2}\int_1^x\sqrt{\frac{s-1}{s}}h_n(s)ds\right] =
                \bigO\left(\exp\left[-\frac{h_0}{3}\sqrt{\frac{x-1}{x}}n(x-1)\right]\right) =
                \bigO(e^{-c(x-1)n^{2/3}})
    \end{align}
    for some $c > 0$. Inserting this relation
 into the previous equation it is straightforward to 
    verify that the last part of the Lemma is satisfied, 
  with a different choice of $c$.
\end{proof}

\subsection{Asymptotic behavior of $\hat\psi_r$}
    \label{subsection: asymptotics Psir}

\subsubsection*{The Bessel region}

Here, we will determine the asymptotics of $\hat\psi_1$ and $\hat\psi_2$ in the
Bessel region $(0,n^{-1}]$ using equation (\ref{psi in U}). Inserting
(\ref{theorem: first column U: bessel}) into (\ref{psi in U}), and using the
fact that $2^{\alpha\sigma_3}R(0)^{-1}R(x)2^{-\alpha\sigma_3}=I+\bigO(x/n)$,
cf.~\cite[Theorem 3.32]{v6}, as $n\to\infty$, uniformly for $x\in(0,\delta]$,
we obtain
\begin{multline}\label{asymptotics psi: bessel: main formula}
    \begin{pmatrix}
        \hat\psi_2(x) \\
        \hat\psi_1(x)
    \end{pmatrix} =
            \frac{(-\tilde f_n(x))^{1/4}n^{-1/2}}{x (1-x)^{1/4} x^{1/4}}
            \begin{pmatrix}
                \frac{\alpha}{4} & \frac{1}{2} \\[1ex]
                -\frac{\alpha}{4} & \frac{1}{2}
            \end{pmatrix}
            (\tilde c_n n^2)^{-\frac{1}{4}\sigma_3} (I+\bigO(x/n))
    \\[1ex]
            \times
            \begin{pmatrix}
                1-\alpha & -i(\alpha+1) \\
                1 & i
            \end{pmatrix}
            \begin{pmatrix}
                \sin\zeta_1(x) & \cos\zeta_1(x) \\
                -i\sin\zeta_2(x) & -i\cos\zeta_2(x)
            \end{pmatrix}
            \begin{pmatrix}
                J_\alpha(2(-\tilde f_n(x))^{1/2}) \\[1ex]
                J_\alpha'(2(-\tilde f_n(x))^{1/2})
            \end{pmatrix},
\end{multline}
as $n\to\infty$, uniformly for $x\in(0,\delta]$. Now, since
$\sin\zeta_1(x)=1+\bigO(x)$, $\sin\zeta_2(x)=-1+\bigO(x)$,
$\cos\zeta_1(x)=(\alpha+1)\sqrt x (1+\bigO(x))$, and
$\cos\zeta_2(x)=(1-\alpha)\sqrt x(1+\bigO(x))$, as $x\to 0$ 
(which follows from
(\ref{remark: eta functions: eq4})) we have
\[
    \begin{pmatrix}
        1-\alpha & -i(\alpha+1) \\
        1 & i
    \end{pmatrix}
    \begin{pmatrix}
        \sin\zeta_1(x) & \cos\zeta_1(x) \\
        -i\sin\zeta_2(x) & -i\cos\zeta_2(x)
    \end{pmatrix} =
            \left[2I+\bigO(x)\right]
            \begin{pmatrix}
                1 & 0 \\
                0 & \sqrt x
            \end{pmatrix},\quad\mbox{as $x\to 0$}.
\]
Inserting this relation into (\ref{asymptotics psi: bessel: main formula})
and using the fact that
\[
    \frac{(-\tilde f_n(x))^{1/4}}{(1-x)^{1/4}x^{1/4}}=\tilde c_n^{1/4}n^{1/2}(1+\bigO(x)),
\]
we then arrive at
\begin{equation}\label{asymptotics psi: bessel: eq1}
    \begin{pmatrix}
        \hat\psi_2(x) \\
        \hat\psi_1(x)
    \end{pmatrix} =
            \tilde c_n^{1/4}
            \begin{pmatrix}
                \frac{\alpha}{2} & 1 \\[1ex]
                -\frac{\alpha}{2} & 1
            \end{pmatrix}
            (\tilde c_n n^2)^{-\frac{1}{4}\sigma_3} (I+\bigO(x))
            \begin{pmatrix}
                \frac{1}{x}J_\alpha(2(-\tilde f_n(x))^{1/2}) \\[1ex]
                \frac{1}{\sqrt x} J_\alpha'(2(-\tilde f_n(x))^{1/2})
            \end{pmatrix},
\end{equation}
as $n\to\infty$, uniformly for $x\in(0,n^{-1}]$.

Now, we split the Bessel region $(0,n^{-1}]$ up into the intervals $(0,n^{-2}]$
and $[n^{-2},n^{-1}]$, and we determine the asymptotics of $\hat\psi_1$
and $\hat\psi_2$ in each of these two intervals. From Corollary \ref{corollary:
bessel functions} and equation (\ref{asymptotics psi: bessel: eq1}) we have,
\[
    \begin{pmatrix}
        \hat\psi_2(x) \\
        \hat\psi_1(x)
    \end{pmatrix} =
            \tilde c_n^{1/4}
            \begin{pmatrix}
                \frac{\alpha}{2} & 1 \\[1ex]
                -\frac{\alpha}{2} & 1
            \end{pmatrix}
            (\tilde c_n n^2)^{-\frac{1}{4}\sigma_3}
            \begin{pmatrix}
                \frac{1}{x}J_\alpha(2(-\tilde f_n(x))^{1/2})
                    + \bigO(n^\alpha x^{\frac{\alpha}{2}}) \\[1ex]
                \frac{1}{\sqrt x}J'_\alpha(2(-\tilde f_n(x))^{1/2})
                    + \bigO(n^\alpha x^{\frac{\alpha}{2}})
            \end{pmatrix},
\]
as $n\to\infty$, uniformly for $x\in(0,n^{-2}]$. Further, from Proposition
\ref{proposition: bessel functions} and the fact that
$J_{\alpha}'(z)=-J_{\alpha+1}(z)+\frac{\alpha}{z}J_\alpha(z)$,
 we then obtain
\begin{multline*}
        \begin{pmatrix}
            \hat\psi_2(x) \\
            \hat\psi_1(x)
        \end{pmatrix}
        \\ =
                \begin{pmatrix}
                    \frac{\alpha}{2} & 1 \\[1ex]
                    -\frac{\alpha}{2} & 1
                \end{pmatrix}
                \begin{pmatrix}
                    n^{-1/2}\frac{1}{x}J_\alpha(2\tilde c_n^{1/2}n\sqrt x)
                        + \bigO(n^{\alpha-\frac{1}{2}} x^{\frac{\alpha}{2}}) \\[1ex]
                    - \frac{\tilde c_n^{1/2}n^{1/2}}{\sqrt x}J_{\alpha+1}(2\tilde c_n^{1/2}n\sqrt x)
                        + \frac{\alpha}{2}n^{-1/2}\frac{1}{x}J_\alpha(2\tilde c_n^{1/2}n\sqrt x)
                        + \bigO(n^{\alpha+\frac{1}{2}} x^{\frac{\alpha}{2}})
                \end{pmatrix},
\end{multline*}
as $n\to\infty$, uniformly for $x\in(0,n^{-2}]$. This gives the asymptotics in
the interval $(0,n^{-2}]$. The derivation in the other interval, i.e.\
$[n^{-2},n^{-1}]$, is analogous and we obtain the following result.

\begin{lemma}\label{lemma: asymptotics psi: bessel}
    As $n\to\infty$,
    \begin{align}\label{lemma: asymptotics psi: bessel: eq1}
        &\hat\psi_1(x) =
                - \frac{\tilde c_n^{1/2}n^{1/2}}{\sqrt x}
                J_{\alpha+1}(2\tilde c_n^{1/2}n\sqrt x) +
                \begin{cases}
                    \bigO(n^{\alpha+\frac{1}{2}}x^{\alpha/2}),
                        & \mbox{uniformly for $x\in(0,n^{-2}]$,} \\
                    \bigO(x^{-1/4}), & \mbox{uniformly for $x\in[n^{-2},n^{-1}]$.}
                \end{cases}
    \end{align}
    \begin{multline}\label{lemma: asymptotics psi: bessel: eq2}
        \hat\psi_2(x) =
                \frac{n^{-1/2}\alpha}{x}J_\alpha(2\tilde c_n^{1/2}n\sqrt x)
                - \frac{\tilde c_n^{1/2}n^{1/2}}{\sqrt x}J_{\alpha+1}(2\tilde c_n^{1/2}n\sqrt x)
        \\
                + \begin{cases}
                    \bigO(n^{\alpha+\frac{1}{2}}x^{\alpha/2}), & \mbox{uniformly for $x\in(0,n^{-2}]$,} \\
                    \bigO(x^{-1/4}), & \mbox{uniformly for $x\in[n^{-2},n^{-1}]$.}
                \end{cases}
    \end{multline}
\end{lemma}

\subsubsection*{The Airy region}

Inserting (\ref{theorem: first column U: airy}) into (\ref{psi in U}) and using
$2^{\alpha\sigma_3}R(0)^{-1}R(x) 2^{-\alpha\sigma_3}=I+\bigO(1/n)$ we obtain
\begin{multline}\label{asymptotics psi: airy: main formula}
    \begin{pmatrix}
        \hat\psi_2(x) \\
        \hat\psi_1(x)
    \end{pmatrix} =
            (-1)^n\frac{n^{-1/2}}{x x^{1/4}}
            \begin{pmatrix}
                \frac{\alpha}{4} & \frac{1}{2} \\[1ex]
                -\frac{\alpha}{4} & \frac{1}{2}
            \end{pmatrix}
            (\tilde c_n n^2)^{-\frac{1}{4}\sigma_3}(I+\bigO(1/n))
            \begin{pmatrix}
                1-\alpha & -i(\alpha+1) \\
                1 & i
            \end{pmatrix}
    \\[1ex]
            \times
            \begin{pmatrix}
                \cos\eta_1(x) & -\frac{\sin\eta_1(x)}{(1-x)^{1/2}} \\[2ex]
                -i\cos\eta_2(x) & i\frac{\sin\eta_2(x)}{(1-x)^{1/2}}
            \end{pmatrix}
            \left|\frac{f_n(x)}{x-1}\right|^{\frac{1}{4}\sigma_3}
            \begin{pmatrix}
                \Ai(f_n(x)) \\
                \Ai'(f_n(x))
            \end{pmatrix},
\end{multline}
as $n\to\infty$, uniformly for $x\in[1-\delta,1+\delta]$. From equation
(\ref{remark: eta functions: eq1}) and Proposition \ref{proposition: airy
functions} we then obtain
\begin{align*}
    \begin{pmatrix}
        \hat\psi_2(x) \\
        \hat\psi_1(x)
    \end{pmatrix} & =
        (-1)^n n^{-1/2}
        \begin{pmatrix}
            \frac{\alpha}{4} & \frac{1}{2} \\[1ex]
            -\frac{\alpha}{4} & \frac{1}{2}
        \end{pmatrix}
        (\tilde c_n n^2)^{-\frac{1}{4}\sigma_3}
        (I+\bigO(n^{\kappa-\frac{2}{3}}))
        \\[1ex]
        & \qquad\qquad\times\begin{pmatrix}
            1-\alpha & -i(\alpha+1) \\
            1 & i
        \end{pmatrix}\begin{pmatrix}
        1 & -(\alpha+1) \\
        -i & i(\alpha-1)
    \end{pmatrix}
    \left|\frac{f_n(x)}{x-1}\right|^{\frac{1}{4}\sigma_3}
    \begin{pmatrix}
        \Ai(f_n(x)) \\
        \Ai'(f_n(x))
    \end{pmatrix} \\[3ex]
    & =
        (-1)^n n^{-1/2}
        \begin{pmatrix}
            \frac{\alpha}{2} & 1 \\[1ex]
            -\frac{\alpha}{2} & 1
        \end{pmatrix}(\tilde c_n n^2)^{-\frac{1}{4}\sigma_3}
        (I+\bigO(n^{\kappa-\frac{2}{3}}))\\[1ex]
    & \qquad\qquad\times\,
    \begin{pmatrix}
            -\alpha & (\alpha^2-1) \\
            1 & -\alpha
        \end{pmatrix}
    \begin{pmatrix}
        c_n^{1/4} n^{1/6} \Ai(c_n n^{2/3}(x-1))+\bigO(n^{-1/2+\frac{9}{4}\kappa}) \\[1ex]
        \bigO(n^{-1/6+\frac{1}{4}\kappa})
    \end{pmatrix},
\end{align*}
as $n\to\infty$, uniformly for
$x\in[1-n^{\kappa-\frac{2}{3}},1+n^{\kappa-\frac{2}{3}}]$, which implies after
a straightforward calculation,
\[
    \begin{pmatrix}
        \hat\psi_2(x) \\
        \hat\psi_1(x)
    \end{pmatrix} = (-1)^n
        \begin{pmatrix}
            \frac{\alpha}{2} & 1 \\[1ex]
            -\frac{\alpha}{2} & 1
        \end{pmatrix}
    \begin{pmatrix}
        \bigO(n^{-5/6}) \\[1ex]
        (c_n \tilde c_n)^{1/4} n^{1/6} \Ai(c_n n^{2/3}(x-1))+\bigO(n^{-1/6+\frac{1}{4}\kappa})
    \end{pmatrix}.
\]
We now have proved the following Lemma.

\begin{lemma} \label{lemma: asymptotics psi: airy}
    Let $r=1$ or $2$. As $n\to\infty$,
    \begin{equation}\label{lemma: asymptotics psi: airy: eq1}
        \hat\psi_r(x) =
                (-1)^n (c_n \tilde c_n)^{1/4} n^{1/6} \Ai\left(c_n n^{2/3}(x-1)\right)
                + \bigO(n^{-1/6+\frac{1}{4}\kappa}),
    \end{equation}
    uniformly for
    $x\in[1-n^{\kappa-\frac{2}{3}},1+n^{\kappa-\frac{2}{3}}]$.
\end{lemma}

\subsubsection*{The bulk region}

From equations (\ref{asymptotics psi: bessel: main formula}) and
(\ref{proposition: matching formulas: eq1}), 
 (\ref{asymptotics
psi: airy: main formula}) and (\ref{proposition: matching formulas: eq2}), 
 (\ref{psi in U}) and (\ref{theorem: first column U: bulk}), we obtain
\begin{multline*}
    \begin{pmatrix}
        \hat\psi_2(x) \\
        \hat\psi_1(x)
    \end{pmatrix}
    =\frac{(-1)^n n^{-1/2}}{\sqrt\pi x (1-x)^{1/4} x^{1/4}}
    \begin{pmatrix}
            \frac{\alpha}{4} & \frac{1}{2} \\[1ex]
            -\frac{\alpha}{4} & \frac{1}{2}
    \end{pmatrix}(\tilde c_n n^2)^{-\frac{1}{4}\sigma_3}
    (I+\bigO(x/n))
        \begin{pmatrix}
            1-\alpha & -i(\alpha+1) \\
            1 & i
        \end{pmatrix} \\[2ex]
    \times
        \begin{cases}
        \begin{pmatrix}
            \cos F_{n,1}(x)+\tau_n(x)\sin F_{n,1}(x) + \bigO(1/n) \\[1ex]
            -i (\cos F_{n,2}(x)+\tau_n(x)\sin F_{n,2}(x)) + \bigO(1/n)
        \end{pmatrix}, \qquad \mbox{uniformly for $x\in[\frac{1}{2}n^{-1},\delta]$,}
         \\[5ex]
         \begin{pmatrix}
            \cos F_{n,1}(x) + \bigO\left(\frac{1}{n (1-x)^{3/2}}\right) \\[2ex]
            -i\cos F_{n,2}(x) + \bigO\left(\frac{1}{n (1-x)^{3/2}}\right)
        \end{pmatrix}, \qquad\mbox{uniformly for $x\in[\delta,1-\frac{1}{2}n^{\kappa-\frac{2}{3}}]$.}
        \end{cases}
\end{multline*}
By (\ref{definition: Fnj})
\[
    \cos\left(\frac{1}{2}F_{n,1}(x)-\frac{1}{2}F_{n,2}(x)\right) =
        \cos\left(\frac{1}{2}\eta_1(x)-\frac{1}{2}\eta_2(x)\right) =
        \cos\left(\frac{1}{2}\arccos(2x-1)\right) =
        \sqrt x,
\]
and hence
\begin{align*}
    &\cos F_{n,1}(x)+\cos F_{n,2}(x)=2\sqrt x\cos G_n(x), \\[1ex]
    &\sin F_{n,1}(x)+\sin F_{n,2}(x)=2\sqrt x\sin G_n(x),
\end{align*}
with $G_n(x)=\frac{1}{2}F_{n,1}(x)+\frac{1}{2}F_{n,2}(x)$. 
Using the fact that
$\tau_n(x)=\bigO\left(\frac{1}{n \sqrt x}\right)$ uniformly
for $x\in[\frac{1}{2}n^{-1},\delta]$, as $n\to\infty$, we obtain
\[
    \begin{pmatrix}
        \hat\psi_2(x) \\
        \hat\psi_1(x)
    \end{pmatrix} =
            \frac{(-1)^n n^{-1/2}}{\sqrt\pi x (1-x)^{1/4} x^{1/4}}
            \begin{pmatrix}
                \frac{\alpha}{2} & 1 \\[1ex]
                -\frac{\alpha}{2} & 1
            \end{pmatrix}
            (\tilde c_n n^2)^{-\frac{1}{4}\sigma_3}
            \begin{pmatrix}
                \bigO(1) \\
                \sqrt x\cos G_n(x) + \bigO\left(\frac{1}{n (1-x)^{3/2}}\right)
            \end{pmatrix},
\]
uniformly for
$x\in[\frac{1}{2}n^{-1},1-\frac{1}{2}n^{\kappa-\frac{2}{3}}]$,
as $n\to\infty$. We then arrive at the following result.

\begin{lemma} \label{lemma: asymptotics psi: bulk}
    Let $r=1$ or $2$. As $n\to\infty$, uniformly for
    $x\in[\frac{1}{2}n^{-1},1-\frac{1}{2}n^{\kappa-\frac{2}{3}}]$,
    \begin{equation}\label{lemma: asymptotics psi: bulk: eq1}
        \hat\psi_r(x) =
                \frac{(-1)^n\tilde c_n^{1/4} }{\sqrt\pi x^{3/4}(1-x)^{1/4}}
                \cos G_n(x) + \bigO\left(\frac{1}{n x^{5/4}(1-x)^{7/4}}\right),
    \end{equation}
    with
    \begin{equation}\label{definition: Gn}
        G_n(x) =
                -\frac{n}{2}\int_1^x\sqrt{\frac{1-s}{s}}h_n(s)ds
                +\frac{1}{2}\alpha\arccos(2x-1)-\frac{\pi}{4}.
    \end{equation}
\end{lemma}

\subsubsection*{The exponential region}

As in the proof of Lemma \ref{lemma: asymptotics phin}(iv) we obtain from
(\ref{psi in U}), (\ref{theorem: first column U: airy}), (\ref{theorem: first
column U: exponential}), Proposition \ref{proposition: matching formulas}(iii)
and (\ref{proof: asymptotics phi: exponential estimate}),
 the following result.

\begin{lemma} \label{lemma: asymptotics psi: exponential}
    Let $r=1$ or $2$. There exists a constant $c>0$ such that
    \begin{equation}\label{lemma: asymptotics psi: exponential: eq1}
        \hat\psi_r(x) =
                \bigO(e^{-c(x-1)n^{2/3}}),
    \end{equation}
    as $n\to\infty$, uniformly for $x\in[1+n^{\kappa-\frac{2}{3}},\infty)$.
\end{lemma}

\section{Asymptotics of the  matrix $B$}
    \label{section: asymptotics B12 matrix}

We determine the asymptotics of the  matrix $B$ by following and occasionally
streamlining the path first developed in \cite[Subsection 4.2]{DeiftGioev}.

The following representations of the entries of $B$ are straightforward to verify.
\begin{align}
    \label{proof: lemma: B12matrix: eq1}
    \langle\varepsilon\phi_q,\phi_p\rangle
        &= \sqrt{\beta_p\beta_q}\left[\frac{1}{2}\int_0^\infty\hat\phi_p(x)dx\int_0^\infty\hat\phi_q(x)dx
            -\int_0^\infty\hat\phi_p(x)\int_{x\frac{\beta_p}{\beta_q}}^\infty\hat\phi_q(y)dydx\right],
        \\[2ex]
    \label{proof: lemma: B12matrix: eq2}
    \langle\varepsilon\psi_r,\phi_p\rangle
        &= \sqrt{\beta_p\beta_n}
            \left[\frac{1}{2}\int_0^\infty\hat\phi_p(x)dx\int_0^\infty\hat\psi_r(x)dx
            - \int_0^\infty\hat\phi_p(x)\int_{x\frac{\beta_p}{\beta_n}}^\infty\hat\psi_r(y)dydx\right],
        \\[2ex]
    \label{proof: lemma: B12matrix: eq3}
    \langle\varepsilon\psi_1,\psi_2\rangle
        &= \beta_n
            \left[\frac{1}{2}\int_0^\infty\hat\psi_1(x)dx\int_0^\infty\hat\psi_2(x)dx
            - \int_0^\infty\hat\psi_2(x)\int_x^\infty\hat\psi_1(y)dydx\right],
\end{align}
with $p,q\in\mathbb{N}$ and $r\in\{1,2\}$, and where $\hat\phi_n$
and $\hat\psi_r$ are defined in (\ref{definition: phinhat
psirhat}). Thus, in order to obtain the asymptotic behavior of the
 matrix $B$ we need to determine the asymptotic behavior of the
single and double integrals appearing in these three equations which 
will be done in Subsections \ref{sec5.1} and \ref{sec5.2} respectively. 
As
noted at the beginning of Section \ref{section: asymptotics
polynomials} we do this by splitting $(0,\infty)$ into four
regions $(0,n^{-1}]$, $[n^{-1},1-n^{\kappa-\frac{2}{3}}]$,
$[1-n^{\kappa-\frac{2}{3}}, 1+n^{\kappa-\frac{2}{3}}]$ and
$[1+n^{\kappa-\frac{2}{3}},\infty)$, with $\kappa=\frac{1}{12}$
fixed, and integrate separately over each of these four regions.
In the final and brief Subsection \ref{sec5.3} we summarize our 
results in such a way that the asymptotic result for the  matrix $B$ stated in 
Lemma \ref{lemma: B12matrix} is apparent.

\subsection{The single integrals}
\label{sec5.1}

We start with the following three auxiliary Propositions, which will also be
used to determine the asymptotic behavior of the double integrals.

\begin{proposition}\label{proposition: derivatives FnGn}
    The first and second derivatives of $F_n$ and $G_n$, defined in {\rm (\ref{definition: Fnj})}
    and {\rm (\ref{definition: Gn})}, satisfy,
    \begin{align}
        \label{proposition: derivatives FnGn: eq1}
        &\frac{1}{Z_n'(x)}=-\frac{2}{h_n(x)} \frac{x^{1/2}}{n(1-x)^{1/2}}
            \left[1+\bigO\left(\frac{1}{n(1-x)}\right)\right], \\[2ex]
        \label{proposition: derivatives FnGn: eq2}
        & Z_n''(x)=\bigO\left(\frac{n}{x^{3/2}(1-x)^{1/2}}\right)\left[1+\bigO\left(\frac{1}{n(1-x)}\right)\right],
    \end{align}
    as $n\to\infty$, uniformly for $x\in(0,1)$, where $Z\in\{F,G\}$.
\end{proposition}

\begin{proof}
    We will prove the result for $F_n$. The result for $G_n$ then
    also follows since $G_n$ equals $F_n$ with $\alpha$ replaced by
    $\alpha-1$. The first derivative of $F_n$ can be explicitly
    determined from the definition (\ref{definition: Fnj}),
    \[
        \frac{1}{F_n'(x)}=-\frac{2}{h_n(x)} \frac{x^{1/2}}{n(1-x)^{1/2}}\left(1-\frac{\alpha+1}{n
        h_n(x)(1-x)+\alpha+1}\right).
    \]
    Since $h_n(x)\geq h_0>0$ for $n$ sufficiently large,
  $x\in[0,\infty)$, see Subsection
    \ref{subsection: relevant result from v6} under (\ref{property: hn: auxiliary results}), we have
    $|nh_n(x)(1-x)+\alpha+1|\geq nh_0(1-x)$ for all $n$ sufficiently large,
  $x\in(0,1)$,
    which proves (\ref{proposition: derivatives FnGn: eq1}). Similarly, it follows from
    \[
        F_n''(x)=\frac{n}{x^{3/2}(1-x)^{1/2}}\left(-\frac{1}{2}h_n'(x)x(1-x)+\frac{1}{4}h_n(x)
            +\frac{1}{4}(\alpha+1)\frac{1-2x}{n(1-x)}\right)
    \]
    that (\ref{proposition: derivatives FnGn: eq2}) is satisfied as well.
\end{proof}

\begin{proposition}\label{proposition2: single integrals}
    As $n\to\infty$, uniformly for
    $x\in[n^{-1},1-\frac{1}{2}n^{\kappa-\frac{2}{3}}]$,
    \begin{align}
        \label{proposition2: single integrals: eq1}
        & \frac{1}{F_n'(x)x^{1/4}(1-x)^{1/4}}
            = \bigO(n^{-1/2-\frac{3}{4}\kappa}),
        \\[2ex]
        \label{proposition2: single integrals: eq2}
        & \frac{1}{G_n'(x)x^{3/4}(1-x)^{1/4}}
            = \bigO(n^{-1/2-\frac{3}{4}\kappa}),
        \\[2ex]
        \label{proposition2: single integrals: eq3}
        & \left(\frac{1}{F_n'(x)x^{1/4}(1-x)^{1/4}}\right)'
            = \bigO\left(\frac{1}{nx^{3/4}(1-x)^{7/4}}\right),
        \\[2ex]
        \label{proposition2: single integrals: eq4}
        & \left(\frac{1}{G_n'(x)x^{3/4}(1-x)^{1/4}}\right)'
            = \bigO\left(\frac{1}{nx^{5/4}(1-x)^{7/4}}\right).
    \end{align}
\end{proposition}

\begin{proof}
    Equations (\ref{proposition2: single integrals: eq1}) and
    (\ref{proposition2: single integrals: eq2}) follow
    from (\ref{proposition: derivatives FnGn: eq1}) and from the fact that
    $h_n(x)\geq h_0>0$ for $n$ sufficiently large, $x\in[0,\infty)$.
    Further, since
    \begin{multline*}
        \left(\frac{1}{F_n'(x)x^{1/4}(1-x)^{1/4}}\right)' \\
            = -\,\frac{1}{F_n'(x)^2}F_n''(x)x^{-1/4}(1-x)^{-1/4}
            - \frac{1}{4}\frac{1}{F_n'(x)}x^{-5/4}(1-x)^{-5/4}(1-2x),
    \end{multline*}
    equation (\ref{proposition2: single integrals: eq3}) follows from equations
    (\ref{proposition: derivatives FnGn: eq1}) and
    (\ref{proposition: derivatives FnGn: eq2}). The proof of the last equation
    of the Proposition is similar.
\end{proof}

%

\begin{proposition}\label{proposition3: single integrals}
    As $n\to\infty$, uniformly for $a,b\in[n^{-1},1-\frac{1}{2}n^{\kappa-\frac{2}{3}}]$,
    \begin{align}
        \label{proposition3: single integrals: eq1}
        &\int_a^b \frac{\cos F_n(y)}{y^{1/4}(1-y)^{1/4}}dy
            = \bigO(n^{-1/2-\frac{3}{4}\kappa}),
        \\[2ex]
        \label{proposition3: single integrals: eq2}
        &\int_a^b \frac{\cos G_n(y)}{y^{3/4}(1-y)^{1/4}}dy
            = \bigO(n^{-1/2-\frac{3}{4}\kappa}).
    \end{align}
\end{proposition}

\begin{proof}
    This is immediate after
integrating by parts and using Proposition \ref{proposition2: single
    integrals}.
\end{proof}

We now have the necessary ingredients to determine the asymptotic
behavior of the single integrals.

\subsubsection{Integrals involving  $\hat\phi_n$}

\begin{proposition}\label{proposition: single integrals phin}
    As $n\to\infty$,
    \begin{enumerate}
    \item[(i)] Bessel, bulk and exponential region: there exists a constant
    $c>0$ such that,
        \begin{align}
            \label{proposition: single integrals phin: eq1}
            &\int_0^x \bigl|\hat\phi_n(y)\bigr| dy =
                \bigO(n^{-3/4}),&&\mbox{uniformly for $x\in(0,n^{-1}]$,}
            \\[2ex]
            \label{proposition: single integrals phin: eq2}
            &\int_{n^{-1}}^{x}\hat\phi_n(y)dy =
                \bigO(n^{-1/2-\frac{3}{4}\kappa}),
                &&\mbox{uniformly for $x\in[n^{-1},1-n^{\kappa-\frac{2}{3}}]$,}
            \\[2ex]
            \label{proposition: single integrals phin: eq3}
            &\int_x^\infty \bigl|\hat\phi_n(y)\bigr| dy =
                \bigO(e^{-cn^\kappa}),
            && \mbox{uniformly for $x\in[1+n^{\kappa-\frac{2}{3}},\infty)$.}
        \end{align}
    \item[(ii)] Airy region:
        \begin{equation} \label{proposition: single integrals phin: eq4}
            \int_{1-n^{\kappa-\frac{2}{3}}}^x\hat\phi_n(y)dy =
                \bigO(n^{-1/2}), \qquad
                \mbox{uniformly for $x\in[1-n^{\kappa-\frac{2}{3}},1+n^{\kappa-\frac{2}{3}}]$,}
        \end{equation}
        \begin{equation}\label{proposition: single integrals phin: eq5}
            \int_{1-n^{\kappa-\frac{2}{3}}}^{1+n^{\kappa-\frac{2}{3}}}
                \hat\phi_n(y)dy = \sqrt 2 c_n^{-3/4}n^{-1/2}+\bigO(n^{-1/2-\frac{3}{4}\kappa}).
        \end{equation}
    \end{enumerate}
\end{proposition}

\begin{proof}
    (i) Equation (\ref{proposition: single integrals phin: eq1}) is immediate
    from (\ref{lemma: asymptotics phin: bessel}), equation (\ref{proposition: single integrals phin:
    eq2}) follows from (\ref{lemma: asymptotics phin: bulk})
    and (\ref{proposition3: single integrals: eq1}),
    and equation (\ref{proposition: single integrals phin: eq3}) follows from
    (\ref{lemma: asymptotics phin: exponential region}).

    (ii) From the asymptotic behavior (\ref{lemma: asymptotics phin: airy}) of
    $\hat\phi_n$ in the Airy region we obtain,
    \begin{equation} \label{proof: proposition: single integrals phin: eq1}
        \int_{1-n^{\kappa-\frac{2}{3}}}^{x}\hat\phi_n(y)dy
            = \sqrt 2 c_n^{-3/4}n^{-1/2}\int_{-c_n n^\kappa}^{c_n n^{\frac{2}{3}}(x-1)}\Ai(u)du
                    + \bigO(n^{-5/6+\frac{5}{4}\kappa}),
    \end{equation}
    as $n\to\infty$, uniformly for $x\in[1-n^{\kappa-\frac{2}{3}},1+n^{\kappa-\frac{2}{3}}]$.
    Since $\int_a^b \Ai(u)du$ is uniformly bounded for $a,b\in\mathbb{R}$, see e.g.\
    \cite[(10.4.82) and (10.4.83)]{AbramowitzStegun},
    this yields (\ref{proposition: single integrals phin: eq4}). Next, note that
    $\int_{-\infty}^\infty\Ai(t)dt=1$, $\int_{-\infty}^{-y}\Ai(t)dt=\bigO(y^{-3/4})$ and
    $\int_y^\infty\Ai(t)dt=\bigO(e^{-cy})$ as $y\to\infty$ for some
    $c>0$, see again \cite[(10.4.82) and (10.4.83)]{AbramowitzStegun},
    implying
    \[
        \int_{-c_n n^\kappa}^{c_n
        n^\kappa}\Ai(u)du=1+\bigO(n^{-\frac{3}{4}\kappa}).
    \]
    Together with (\ref{proof: proposition: single integrals phin: eq1}) this proves
    the remaining statement (\ref{proposition: single integrals phin: eq5}) of the 
Proposition.
\end{proof}

\begin{lemma}\label{lemma: single integrals phin}
    There exists $0<\tau=\tau(m,\alpha)<1$ such that
    \begin{align}
        \label{lemma: single integrals phin: eq1}
        &\int_0^\infty \hat\phi_n(y)dy
            = \left(\frac{1}{\sqrt m}+\bigO(n^{-\tau})\right)n^{-1/2},
            \qquad\mbox{as $n\to\infty$,}
        \\[1ex]
        \label{lemma: single integrals phin: eq2}
        &\int_a^b \hat\phi_n(y)dy = \bigO(n^{-1/2}),
        \qquad\mbox{as $n\to\infty$, uniformly for $a,b\in[0,\infty]$.}
    \end{align}
\end{lemma}

\begin{proof}
    The Lemma is immediate from the previous Proposition together with the fact
    that $c_n=(2m)^{2/3}+\bigO(n^{-1/m})$ as $n\to\infty$, see (\ref{definition: cn auxiliary results}).
\end{proof}

\subsubsection{Integrals involving $\hat\psi_r$}

\begin{proposition}\label{proposition: single integrals psi}
    Let $r\in\{1,2\}$. As $n\to\infty$,
    \begin{enumerate}
        \item[(i)] Bulk and exponential region: there exists a constant $c>0$
            such that,
            \begin{align}
                \label{proposition: single integrals psi: eq1}
                & \int_{n^{-1}}^{x}\hat\psi_r(y)dy = \bigO(n^{-1/2-\frac{3}{4}\kappa}),
                    \quad\mbox{uniformly for $x\in[n^{-1},1-n^{\kappa-\frac{2}{3}}]$,}
                \\[1ex]
                \label{proposition: single integrals psi: eq2}
                & \int_x^\infty\bigl|\hat\psi_r(y)\bigr| dy = \bigO(e^{-cn^\kappa}),
                    \qquad\mbox{uniformly for $x\in[1+n^{\kappa-\frac{2}{3}},\infty]$.}
            \end{align}
        \item[(ii)] Bessel region:
            \begin{align}
                \label{proposition: single integrals psi: eq3}
                & \int_0^x \hat\psi_r(y)dy = \bigO(n^{-1/2}),
                        \qquad\mbox{uniformly for $x\in(0,n^{-1}]$,}
                \\[1ex]
                \label{proposition: single integrals psi: eq4}
                & \int_0^{n^{-1}}\hat\psi_r(y)dy = (-1)^r n^{-1/2}+\bigO(n^{-3/4}).
            \end{align}
        \item[(iii)] Airy region:
            \begin{align}
                \label{proposition: single integrals psi: eq5}
                & \int_{1-n^{\kappa-\frac{2}{3}}}^x \hat\psi_r(y)dy = \bigO(n^{-1/2}),
                        \qquad \mbox{uniformly for
                        $x\in[1-n^{\kappa-\frac{2}{3}},1+n^{\kappa-\frac{2}{3}}]$,}
                \\[1ex]
                \label{proposition: single integrals psi: eq6}
                & \int_{1-n^{\kappa-\frac{2}{3}}}^{1+n^{\kappa-\frac{2}{3}}}\hat\psi_r(y)dy =
                        (-1)^n \tilde c_n^{1/4}c_n^{-3/4}n^{-1/2} + \bigO(n^{-1/2-\frac{3}{4}\kappa}).
            \end{align}
    \end{enumerate}
\end{proposition}

\begin{proof}
    (i) Equation (\ref{proposition: single integrals psi: eq1}) is immediate from
    (\ref{lemma: asymptotics psi: bulk: eq1}) and
    (\ref{proposition3: single integrals: eq2}), and equation (\ref{proposition: single integrals psi:
    eq2}) follows from (\ref{lemma: asymptotics psi: exponential: eq1}).

    (ii) From the asymptotic behavior
    (\ref{lemma: asymptotics psi: bessel: eq1}) of $\hat\psi_1$ in the Bessel region we
    obtain,
    \begin{equation} \label{proof: proposition: single integrals psi: eq1}
            \int_0^x\hat\psi_1(y)dy =
                - n^{-1/2} \int_0^{2\tilde c_n^{1/2}n\sqrt x} J_{\alpha+1}(t)dt
                + \bigO(n^{-3/4}),
    \end{equation}
    as $n\to\infty$, uniformly for $x\in(0,n^{-1}]$.
 From \cite[(9.2.1) and (11.4.17)]{AbramowitzStegun}
    we learn that $\int_0^\infty J_{\alpha+1}(t)dt=1$ and $\int_y^\infty
    J_{\alpha+1}(t)dt=\bigO(y^{-1/2})$ as $y\to\infty$. Together with
    (\ref{proof: proposition: single integrals psi: eq1}) this yields
    (\ref{proposition: single integrals psi: eq3}) as well as (\ref{proposition: single integrals psi: eq4})
    for the case $r=1$. The
    case $r=2$ can be proven similarly using the asymptotic behavior
    (\ref{lemma: asymptotics psi: bessel: eq2}) of $\hat\psi_2$ in the Bessel region
    together with the previous facts about Bessel integrals
    as well as the fact $\int_0^\infty t^{-1} J_\alpha(t) dt=1/\alpha$, see 
\cite[(11.4.16)]{AbramowitzStegun}.

    (iii) Finally, the proof of the last part of the Proposition is analogous to the proof of
    Proposition \ref{proposition: single integrals phin}(ii) using the
    asymptotic behavior (\ref{lemma: asymptotics psi: airy: eq1}) of $\hat\psi_r$
    in the Airy region.
\end{proof}

\begin{lemma}\label{lemma: single integrals psin}
    Let $r\in\{1,2\}$. There exists $0<\tau=\tau(m,\alpha)<1$ such that,
    \begin{align}
        \label{lemma: single integrals psin: eq1}
        & \int_0^\infty \hat\psi_r(x)dx=\left((-1)^r +\frac{(-1)^n}{\sqrt{2m-1}}+\bigO(n^{-\tau})\right)n^{-1/2},
            \qquad\mbox{as $n\to\infty$,}
        \\[1ex]
        \label{lemma: single integrals psin: eq2}
        & \int_a^b\hat\psi_r(x)dx=\bigO(n^{-1/2}),\qquad\mbox{as $n\to\infty$
        uniformly for $a,b\in[0,\infty]$.}
    \end{align}
\end{lemma}

\begin{proof}
    The Lemma is immediate from the previous Proposition together with the
    facts that $c_n=(2m)^{2/3}+\bigO(n^{-1/m})$ and $\tilde
    c_n=\left(\frac{2m}{2m-1}\right)^2+\bigO(n^{-1/m})$ as $n\to\infty$, see
    (\ref{definition: cn auxiliary results}) and
    (\ref{definition: cntilde auxiliary results}).
\end{proof}

\subsection{The double integrals}
\label{sec5.2}

The goal of this subsection is to determine the asymptotic behaviour of the 
double integrals appearing in 
(\ref{proof: lemma: B12matrix: eq1})-(\ref{proof: lemma: B12matrix: eq3}). 
Following  \cite{DeiftGioev} we decompose the 
range of integration $\mathbb R_+$ of the outer integral into two regions, namely into 
the bulk region which is essentially given by $(n^{-1}, 1 - n^{\kappa - \frac{2}{3}})$
and its complement. We first determine the contribution from the region outside the bulk
in Subsection \ref{sec5.2.1}. As in \cite{DeiftGioev} a more subtle argument is needed
to determine the leading order asymptotics in the oscillatory bulk region in 
Subsection \ref{sec5.2.2}. An important ingredient in the argument is Proposition
\ref{proposition2: double integrals bulk} which provides a surprisingly simple description of the phase deviations
of orthogonal polynomials with different degrees in the oscillatory region. Such a 
formula was first presented in \cite[Lemma 4.7]{DeiftGioev}. The formula follows from a special property
of the equilibrium measure stated in Proposition 
\ref{proposition1: double integrals bulk} (see \cite[Lemma 4.8]{DeiftGioev} for the
corresponding property in the Hermite case). Our results on the double integrals are summarized in
Subsection \ref{sec5.2.3}.

\subsubsection{The double integrals outside the bulk}
\label{sec5.2.1}

We start with the following technical Propositions.

\begin{proposition} \label{proposition1: double integrals out bulk}
    Let $p=n+i$ and $q=n+j$ with $i,j$ some fixed integers. Then,
    \begin{equation}\label{proposition1: double integrals out bulk: eq1}
        \frac{\beta_p}{\beta_q}=1+\frac{1}{m}\frac{p-q}{q}+\bigO(n^{-1-1/m}),\qquad\mbox{as
        $n\to\infty$.}
    \end{equation}
    In particular, $\frac{\beta_p}{\beta_q}=1+\bigO(1/n)$ as $n\to\infty$.
\end{proposition}

\begin{proof}
    The proof is similar to the proof of \cite[Lemma 4.4]{DeiftGioev}. Recall
    from (\ref{definition: betan auxiliary results})
    that $\beta_n=\sum_{k=-1}^\infty \beta_{(k)}n^{-k/m}$.
    Since $p^{-c}-q^{-c}=\bigO(n^{-1-c})$ as $n\to\infty$ (for $c>0$) we then
    obtain,
    \begin{align*}
        \beta_p-\beta_q
            &= \beta_{(-1)}(p^{1/m}-q^{1/m})
                    +\sum_{k=1}^m \beta_{(k)}(p^{-k/m}-q^{-k/m})+\bigO(n^{-1-1/m})
        \\[1ex]
            &= \beta_{(-1)}(p^{1/m}-q^{1/m})+\bigO(n^{-1-1/m}).
    \end{align*}
    This implies,
    \begin{align*}
        \frac{\beta_p}{\beta_q}-1 &=\frac{\beta_p-\beta_q}{\beta_q}
            =\left[\frac{p^{1/m}-q^{1/m}}{q^{1/m}}+\bigO(n^{-1-\frac{2}{m}})\right]
                \left(1+\bigO(n^{-1/m})\right) \\[1ex]
            &=
            \left[\left(1+\frac{p-q}{q}\right)^{1/m}-1+\bigO(n^{-1-\frac{2}{m}})\right]\left(1+\bigO(n^{-1/m})\right).
    \end{align*}
    The Proposition now follows by expanding the $1/m$-th power at 1.
\end{proof}

\begin{proposition} \label{proposition2: double integrals out bulk}
    Let $p=n+i$ and $q=n+j$ with $i,j$ some fixed integers and define for $u\in\mathbb{R}$,
    \begin{equation}\label{definition: u_pq}
        u_{p,q}=
        c_q q^{2/3}\left(\frac{\beta_p}{\beta_q}-1\right)
        +\frac{c_q q^{2/3}}{c_p p^{2/3}}\frac{\beta_p}{\beta_q}u.
    \end{equation}
    Then,
    \begin{equation}
        2\int_{-c_p p^\kappa}^{c_p p^\kappa}
        \Ai(u)\int_{u_{p,q}}^{c_q q^\kappa}\Ai(v)dvdu
        = 1+\bigO(n^{-\frac{3}{4}\kappa}),\qquad\mbox{as $n\to\infty$.}
    \end{equation}
\end{proposition}

\begin{proof}
    As in the previous Proposition one can verify that $\frac{c_q}{c_p}=1+\bigO(n^{-1-1/m})$ as $n\to\infty$.
    By (\ref{proposition1: double integrals out bulk: eq1}) we then obtain
    $u_{p,q}=u+\bigO(n^{-1/3})+\bigO(un^{-1})$ as $n\to\infty$.
    Together with the boundedness (on the real line) of the Airy
    function, this yields
    \[
        \int_{u_{p,q}}^{c_q q^\kappa}\Ai(v)dv
            = \int_u^{c_q q^\kappa}\Ai(v)dv + \bigO(n^{-1/3})+\bigO(un^{-1}).
    \]
    Then, since $|\Ai(t)|\leq C(1+|t|)^{-1/4}$ and $|t\Ai(t)|\leq C|t|^{3/4}$ for
    $t\in\mathbb{R}$ and $C>0$ some constant, we obtain,
    \begin{multline*}
        2\int_{-c_p p^\kappa}^{c_p p^\kappa}
            \Ai(u)\int_{u_{p,q}}^{c_q q^\kappa}\Ai(v)dvdu
        =
            2\int_{-c_p p^\kappa}^{c_p p^\kappa}
                \Ai(u)\int_{u}^{c_q q^\kappa}\Ai(v)dvdu +
                \bigO(n^{-\frac{1}{3}+\frac{3}{4}\kappa}) \\[1ex]
        = \left(\int_{-c_p p^\kappa}^{c_q q^\kappa}\Ai(v)dv\right)^2
                - \left(\int_{c_p p^\kappa}^{c_q q^\kappa}\Ai(v)dv\right)^2
                + \bigO(n^{-\frac{1}{3}+\frac{3}{4}\kappa}).
    \end{multline*}
    Since the Airy function is bounded on the real line we have
    \[
        \int_{c_p p^\kappa}^{c_q q^\kappa}\Ai(v)dv =
        \bigO(n^{\kappa-1}).
    \]
    As in the proof of Proposition \ref{proposition: single integrals phin}(ii) we obtain
    \[
        \int_{-c_p p^\kappa}^{c_q
                q^\kappa}\Ai(v)dv=1+\bigO(n^{-\frac{3}{4}\kappa}).
    \]
    This proves the Proposition.
\end{proof}

\begin{proposition}\label{proposition3: double integrals out bulk}
    As $n\to\infty$,
    \begin{align}
        \label{proposition3: double integrals out bulk: eq1}
        &\int_0^{2\tilde c_n^{1/2}\sqrt n} J_\alpha'(u)
            \int_u^{2\tilde c_n^{1/2}\sqrt n}J_{\alpha+1}(v)dvdu
            = \frac{1}{2}+\bigO(n^{-1/2}),
        \\[1ex]
        \label{proposition3: double integrals out bulk: eq2}
        & \int_0^{2\tilde c_n^{1/2}\sqrt n}J_{\alpha+1}(u)
            \int_u^{2\tilde c_n^{1/2}\sqrt n}J_{\alpha+1}(v)dvdu=\frac{1}{2}+\bigO(n^{-1/4}).
    \end{align}
\end{proposition}

\begin{proof}
    Integrating by parts and using $J_\alpha(0)=0$ for
    $\alpha>0$ and $\int_0^\infty J_\alpha(u)J_{\alpha+1}(u)du=1/2$, see e.g.\
    \cite[(11.4.42)]{AbramowitzStegun}, we obtain
    \begin{align}
        \nonumber
        \int_0^{2\tilde c_n^{1/2}\sqrt n}J_\alpha'(u)\int_u^{2\tilde c_n^{1/2}\sqrt
        n}J_{\alpha+1}(v)dvdu &= \int_0^{2\tilde c_n^{1/2}\sqrt
        n}J_\alpha(u)J_{\alpha+1}(u)du \\[1ex]
        \label{proof: proposition3: double integrals out bulk: eq1}
        &= \frac{1}{2}-\int_{2\tilde c_n^{1/2}\sqrt n}^\infty J_\alpha(u)J_{\alpha+1}(u)du.
    \end{align}
    From \cite[(9.2.1)]{AbramowitzStegun} we have
    $J_\alpha(u)J_{\alpha+1}(u)=-\frac{\cos(2u-\alpha\pi)}{\pi u}+\bigO(u^{-2})$ as $u\to\infty$.
    Integrating by parts one can verify that
    \[
        \int_{2\tilde c_n^{1/2}\sqrt
        n}^\infty\frac{\cos(2u-\alpha\pi)}{\pi u}du=\bigO(n^{-1/2}),\qquad\mbox{as
        $n\to\infty$,}
    \]
    so that also
    \[
        \int_{2\tilde c_n^{1/2}\sqrt n}^\infty J_\alpha(u)J_{\alpha+1}(u)du=\bigO(n^{-1/2}),\qquad\mbox{as
        $n\to\infty$.}
    \]
    Inserting these estimates
 into (\ref{proof: proposition3: double integrals out bulk:
    eq1}) the proof of the first part of the Proposition follows. Next,
    \[
        \int_0^{2\tilde c_n^{1/2}\sqrt n}J_{\alpha+1}(u)
            \int_u^{2\tilde c_n^{1/2}\sqrt n}J_{\alpha+1}(v)dvdu=\frac{1}{2}
            \int_0^{2\tilde c_n^{1/2}\sqrt n}J_{\alpha+1}(v)dv.
    \]
    Since $\int_0^\infty J_{\alpha+1}(u)du=1$ and $\int_x^\infty
    J_{\alpha+1}(u)du=\bigO(x^{-1/2})$ as $x\to\infty$, see Proof of Proposition
    \ref{proposition: single integrals psi}(ii), this yields
    (\ref{proposition3: double integrals out bulk: eq2}), and the 
Proposition is proven.
\end{proof}

Now, we have the necessary ingredients to determine the asymptotic behavior of
the double integrals in (\ref{proof: lemma: B12matrix: eq1})--(\ref{proof:
lemma: B12matrix: eq3}), except for the part of the outer integral which lies
in the bulk.

\begin{proposition} \label{proposition: double integrals out bulk}
    Let $p=n+i$ and $q=n+j$ with $i,j$ some fixed integers and let $r\in\{1,2\}$.
    There exists $0<\tau=\tau(m,\alpha)<1$ such that as $n\to\infty$,
    \begin{align}
        \nonumber
        & \int_0^\infty \hat\phi_p(x)
            \int_{x\frac{\beta_p}{\beta_q}}^\infty\hat\phi_q(y) dy dx \\
        \label{proposition: double integrals out bulk: eq1}
            &\qquad\qquad = \frac{1}{2m}n^{-1}
            +\int_{p^{-1}}^{1-p^{\kappa-\frac{2}{3}}} \hat\phi_p(x)
            \int_{x\frac{\beta_p}{\beta_q}}^\infty \hat\phi_q(y) dy dx+\bigO(n^{-1-\tau}), \\[3ex]
        \nonumber
        & \int_0^\infty \hat\phi_p(x)
            \int_{x\frac{\beta_p}{\beta_n}}^\infty \hat\psi_r(y) dy dx \\
        \label{proposition: double integrals out bulk: eq2}
            &\qquad\qquad = (-1)^n\frac{1}{2m}\sqrt\frac{m}{2m-1} n^{-1}
            +\int_{p^{-1}}^{1-p^{\kappa-\frac{2}{3}}} \hat\phi_p(x)
            \int_{x\frac{\beta_p}{\beta_n}}^\infty \hat\psi_r(y) dy dx+\bigO(n^{-1-\tau}),
    \end{align}
    and
    \begin{align}
        \nonumber
        & \int_0^\infty \hat\psi_2(x)\int_x^\infty \hat\psi_1(y) dy dx = \left(-\frac{3}{2}+
            \frac{(-1)^n}{\sqrt{2m-1}}+\frac{1}{2}\frac{1}{2m-1}\right)n^{-1} \\
        \label{proposition: double integrals out bulk: eq3}
            &\qquad\qquad +\int_{n^{-1}}^{1-n^{\kappa-\frac{2}{3}}}
            \hat\psi_2(x)\int_x^\infty\hat\psi_1(y)dydx+\bigO(n^{-1-\tau}).
    \end{align}
\end{proposition}

\begin{proof}
    From (\ref{proposition: single integrals phin: eq1}),
    (\ref{proposition: single integrals phin: eq3}) and (\ref{lemma: single integrals phin: eq2})
    one concludes,
    \begin{multline}\label{proof: proposition: double integrals out bulk: eq1a}
        \int_0^\infty \hat\phi_p(x) \int_{x\frac{\beta_p}{\beta_q}}^\infty\hat\phi_q(y) dy dx
            = \int_{p^{-1}}^{1-p^{\kappa-\frac{2}{3}}} \hat\phi_p(x)
                \int_{x\frac{\beta_p}{\beta_q}}^\infty \hat\phi_q(y) dy dx
        \\[1ex]
            + \int_{1-p^{\kappa-\frac{2}{3}}}^{1+p^{\kappa-\frac{2}{3}}} \hat\phi_p(x)
                \int_{x\frac{\beta_p}{\beta_q}}^\infty \hat\phi_q(y) dy dx
            + \bigO(n^{-1-\frac{1}{4}}).
    \end{multline}
    For notational convenience we denote the second double integral on the 
right hand side of
    (\ref{proof: proposition: double integrals out bulk: eq1a}) by $J$.
    From equations (\ref{proposition: single integrals phin: eq3})
    and (\ref{lemma: single integrals phin: eq2}), and from the asymptotic behavior
    (\ref{lemma: asymptotics phin: airy}) of
    $\hat\phi_p$ in the Airy region, we have
    \begin{align*}
        J &=
        \int_{1-p^{\kappa-\frac{2}{3}}}^{1+p^{\kappa-\frac{2}{3}}}
            \hat\phi_p(x) \int_{x\frac{\beta_p}{\beta_q}}^{1+q^{\kappa-\frac{2}{3}}}
                \hat\phi_q(y)dy dx + \bigO(e^{-cn^\kappa})
                \\[2ex]
        &=
        \sqrt 2 c_p^{1/4} p^{\frac{1}{6}}
        \int_{1-p^{\kappa-\frac{2}{3}}}^{1+p^{\kappa-\frac{2}{3}}}
            \Ai(c_p p^{\frac{2}{3}}(x-1))
            \int_{x\frac{\beta_p}{\beta_q}}^{1+q^{\kappa-\frac{2}{3}}}
                \hat\phi_q(y)dy dx + \bigO(n^{-4/3+\frac{5}{4}\kappa}).
    \end{align*}
    Using Proposition \ref{proposition1: double integrals out bulk} one can 
verify that
    $x\frac{\beta_p}{\beta_q}\in[1-2q^{\kappa-\frac{2}{3}}, 1+2q^{\kappa-\frac{2}{3}}]$
    for $n$ large enough, so that, from (\ref{lemma: asymptotics phin: airy}) and
    from the fact that the Airy function is bounded on the real line,
    \begin{align}
        \nonumber
        J &=
        2 (c_p c_q)^{\frac{1}{4}} (pq)^{\frac{1}{6}}
        \int_{1-p^{\kappa-\frac{2}{3}}}^{1+p^{\kappa-\frac{2}{3}}}
            \Ai(c_p p^{\frac{2}{3}}(x-1))
            \int_{x\frac{\beta_p}{\beta_q}}^{1+q^{\kappa-\frac{2}{3}}}
                \Ai(c_q q^{\frac{2}{3}}(y-1))dy dx +
                \bigO(n^{-\frac{4}{3}+\frac{9}{4}\kappa})\\[2ex]
        &=
        \frac{(c_p c_q)^{-3/4}}{(pq)^{1/2}}
        \left(2\int_{-c_p p^\kappa}^{c_p p^\kappa}
            \Ai(u)\int_{u_{p,q}}^{c_q q^\kappa}\Ai(v)dv du\right)
            +\bigO(n^{-\frac{4}{3}+\frac{9}{4}\kappa}),
    \end{align}
    with $u_{p,q}$ defined by (\ref{definition: u_pq}). Proposition
    \ref{proposition2: double integrals out bulk} and (\ref{definition: cn auxiliary
    results}) yield (\ref{proposition: double integrals out bulk: eq1}). The proof of
    (\ref{proposition: double integrals out bulk: eq2}) is analogous.

    It now remains to prove (\ref{proposition: double integrals out bulk: eq3}).
    Note that as in the proof of (\ref{proposition: double integrals out bulk: eq1}) and
    (\ref{proposition: double integrals out bulk: eq2}), the reader can verify that
    \[
        \int_{1-n^{\kappa-\frac{2}{3}}}^{1+n^{\kappa-\frac{2}{3}}}\hat\psi_2(x)\int_x^\infty\hat\psi_1(y)dydx
            = \frac{1}{2}\tilde c_n^{1/2}c_n^{-3/2}n^{-1}
            + \bigO(n^{-1-\frac{3}{4}\kappa}).
    \]
    Further, from Proposition \ref{proposition: single integrals psi} one has,
    \[
        \int_0^{n^{-1}}\hat\psi_2(x)dx \int_{n^{-1}}^\infty\hat\psi_1(y)dy
            = (-1)^n\tilde c_n^{1/4}c_n^{-3/4}n^{-1}
                    + \bigO(n^{-1-\frac{3}{4}\kappa}).
    \]
    The previous two equations together with (\ref{proposition: single integrals
    psi: eq2}) yield
    \begin{multline} \label{proof: proposition: double integrals out bulk: eq1}
        \int_0^\infty\hat\psi_2(x)\int_x^\infty\hat\psi_1(y)dydx= \left((-1)^n\tilde
            c_n^{1/4}c_n^{-3/4}+\frac{1}{2}\tilde
            c_n^{1/2}c_n^{-3/2}\right)\frac{1}{n} + \bigO(n^{-1-\frac{3}{4}\kappa}) \\
            + \int_{n^{-1}}^{1-n^{\kappa-\frac{2}{3}}} \hat\psi_2(x)\int_x^\infty\hat\psi_1(y)dydx
            + \int_0^{n^{-1}}\hat\psi_2(x)\int_x^{n^{-1}}\hat\psi_1(y)dydx.
    \end{multline}
    For notational convenience let us denote the last double integral of this
    equation again by $J$. Changing
    the order of integration, using the asymptotic behavior of $\hat\psi_1$ in the
    Bessel region given by (\ref{lemma: asymptotics psi: bessel: eq1}), and
    using (\ref{lemma: single integrals psin: eq2}), we obtain
    \begin{align*}
        J  &=\int_0^{n^{-1}}\hat\psi_1(y)\int_0^y\hat\psi_2(x)dxdy \\[1ex]
            &=-\int_0^{n^{-1}}\frac{\tilde c_n^{1/2}n^{1/2}}{\sqrt
            y}J_{\alpha+1}(2\tilde c_n^{1/2}n\sqrt
            y)\int_0^y\hat\psi_2(x)dxdy+\bigO(n^{-1-\frac{1}{4}}).
    \end{align*}
    Changing back the order of integration, using the asymptotic behavior (\ref{lemma: asymptotics psi: bessel: eq2})
    of $\hat\psi_2$ in the Bessel region, and using the fact that $\int_a^b J_{\alpha+1}(u)du$ is uniformly
    bounded for $a,b\in[0,\infty]$, we arrive at
    \begin{align}
        \nonumber
        J &= -n^{-1/2}\int_0^{n^{-1}}\hat\psi_2(x)\int_{2\tilde c_n^{1/2}n\sqrt x}^{2\tilde c_n^{1/2}\sqrt
        n}J_{\alpha+1}(v)dvdx+\bigO(n^{-1-\frac{1}{4}})
        \\[1ex]
        &=n^{-1}\int_0^{2\tilde c_n^{1/2}\sqrt
        n}\left(-2\frac{\alpha}{u}J_\alpha(u)+J_{\alpha+1}(u)\right)\int_u^{2\tilde c_n^{1/2}\sqrt
        n}J_{\alpha+1}(v)dvdu+\bigO(n^{-1-\frac{1}{4}}).
    \end{align}
    Since $ \frac{\alpha}{u}J_\alpha(u)=J_\alpha'(u)+J_{\alpha+1}(u)$, see
    e.g.\ \cite[(9.1.27)]{AbramowitzStegun}, we then have from Proposition
    \ref{proposition3: double integrals out bulk},
    \begin{align}
        \nonumber
        J &= -2n^{-1}\int_0^{2\tilde c_n^{1/2}\sqrt n} J_\alpha'(u)
                \int_u^{2\tilde c_n^{1/2}\sqrt n}J_{\alpha+1}(v)dvdu
        \\[1ex]
        \nonumber
            &\qquad\qquad- n^{-1}\int_0^{2\tilde c_n^{1/2}\sqrt n} J_{\alpha+1}(u)
                \int_u^{2\tilde c_n^{1/2}\sqrt n}J_{\alpha+1}(v)dvdu
            + \bigO(n^{-1-\frac{1}{4}})
        \\[1ex]
        &= -\frac{3}{2}n^{-1}+\bigO(n^{-1-\frac{1}{4}}).
    \end{align}
    Inserting this into (\ref{proof: proposition: double integrals out bulk:
    eq1}) and using (\ref{definition: cn auxiliary results}) and
    (\ref{definition: cntilde auxiliary results}) the Proposition is now proven.
\end{proof}

\subsubsection{The double integrals in the bulk}
\label{sec5.2.2}

Here we will determine the asymptotic behavior (as $n\to\infty$) of the
following three double integrals which appear in Proposition \ref{proposition:
double integrals out bulk},
\[
    J_1 \equiv \int_{p^{-1}}^{1-p^{\kappa-\frac{2}{3}}} \hat\phi_p(x)
                    \int_{x\frac{\beta_p}{\beta_q}}^\infty \hat\phi_q(y) dy dx,
    \qquad
    J_2 \equiv \int_{p^{-1}}^{1-p^{\kappa-\frac{2}{3}}} \hat\phi_p(x)
                    \int_{x\frac{\beta_p}{\beta_n}}^\infty \hat\psi_r(y) dy dx,
\]
and
\[
    J_3 \equiv \int_{n^{-1}}^{1-n^{\kappa-\frac{2}{3}}}
                    \hat\psi_2(x)\int_x^\infty\hat\psi_1(y)dydx,
\]
with $p=n+i$ and $q=n+j$ for some fixed integers $i,j$, and with
$r\in\{1,2\}$. In order to determine the asymptotics
we proceed as in the derivation of the asymptotics of the
double integral $J_3$ under equation (4.120) in \cite{DeiftGioev}. 
We will
need the following auxiliary results.

\begin{proposition}\label{proposition1: double integrals bulk}
    The scalar function
    \begin{equation}\label{definition: theta}
        \theta(x)=\frac{1}{2}\int_0^x\sqrt{\frac{1-s}{s}}h(s)ds,\qquad\mbox{for $x\in[0,1]$,}
    \end{equation}
    satisfies the following differential equation,
    \begin{equation}
        \theta(x)-\frac{1}{m}x\theta'(x)-\pi=-\arccos(2x-1).
    \end{equation}
\end{proposition}

\begin{proof}
    The proof is similar to the proof of \cite[Lemma 4.8]{DeiftGioev}. We will need the first and
    second derivative of $\theta$. From (\ref{definition: theta})
    we have
    \begin{align}\label{derivatives theta: eq1}
        & \theta'(x)=\frac{1}{2}(1-x)^{1/2}x^{-1/2}h(x),\\[1ex]
        \label{derivatives theta: eq2}
        & \theta''(x)=-\frac{1}{4}(1-x)^{-1/2}x^{-3/2}\Bigl(h(x)-2x(1-x)h'(x)\Bigr).
    \end{align}
    Now, we will obtain a convenient expression for $\theta''$
    by deriving a differential 
equation for $h$, cf.~\cite[Proposition 6.2]{DeiftGioev}.
    Since $h(x)=\frac{4m}{2m-1}\, {}_2F_1(1,1-m,3/2-m,x)$, it
    satisfies the following hypergeometric equation (see \cite[(15.5.1)]{AbramowitzStegun}),
    \[
        x(1-x)h''(x)+\left((-m+\frac{3}{2})+(m-3)x\right)h'(x)+(m-1)h(x)=0,
    \]
    which in turn implies that
    \[
        \frac{d}{dx}\Bigl(x(1-x)h'(x)-[(m-1/2)-(m-1)x)]h(x)\Bigr)=0.
    \]
    Therefore, the function inside the outer brackets is a constant,
    which can be determined by letting $x\to 1$. We then obtain the
    following differential equation for $h$,
    \begin{equation}\label{differential equation for h}
        x(1-x)h'(x)-[(m-1/2)-(m-1)x)]h(x)=-\frac{1}{2}h(1)=-2m.
    \end{equation}
    Inserting \eqref{differential equation for h} into (\ref{derivatives theta: eq2})
    we obtain
    \[
        \theta''(x)=-\frac{1}{4}(1-x)^{-1/2}x^{-3/2}\Bigl(4m-2(m-1)(1-x)h(x)\Bigr),
    \]
    which implies, together with (\ref{derivatives theta: eq1}), that
    \[
        \frac{d}{dx}\left(\theta(x)-\frac{1}{m}x\theta'(x)\right)=\frac{(m-1)\theta'(x)-x\theta''(x)}{m}
        =(1-x)^{-1/2}x^{-1/2}.
    \]
    Therefore,
    \begin{equation}
        \theta(x)-\frac{1}{m}x\theta'(x)=\int_0^x\frac{dy}{\sqrt{y(1-y)}}=\pi-\arccos(2x-1),
    \end{equation}
    and the Proposition is proven.
\end{proof}

\begin{proposition} \label{proposition2: double integrals bulk}
    Let $p=n+i$ and $q=n+j$ for some fixed integers $i,j$. 
    Uniformly for $x\in(0,1-p^{\kappa-2/3}]$, as $n\to\infty$,
    \begin{equation} \label{proposition2: double integrals bulk: eq1}
        F_q\left(x\frac{\beta_p}{\beta_q}\right)-F_p(x)
            =-(p-q)\arccos(2x-1)+\bigO(n^{-\frac{1}{3m}}).
    \end{equation}
\end{proposition}

\begin{proof}
    The proof of this Proposition is similar to
the proof of \cite[Lemma 4.7]{DeiftGioev}.
    We write
    the left hand side of (\ref{proposition2: double integrals bulk: eq1}) as,
    \begin{equation} \label{proof: proposition2: double integrals bulk: eq1}
        F_q\left(x\frac{\beta_p}{\beta_q}\right)-F_p(x) =
            \Bigl[F_q\Bigl(x\frac{\beta_p}{\beta_q}\Bigr)-F_q(x)\Bigr] + \Bigl[F_q(x)-F_p(x)\Bigr],
    \end{equation}
    and we treat each of the terms inside the brackets separately.
    First, there exists a number
$\xi_{n,x}$ between $x$ and $x\frac{\beta_p}{\beta_q}$ such that,
    \begin{equation} \label{proof: proposition2: double integrals bulk: eq2}
        F_q\left(x\frac{\beta_p}{\beta_q}\right) - F_q(x) =
            x F_q'(x) \left(\frac{\beta_p}{\beta_q}-1\right)
            + \frac{1}{2}x^2 F_q''(\xi_{n,x}) \left(\frac{\beta_p}{\beta_q}-1\right)^2.
    \end{equation}
    From (\ref{proposition: derivatives FnGn: eq1}), from the fact that
    $h_q(x)=h(x)+\bigO(n^{-1/m})$, and from (\ref{derivatives theta: eq1}) we have
    \[
        xF_q'(x)=-q x\theta'(x)+\bigO(n^{1-1/m})+\bigO(n^{2/3-\kappa}).
    \]
    Further, from (\ref{proposition: derivatives FnGn: eq2}) and from the fact that
    $\xi_{n,x}=x(1+\bigO(1/n))$, we obtain,
    \[
        x^2F_q''(\xi_{n,x})=\bigO(n^{4/3-\frac{1}{2}\kappa}).
    \]
    Inserting these two equations into (\ref{proof: proposition2: double integrals bulk: eq2})
    and using Proposition \ref{proposition1: double integrals out bulk}
    we arrive at
    \begin{equation} \label{proof: proposition2: double integrals bulk: eq3}
        F_q\left(x\frac{\beta_p}{\beta_q}\right)-F_q(x)=-(p-q)\frac{1}{m}x\theta'(x)+\bigO(n^{-1/m})
        +\bigO(n^{-1/3-\kappa}).
    \end{equation}
    Next, we determine the asymptotic behavior of the second term in
    (\ref{proof: proposition2: double integrals bulk: eq1}). Note that by (\ref{definition: Fnj})
    and (\ref{property: equilibrium measure}),
    \[
        F_p(x) =
        p\pi-\frac{p}{2}\int_0^x\sqrt{\frac{1-s}{s}}h_p(s)ds+\frac{1}{2}(\alpha+1)\arccos(2x-1)-\frac{\pi}{4},
    \]
    which implies that
    \[
        F_q(x)-F_p(x)=(q-p)\pi+\frac{1}{2}\int_0^x
        \sqrt{\frac{1-s}{s}}\Bigl(ph_p(s)-qh_q(s)\Bigr)ds.
    \]
    Now,
    \begin{align*}
        ph_p(s)-qh_q(s) &=(p-q)h(s)+\sum_{\ell=1}^m
        h_{(\ell)}(s)(p^{1-\ell/m}-q^{1-\ell/m})+\bigO(n^{-1/m}) \\[1ex]
        &= (p-q) h(s) + \bigO(n^{-1/m}),\qquad\mbox{as $n\to\infty$,}
    \end{align*}
    uniformly for $s\in[0,1]$, so that
    \begin{equation} \label{proof: proposition2: double integrals bulk: eq4}
        F_q(x)-F_p(x)=(p-q)(\theta(x)-\pi)+\bigO(n^{-1/m}).
    \end{equation}
    Inserting equations (\ref{proof: proposition2: double integrals bulk: eq3}) and
    (\ref{proof: proposition2: double integrals bulk: eq4}) into equation
    (\ref{proof: proposition2: double integrals bulk: eq1}), the relation
    (\ref{proposition2: double integrals bulk: eq1}) follows from
    the previous Proposition.
\end{proof}

\subsubsection*{Asymptotics of $J_1$:}

We start with the asymptotic behavior of the double integral $J_1$. From
equations (\ref{proposition: single integrals phin: eq2}) and (\ref{lemma:
single integrals phin: eq2}), and from the asymptotic behavior (\ref{lemma:
asymptotics phin: bulk}) of $\hat\phi_p$ in the bulk region, we obtain
\begin{align}
    \nonumber
    J_1 &= \int_{p^{-1}}^{1-p^{\kappa-\frac{2}{3}}} \hat\phi_p(x)
                \int_{x\frac{\beta_p}{\beta_q}}^{1-q^{\kappa-\frac{2}{3}}}\hat\phi_q(y)dydx
                + \bigO(n^{-1-\frac{3}{4}\kappa})
    \\[1ex]
    \label{determination J1: eq1}
        &= \sqrt{\frac{2}{\pi}}\int_{p^{-1}}^{1-p^{\kappa-\frac{2}{3}}}
            \frac{\cos F_p(x)}{x^{1/4}(1-x)^{1/4}}
            \int_{x\frac{\beta_p}{\beta_q}}^{1-q^{\kappa-\frac{2}{3}}}
            \hat\phi_q(y)dydx + \bigO(n^{-1-\frac{3}{4}\kappa}).
\end{align}
Observe that
$x\frac{\beta_p}{\beta_q}\in[\frac{1}{2}q^{-1},1-\frac{1}{2}q^{\varepsilon-\frac{2}{3}}]$
if $x\in[p^{-1},1-p^{\kappa-\frac{2}{3}}]$ and $n$ is sufficiently large. By
changing the order of integration and using (\ref{proposition3: single
integrals: eq1}) we derive the estimate
\begin{equation}\label{determination J1: change order of integration}
    \int_{p^{-1}}^{1-p^{\kappa-\frac{2}{3}}}
        \frac{\cos
        F_p(x)}{x^{1/4}(1-x)^{1/4}}\int_{x\frac{\beta_p}{\beta_q}}^{1-q^{\kappa-\frac{2}{3}}}
        \bigO\left(\frac{1}{qy^{3/4}(1-y)^{7/4}}\right) dy
        dx
     = \bigO(n^{-1-\frac{3}{2}\kappa}).
\end{equation}
The asymptotic behavior of $\hat\phi_q$ in the bulk region, given by
(\ref{lemma: asymptotics phin: bulk}), together with (\ref{determination J1:
eq1}) and (\ref{determination J1: change order of integration}), leads to
\[
    J_1=\frac{2}{\pi}\int_{p^{-1}}^{1-p^{\kappa-\frac{2}{3}}}
        \frac{\cos
        F_p(x)}{x^{1/4}(1-x)^{1/4}}\int_{x\frac{\beta_p}{\beta_q}}^{1-q^{\kappa-\frac{2}{3}}}
        \frac{\cos F_q(y)}{y^{1/4}(1-y)^{1/4}}dy
        dx+\bigO(n^{-1-\frac{3}{4}\kappa}).
\]
Integrating by parts the inner integral of this expression and using
(\ref{proposition2: single integrals: eq3}) we obtain
    \begin{align*}
        J_1 &= -\,\frac{2}{\pi}\int_{p^{-1}}^{1-p^{\kappa-\frac{2}{3}}}
        \frac{\cos
        F_p(x)}{x^{1/4}(1-x)^{1/4}}
        \frac{\sin F_q(x\frac{\beta_p}{\beta_q})}{F_q'(x\frac{\beta_p}{\beta_q})
        (x\frac{\beta_p}{\beta_q})^{1/4}(1-x\frac{\beta_p}{\beta_q})^{1/4}}
        dx \\[2ex]
        &\qquad\quad +\, \frac{2}{\pi}\int_{p^{-1}}^{1-p^{\kappa-\frac{2}{3}}}
        \frac{\cos
        F_p(x)}{x^{1/4}(1-x)^{1/4}}dx \left.\frac{\sin
        F_q(y)}{F_q'(y)y^{1/4}(1-y)^{1/4}}\right|_{y=1-q^{\kappa-\frac{2}{3}}} \\[2ex]
        &\qquad\quad
        -\, \frac{2}{\pi}\int_{p^{-1}}^{1-p^{\kappa-\frac{2}{3}}}
        \frac{\cos
        F_p(x)}{x^{1/4}(1-x)^{1/4}} \int_{x\frac{\beta_p}{\beta_q}}^{1-q^{\kappa-\frac{2}{3}}}
        \bigO\left(\frac{1}{qy^{3/4}(1-y)^{7/4}}\right)dydx+\bigO(n^{-1-\frac{3}{4}\kappa}).
    \end{align*}
From equations (\ref{proposition2: single integrals: eq1}), (\ref{proposition3:
single integrals: eq1}) and (\ref{determination J1: change order of
integration}), we then have
    \begin{equation}\label{determination J1: eq2}
        J_1= -\frac{2}{\pi}\int_{p^{-1}}^{1-p^{\kappa-\frac{2}{3}}}
        \frac{\cos
        F_p(x)}{x^{1/4}(1-x)^{1/4}}
        \frac{\sin F_q(x\frac{\beta_p}{\beta_q})}{F_q'(x\frac{\beta_p}{\beta_q})
        (x\frac{\beta_p}{\beta_q})^{1/4}(1-x\frac{\beta_p}{\beta_q})^{1/4}}
        dx+\bigO(n^{-1-\frac{3}{4}\kappa}).
    \end{equation}

Now we will determine a convenient expression for the integrand. Note that,
for some $\xi_{n,x}$ between $x$ and $x\frac{\beta_p}{\beta_q}$,
    \[
        \frac{1}{F_q'(x\frac{\beta_p}{\beta_q})}=\frac{1}{F_q'(x)}
            \left[1+x\frac{F_q''(\xi_{n,x})}{F_q'(x)}\left(\frac{\beta_p}{\beta_q}-1\right)\right]^{-1}.
    \]
Since $\xi_{n,x}=x(1+\bigO(1/n))$, one has by Propositions \ref{proposition:
derivatives FnGn} and \ref{proposition1: double integrals out bulk}
    \[
        x\frac{F_q''(\xi_{n,x})}{F_q'(x)}\left(\frac{\beta_p}{\beta_q}-1\right)
        =\bigO\left(\frac{1}{n(1-x)}\right),
    \]
    so that by (\ref{proposition: derivatives FnGn: eq1}),
    \begin{align}
        \nonumber
        \frac{1}{F_q'(x\frac{\beta_p}{\beta_q})
        (x\frac{\beta_p}{\beta_q})^{1/4}(1-x\frac{\beta_p}{\beta_q})^{1/4}}
        &=
        \frac{1}{F_q'(x)
        x^{1/4}(1-x)^{1/4}}\left[1+\bigO\left(\frac{1}{n(1-x)}\right)\right]
        \\[1ex]
        \label{determination J1: eq3}
        &=  \frac{-2x^{1/4}}{q
        h_q(x)(1-x)^{3/4}}\left[1+\bigO\left(\frac{1}{n(1-x)}\right)\right].
    \end{align}
    Inserting this expression into equation (\ref{determination J1: eq2}) we arrive at,
    \begin{align}
        \nonumber
        J_1
        &= \frac{2}{\pi}\int_{p^{-1}}^{1-p^{\kappa-\frac{2}{3}}}
                \frac{2\cos F_p(x)\sin F_q(x\frac{\beta_p}{\beta_q})}{q h_q(x)(1-x)}dx
                + \bigO(n^{-1-\frac{3}{4}\kappa})
        \\[2ex]
        \nonumber
        &= \frac{2}{\pi}\int_{p^{-1}}^{1-p^{\kappa-\frac{2}{3}}}
                \frac{\sin\left(F_q(x\frac{\beta_p}{\beta_q})-F_p(x)\right)}{qh_q(x)(1-x)}dx
            + \frac{2}{\pi}\int_{p^{-1}}^{1-p^{\kappa-\frac{2}{3}}}
                \frac{\sin\left(F_q(x\frac{\beta_p}{\beta_q})+F_p(x)\right)}{qh_q(x)(1-x)}dx
        \\[1ex]
        \nonumber
        & \hspace{11cm} + \bigO(n^{-1-\frac{3}{4}\kappa})
        \\
        & \equiv J_1'+J_1''+ \bigO(n^{-1-\frac{3}{4}\kappa}).
    \end{align}

    It remains to determine the asymptotic behavior of $J_1'$ and $J_1''$.
    Using partial integration and using calculations similar to those used in
    proving (\ref{proposition2: single integrals: eq3}) we can show that
    \[
        J_1''=\bigO(n^{-1-\frac{3}{2}\kappa}).
    \]
    From Proposition \ref{proposition2: double integrals bulk} and from
    $1/h_q(x)=1/h(x)+\bigO(n^{-1/m})$, see (\ref{property: hn: auxiliary results}), we have uniformly
    for $x\in(0,1-p^{\kappa-\frac{2}{3}}]$,
    \[
        \frac{1}{h_q(x)}\sin\left(
        F_q(x\frac{\beta_p}{\beta_q})-F_p(x)\right)=
        -\frac{1}{h(x)}\sin((p-q)\arccos(2x-1))+\bigO(n^{-\frac{1}{3m}}),
    \]
    so that
    \[
        J_1' = -\frac{2}{\pi}\int_{p^{-1}}^{1-p^{\kappa-\frac{2}{3}}}
            \frac{\sin((p-q)\arccos(2x-1))}{q h(x)(1-x)}dx
            +\bigO(n^{-1-\frac{1}{3m}}\log n).
    \]
    In conclusion we have shown that that there exists $0<\tau<1$ such that as
    $n\to\infty$,
    \begin{equation}\label{determination J1: result}
        J_1=-\widehat I(p-q)n^{-1}+\bigO(n^{-1-\tau}),
    \end{equation}
    with $\widehat I$ given by (\ref{definition: Ihat}).

\subsubsection*{Asymptotics of $J_2$:}

Next, we determine the asymptotics of $J_2$. From equations (\ref{proposition:
single integrals phin: eq2}) and (\ref{lemma: single integrals psin: eq2}), and
from the asymptotic behavior of $\hat\phi_p$ in the bulk region given by
(\ref{lemma: asymptotics phin: bulk}), we have,
\begin{align*}
    J_2 &= \int_{p^{-1}}^{1-p^{\kappa-\frac{2}{3}}} \hat\phi_p(x)
                \int_{x\frac{\beta_p}{\beta_n}}^{1-n^{\kappa-\frac{2}{3}}}\hat\psi_r(y)dydx
                + \bigO(n^{-1-\frac{3}{4}\kappa})
    \\[2ex]
        &= \sqrt{\frac{2}{\pi}} \int_{p^{-1}}^{1-p^{\kappa-\frac{2}{3}}}
                \frac{\cos F_p(x)}{x^{1/4}(1-x)^{1/4}}
                \int_{x\frac{\beta_p}{\beta_n}}^{1-n^{\kappa-\frac{2}{3}}}\hat\psi_r(y)dydx
                + \bigO(n^{-1-\frac{3}{4}\kappa}).
\end{align*}
By changing the order of integration and using equation (\ref{proposition3:
single integrals: eq1}) we obtain the analog of equation
(\ref{determination J1: change order of integration}),
\begin{equation}\label{determination J2: change order of integration}
    \int_{p^{-1}}^{1-p^{\kappa-\frac{2}{3}}}
        \frac{\cos F_p(x)}{x^{1/4}(1-x)^{1/4}}
        \int_{x\frac{\beta_p}{\beta_n}}^{1-n^{\kappa-\frac{2}{3}}}
        \bigO\left(\frac{1}{n y^{5/4}(1-y)^{7/4}}\right)dydx
        = \bigO\left(n^{-1-\frac{3}{2}\kappa}\right).
\end{equation}
Using the asymptotic behavior (\ref{lemma: asymptotics psi: bulk: eq1}) of
$\hat\psi_r$ in the bulk region we then obtain,
\begin{align}
    \nonumber
    J_2 &= \frac{(-1)^n \tilde c_n^{1/4}}{\sqrt 2}\frac{2}{\pi}
                \int_{p^{-1}}^{1-p^{\kappa-\frac{2}{3}}}
                \frac{\cos F_p(x)}{x^{1/4}(1-x)^{1/4}}
                \int_{\frac{\beta_p}{\beta_n}x}^{1-n^{\kappa-\frac{2}{3}}}
                \frac{\cos G_n(y)}{y^{3/4}(1-y)^{1/4}}dydx
                + \bigO(n^{-1-\frac{3}{4}\kappa})
    \\[1ex]
        &\equiv \frac{(-1)^n\tilde c_n^{1/4}}{\sqrt 2}\widehat J_2
                + \bigO(n^{-1-\frac{3}{4}\kappa}).
\end{align}
Here we have introduced the notation $\widehat J_2$ for notational convenience.
Integrating the inner integral of $\widehat J_2$ by parts, and using
(\ref{proposition2: single integrals: eq4}) we have,
\begin{align*}
    \widehat J_2 &= -\frac{2}{\pi} \int_{p^{-1}}^{1-p^{\kappa-\frac{2}{3}}}
                \frac{\cos F_p(x)}{x^{1/4}(1-x)^{1/4}}
                \frac{\sin G_n(\frac{\beta_p}{\beta_n}x)}
                {G_n'(\frac{\beta_p}{\beta_n}x)(\frac{\beta_p}{\beta_n}x)^{3/4}
                (1-\frac{\beta_p}{\beta_n}x)^{1/4}}dx
    \\[2ex]
        &\qquad\quad +\, \frac{2}{\pi} \int_{p^{-1}}^{1-p^{\kappa-\frac{2}{3}}}
                \frac{\cos F_p(x)}{x^{1/4}(1-x)^{1/4}}dx
                \left. \frac{\sin G_n(y)}{G_n'(y)y^{3/4}(1-y)^{1/4}}
                \right|_{y=1-n^{\kappa-\frac{2}{3}}}
    \\[2ex]
        &\qquad\quad -\, \frac{2}{\pi} \int_{p^{-1}}^{1-p^{\kappa-\frac{2}{3}}}
                \frac{\cos F_p(x)}{x^{1/4}(1-x)^{1/4}}
                \int_{x\frac{\beta_p}{\beta_n}}^{1-n^{\kappa-\frac{2}{3}}}
                \bigO\left(\frac{1}{n y^{5/4}(1-y)^{7/4}}\right)dydx
                + \bigO(n^{-1-\frac{3}{4}\kappa}).
\end{align*}
From (\ref{proposition2: single integrals: eq2}), (\ref{proposition3: single
integrals: eq1}) and (\ref{determination J2: change order of integration}) we
arrive at,
\[
    \widehat J_2 = - \frac{2}{\pi}\int_{p^{-1}}^{1-p^{\kappa-\frac{2}{3}}}
            \frac{\cos F_p(x)}{x^{1/4}(1-x)^{1/4}}
            \frac{\sin G_n(\frac{\beta_p}{\beta_n}x)}
            {G_n'(\frac{\beta_p}{\beta_n}x)(\frac{\beta_p}{\beta_n}x)^{3/4}
            (1-\frac{\beta_p}{\beta_n}x)^{1/4}}dx
            + \bigO(n^{-1-\frac{3}{4}\kappa}).
\]
As in (\ref{determination J1: eq3}) we are led to
\[
    \frac{1}{G_n'(\frac{\beta_p}{\beta_n}x)(\frac{\beta_p}{\beta_n}x)^{3/4}
    (1-\frac{\beta_p}{\beta_n}x)^{1/4}}=\frac{-2}{nh_n(x)
    x^{1/4}(1-x)^{3/4}}\left[1+\bigO\left(\frac{1}{n(1-x)}\right)\right],
\]
which yields
\begin{align*}
    \widehat J_2 &= \frac{2}{\pi}\int_{p^{-1}}^{1-p^{\kappa-\frac{2}{3}}}
                \frac{2\cos F_p(x)\sin G_n(\frac{\beta_p}{\beta_n}x)}{n h_n(x) x^{1/2}(1-x)}dx
                + \bigO(n^{-1-\frac{3}{4}\kappa})
    \\[2ex] &= \frac{2}{\pi}\int_{p^{-1}}^{1-p^{\kappa-\frac{2}{3}}}
                \frac{\sin\left(G_n(\frac{\beta_p}{\beta_n}x)-F_p(x)\right)}{n h_n(x) x^{1/2}(1-x)}dx
                +
                \frac{2}{\pi}\int_{p^{-1}}^{1-p^{\kappa-\frac{2}{3}}}
                \frac{\sin\left(G_n(\frac{\beta_p}{\beta_n}x)+F_p(x)\right)}{n h_n(x) x^{1/2}(1-x)}dx
    \\[1ex]
      &\hspace{12cm} + \bigO(n^{-1-\frac{3}{4}\kappa})
    \\
                &\equiv \widehat J_2' + \widehat J_2''+ \bigO(n^{-1-\frac{3}{4}\kappa}).
\end{align*}
As before one can show that $\widehat J_2''=\bigO(n^{-1-\frac{3}{2}\kappa})$. We will
now determine the asymptotic behavior of $\widehat J_2'$. Note that by
Proposition \ref{proposition2: double integrals bulk},
\begin{align}
    \nonumber
    G_n(\frac{\beta_p}{\beta_n}x)-F_p(x) &=
        F_n(\frac{\beta_p}{\beta_n}x)-F_p(x) -\frac{1}{2}\arccos\left(2\frac{\beta_p}{\beta_n}x-1\right)
    \\[1ex]
    \nonumber
        &= F_n(\frac{\beta_p}{\beta_n}x)-F_p(x)-\frac{1}{2}\arccos(2x-1)
            + \bigO\left(\frac{x^{1/2}}{n(1-x)^{1/2}}\right)
    \\[1ex]
        &= -(p-n+\frac{1}{2})\arccos(2x-1)+\bigO(n^{-\frac{1}{3m}}),
\end{align}
so that uniformly for $x\in(0,1-p^{\kappa-\frac{2}{3}}]$,
\[
    \frac{1}{h_n(x)}\sin\left(G_n(\frac{\beta_p}{\beta_n}x)-F_p(x)\right)
        =
        -\,\frac{1}{h(x)}\sin\left((p-n+\frac{1}{2})\arccos(2x-1)\right)
        +\bigO(n^{-\frac{1}{3m}}).
\]
Therefore,
\[
    \widehat J_2' = -\frac{2}{\pi} \int_{p^{-1}}^{1-p^{\kappa-\frac{2}{3}}}
                \frac{\sin\left((p-n+\frac{1}{2})\arccos(2x-1)\right)}{n h(x)x^{1/2}(1-x)}dx
        + \bigO(n^{-1-\frac{1}{3m}}\log n).
\]
Using (\ref{definition: cntilde auxiliary results}) we then have shown that
there exists $0<\tau<1$ such that as $n\to\infty$,
\begin{equation}\label{determination J2: result}
    J_2= -(-1)^n\sqrt\frac{m}{2m-1}
    I(p-n+1)n^{-1} + \bigO(n^{-1-\tau}),
\end{equation}
with $I$ given by (\ref{definition: I}).

\subsubsection*{Asymptotics of $J_3$:}

Finally, we will determine the asymptotic behavior of the double integral
$J_3$. From equations (\ref{proposition: single integrals psi: eq1}) and
(\ref{lemma: single integrals psin: eq2}), and from the asymptotic behavior of
$\hat\psi_2$ in the bulk region, given by (\ref{lemma: asymptotics psi: bulk:
eq1}), we have,
\begin{align*}
    J_3 &= \int_{n^{-1}}^{1-n^{\kappa-\frac{2}{3}}} \hat\psi_2 (x)
                \int_x^{1-n^{\kappa-\frac{2}{3}}}\hat\psi_1(y)dydx
                + \bigO(n^{-1-\frac{3}{4}\kappa})
    \\[2ex]
        &= \frac{(-1)^n\tilde c_n^{1/4}}{\sqrt\pi}
                \int_{n^{-1}}^{1-n^{\kappa-\frac{2}{3}}} \frac{\cos G_n(x)}{x^{3/4}(1-x)^{1/4}}
                \int_x^{1-n^{\kappa-\frac{2}{3}}}\hat\psi_1(y)dydx
                + \bigO(n^{-1-\frac{3}{4}\kappa}).
\end{align*}
Now, by changing the order of integration, using the asymptotic behavior
(\ref{lemma: asymptotics psi: bulk: eq1}) of $\hat\psi_1$ in the bulk region,
and using equation (\ref{proposition3: single integrals: eq2}), we arrive at
\begin{align*}
    J_3 &= \frac{(-1)^n\tilde c_n^{1/4}}{\sqrt\pi}
                \int_{n^{-1}}^{1-n^{\kappa-\frac{2}{3}}} \hat\psi_1(y)
                \int_{n^{-1}}^y \frac{\cos G_n(x)}{x^{3/4}(1-x)^{1/4}}dx dy
                + \bigO(n^{-1-\frac{3}{4}\kappa})
    \\[2ex]
        &= \frac{\tilde c_n^{1/2}}{\pi}
                \int_{n^{-1}}^{1-n^{\kappa-\frac{2}{3}}} \frac{\cos G_n(y)}{y^{3/4}(1-y)^{1/4}}
                \int_{n^{-1}}^y \frac{\cos G_n(x)}{x^{3/4}(1-x)^{1/4}}dxdy
                + \bigO(n^{-1-\frac{3}{4}\kappa}).
\end{align*}
Integrating by parts the inner integral and using (\ref{proposition2: single
integrals: eq4}) we then obtain,
\begin{align*}
    J_3 &= \frac{\tilde c_n^{1/2}}{\pi}
                \int_{n^{-1}}^{1-n^{\kappa-\frac{2}{3}}}
                \frac{\cos G_n(y)\sin G_n(y)}{G_n'(y)y^{3/2}(1-y)^{1/2}}dy
    \\[2ex]
        & \qquad\quad -\, \frac{\tilde c_n^{1/2}}{\pi}
                \int_{n^{-1}}^{1-n^{\kappa-\frac{2}{3}}}
                \frac{\cos G_n(y)}{y^{3/4}(1-y)^{1/4}}dy
                \left.\frac{\sin G_n(x)}{G_n'(x)x^{3/4}(1-x)^{1/4}}\right|_{x=n^{-1}}
    \\[2ex]
        & \qquad\quad -\, \frac{\tilde c_n^{1/2}}{\pi}
                \int_{n^{-1}}^{1-n^{\kappa-\frac{2}{3}}}
                \frac{\cos G_n(y)}{y^{3/4}(1-y)^{1/4}}
                \int_{n^{-1}}^y \bigO\left(\frac{1}{nx^{5/4}(1-x)^{7/4}}\right)
                dxdy
                + \bigO(n^{-1-\frac{3}{4}\kappa}).
\end{align*}
From (\ref{proposition2: single integrals: eq2}), (\ref{proposition3: single
integrals: eq2}) and from the fact that
\[
    \int_{n^{-1}}^{1-n^{\kappa-\frac{2}{3}}}
        \frac{\cos G_n(y)}{y^{3/4}(1-y)^{1/4}} \int_{n^{-1}}^y
        \bigO\left(\frac{1}{nx^{5/4}(1-x)^{7/4}}\right)dxdy
        = \bigO(n^{-1-\frac{3}{4}\kappa}),
\]
which follows from changing the order of integration together with equation
(\ref{proposition3: single integrals: eq2}), we then obtain,
\begin{equation}
    J_3 = \frac{\tilde c_n^{1/2}}{2\pi}
                \int_{n^{-1}}^{1-n^{\kappa-\frac{2}{3}}}
                \frac{\sin (2G_n(y))}{G_n'(y)y^{3/2}(1-y)^{1/2}}dy
                + \bigO(n^{-1-\frac{3}{4}\kappa}).
\end{equation}
Integrating by parts once more we have,
\begin{multline*}
    J_3 = \frac{\tilde c_n^{1/2}}{2\pi} \int_{n^{-1}}^{1-n^{\kappa-\frac{2}{3}}}
            \frac{\cos(2G_n(y))}{G_n'(y) y^{3/4}(1-y)^{1/4}}
            \left(\frac{1}{G_n'(y) y^{3/4}(1-y)^{1/4}}\right)'dy
    \\[2ex]
        -\, \frac{\tilde c_n^{1/2}}{4\pi}\cos(2G_n(y))
            \left. \left(\frac{1}{G_n'(y) y^{3/4}(1-y)^{1/4}}\right)^2\
            \right|_{y=n^{-1}}^{1-n^{\kappa-\frac{2}{3}}}
        + \bigO(n^{-1-\frac{3}{4}\kappa}).
\end{multline*}
Using (\ref{proposition2: single integrals: eq2}) and (\ref{proposition2:
single integrals: eq4}) we finally arrive at,
\begin{equation}\label{determination J3: result}
    J_3 = \bigO(n^{-1-\frac{3}{4}\kappa}),
        \qquad\mbox{as $n\to\infty$.}
\end{equation}

\subsubsection{The result}
\label{sec5.2.3}

\begin{lemma}\label{lemma: double integrals}
    Let $p=n+i$ and $q=n+j$ with $i,j$ some fixed integers and let $r\in\{1,2\}$.
    There exists $0<\tau=\tau(m,\alpha)<1$ such that as $n\to\infty$,
    \begin{align}
        \label{lemma: double integrals: eq1}
        & \int_0^\infty \hat\phi_p(x)
            \int_{x\frac{\beta_p}{\beta_q}}^\infty\hat\phi_q(y) dy dx
            = \left(\frac{1}{2m} - \widehat I(p-q)+ \bigO(n^{-\tau})\right)\frac{1}{n}, \\[2ex]
        \label{lemma: double integrals: eq2}
        & \int_0^\infty \hat\phi_p(x)
            \int_{x\frac{\beta_p}{\beta_n}}^\infty \hat\psi_r(y) dy dx
            = (-1)^n\sqrt{\frac{m}{2m-1}}\left( \frac{1}{2m}
            - I(p-n+1)+\bigO(n^{-\tau})\right)\frac{1}{n},
    \end{align}
    and
    \begin{equation}\label{lemma: double integrals: eq3}
        \int_0^\infty \hat\psi_2(x)\int_x^\infty \hat\psi_1(y) dy dx
            = \left(-\frac{3}{2}+\frac{(-1)^n}{\sqrt{2m-1}}
                + \frac{1}{2}\frac{1}{2m-1}+\bigO(n^{-\tau})\right)\frac{1}{n}.
    \end{equation}
\end{lemma}

\begin{proof}
    The Lemma is immediate from Proposition \ref{proposition: double integrals out bulk} and
    from equations
    (\ref{determination J1: result}), (\ref{determination J2: result}) and
    (\ref{determination J3: result}).
\end{proof}

\subsection{Asymptotics of the  matrix $B$}
\label{sec5.3}

Let $p=n+i$ and $q=n+j$ with $i,j$ some fixed integers and let $r\in\{1,2\}$.
From equations (\ref{proof: lemma: B12matrix: eq1})--(\ref{proof: lemma:
B12matrix: eq3}), from Lemmas \ref{lemma: single integrals phin}, \ref{lemma:
single integrals psin} and \ref{lemma: double integrals}, and from Proposition
\ref{proposition1: double integrals out bulk}, it is immediate that there exists
$0<\tau<1$ such that as $n\to\infty$, $n$ even,
\begin{align}
    \label{proof: lemma B12 matrix: innerproducts: eq1}
    & \langle\varepsilon\phi_q,\phi_p\rangle=\frac{\beta_n}{n}\left(\widehat
    I(p-q)+\bigO(n^{-\tau})\right), \\[2ex]
    \label{proof: lemma B12 matrix: innerproducts: eq2}
    & \langle\varepsilon\psi_r,\phi_p\rangle=\frac{\beta_n}{n}\left(
    \sqrt\frac{m}{2m-1}I(p-n+1)+\frac{(-1)^r}{2\sqrt
    m}+\bigO(n^{-\tau})\right), \\[2ex]
    \label{proof: lemma B12 matrix: innerproducts: eq3}
    & \langle\varepsilon\psi_1,\psi_2\rangle
        =\frac{\beta_n}{n}\left(1-\frac{1}{\sqrt{2m-1}}+\bigO(n^{-\tau})\right).
\end{align}
These equations prove Lemma \ref{lemma: B12matrix}.

\section{Proof of the main results}
\label{sec6}

Based on the results of the previous sections we will now prove our main 
results stated in the Introduction to this paper. Recall that the strategy of 
the proofs was outlined in Remark \ref{remark:important}. We will treat the 
different spectral regions (bulk, hard and soft edge) each in a seperate 
subsection. Full proofs are provided for the hard edge which has no
analogue in the Hermite case. For the soft edge and the bulk we 
do not repeat arguments already presented in \cite{DeiftGioev2,DeiftGioev}.

\subsection{The hard edge of the spectrum}
\label{sec6.1}

\begin{proof}[Proof of Theorem \ref{theorem: universality hard edge}(i).]
This result for $\beta=2$ has been proven by one of the authors in \cite[Theorem 2.8(c)]{v6}, see also Proposition \ref{proposition: universality hard edge:
derivates Kn} below.
\end{proof}

In order to prove Theorem \ref{theorem: universality hard edge}
for $\beta=1$, $4$ we proceed as in the proof of \cite[Theorem
1.1]{DeiftGioev2}. We need the following six auxiliary
propositions (Propositions \ref{proposition: universality hard
edge: derivates Kn}--\ref{corollary: hard universality: intPhi
epsilon}).

\begin{proposition}\label{proposition: universality hard edge:
derivates Kn}
    Let $k,j\in\mathbb{N}$. As $n\to\infty$, uniformly for $\xi,\eta$ in 
bounded subsets of
    $(0,\infty)$,
    \begin{equation}
        \frac{\partial^{k+j}}{\partial\xi^k\partial\eta^j}
        \left[\frac{1}{\nu_n^2}K_n(\tilde\xi^{(n)},\tilde\eta^{(n)})\right]=
        \frac{\partial^{k+j}}{\partial\xi^k\partial\eta^j}
        K_J(\xi,\eta)+\bigO\left(\frac{\xi^{\frac{\alpha}{2}-k}\eta^{\frac{\alpha}{2}-j}}{n}\right).
    \end{equation}
\end{proposition}

\begin{proof}
    For the sake of brevity, we introduce the following notation,
    \begin{align}
        & z_n=\frac{z}{4\tilde c_n n^2},
            && \tilde z_n=2(-\tilde f_n(z_n))^{1/2},
        \\[1ex]
        & \chi_{1,n}(z)=z^{-\alpha/2}\tilde z_n J_\alpha'(\tilde z_n),
            && \chi_1(z)=z^{-\alpha/2}z^{1/2}J_\alpha'(z^{1/2}),
            && \widehat\chi_{1,n}=\chi_{1,n}-\chi_1,
        \\[2ex]
        & \chi_{2,n}(z)=z^{-\alpha/2}J_\alpha(\tilde z_n),
            && \chi_2(z)=z^{-\alpha/2}J_\alpha(z^{1/2}),
            && \widehat\chi_{2,n}=\chi_{2,n}-\chi_2.
    \end{align}
    With this notation we obtain from \cite[(6.1), (6.4) 
and (6.5)]{v6}
    \begin{align}
        \nonumber
        &\xi^{-\frac{\alpha}{2}}\eta^{-\frac{\alpha}{2}}\left(\frac
{1}{\nu_n^2}K_n(\tilde\xi^{(n)},\tilde\eta^{(n)})
            -K_J(\xi,\eta)\right)\\[3ex]
        \nonumber
        &\quad =
            \frac{1}{2(\xi-\eta)}
            \begin{pmatrix}
                \chi_{1,n}(\eta) & \chi_{2,n}(\eta)
            \end{pmatrix}
            \begin{pmatrix}
                \chi_{2,n}(\xi) \\
                -\chi_{1,n}(\xi)
            \end{pmatrix}
            -\frac{1}{2(\xi-\eta)}
            \begin{pmatrix}
                \chi_1(\eta) & \chi_2(\eta)
            \end{pmatrix}
            \begin{pmatrix}
                \chi_2(\xi) \\
                -\chi_1(\xi)
            \end{pmatrix}
        \\[1ex]
        \nonumber
        &\qquad\qquad + \frac{1}{2\pi i(\xi-\eta)}
            \begin{pmatrix}
                \pi i\chi_{1,n}(\eta) & \chi_{2,n}(\eta)
            \end{pmatrix}
            \left(L_n^{-1}(\eta_n)L_n(\xi_n)-I\right)
            \begin{pmatrix}
                \chi_{2,n}(\xi) \\
                -\pi i\chi_{1,n}(\xi)
            \end{pmatrix}
        \\[4ex]
        \nonumber
        &\quad=
        \begin{pmatrix}
            {\ds \frac{\widehat\chi_{1,n}(\eta)-\widehat\chi_{1,n}
(\xi)}{2(\xi-\eta)}} &
                {\ds
                \frac{\widehat\chi_{2,n}(\eta)-\widehat\chi_{2,n}
(\xi)}{2(\xi-\eta)}}
        \end{pmatrix}
        \begin{pmatrix}
            \chi_{2,n}(\xi) \\-\chi_{1,n}(\xi)
        \end{pmatrix}\\[2ex]
        \nonumber
        &\qquad\qquad+\begin{pmatrix}
            {\ds \frac{\chi_1(\eta)-\chi_1(\xi)}{2(\xi-\eta)}} &
            {\ds \frac{\chi_2(\eta)-\chi_2(\xi)}{2(\xi-\eta)}}
        \end{pmatrix}
        \begin{pmatrix}
            \widehat\chi_{2,n}(\xi) \\
            -\widehat\chi_{1,n}(\xi)
        \end{pmatrix}\\[1ex]
        &\qquad\qquad + \frac{1}{2\pi i(\xi-\eta)}
            \begin{pmatrix}
                \pi i\chi_{1,n}(\eta) & \chi_{2,n}(\eta)
            \end{pmatrix}
            \left(L_n^{-1}(\eta_n)L_n(\xi_n)-I\right)
            \begin{pmatrix}
                \chi_{2,n}(\xi) \\
                -\pi i\chi_{1,n}(\xi)
            \end{pmatrix},
         \label{eqn:6.23}   
    \end{align}
    where $L_n$ is the $2\times 2$ matrix valued function defined 
in \cite[Lemma 6.1]{v6}.
    We will now denote the first term of the right hand side of
    equation (\ref{eqn:6.23}) by $H_{n,1}(\xi,\eta)$, the second term by
    $H_{n,2}(\xi,\eta)$, and the third term by
    $H_{n,3}(\xi,\eta)$.

    Observe that it is sufficient to show
    that the following estimates hold as $n\to\infty$, uniformly for
    $\xi,\eta$ in bounded subsets of $(0,\infty)$,
    \begin{equation}\label{proof: besselderivaties: sufficiency}
        \frac{\partial^{k+j}}{\partial\xi^k\partial\eta^j}
        H_{n,i}(\xi,\eta)=\bigO(1/n),\qquad i=1,2,3.
    \end{equation}
    Since $z^{-\alpha}J_\alpha(z)$ is even and entire \cite[(9.1.10)]{AbramowitzStegun} it follows
    that $\chi_1$ and $\chi_2$ are also entire. 
Further, from the 
form (\ref{definition: fn fntilde})
    of $\tilde f_n$ we have that
    $\chi_{1,n}(z)$ and $\chi_{2,n}(z)$ (and hence also $\widehat\chi_
{1,n}$ and $\widehat\chi_{2,n}$) are
    analytic for $z$ in compact subsets of $\mathbb{C}$ and $n$ 
sufficiently
    large, and that
        $\widehat\chi_{i,n}(z)=\chi_{i,n}(z)-\chi_i(z)=\bigO(1/n^2)$,
        for $i=1,2$,
    as $n\to\infty$, uniformly for $z$ in compact subsets of
    $\mathbb{C}$. 
Using the above properties we observe for $i=1,2$ and
    $\ell_1,\ell_2\in\mathbb{N}$ that all derivatives
    \begin{equation}\label{proof: besselderivatives: eq1}
        \frac{\partial^{\ell_1+\ell_2}}{\partial\xi^{\ell_1}
\partial\eta^{\ell_2}}\frac{\chi_i(\xi)-\chi_i(\eta)}{\xi-\eta},
        \qquad\mbox{remain bounded for $\xi,\eta$ in compact 
subsets of $\mathbb{C}$,}
    \end{equation}
    and that,
    \begin{align}\label{proof: besselderivatives: eq2}
        &\frac{\partial^{\ell_1+\ell_2}}{\partial\xi^{\ell_1}
\partial\eta^{\ell_2}}
        \frac{\widehat\chi_{i,n}(\xi)-\widehat\chi_{i,n}(\eta)}{\xi-
\eta}=\bigO(1/n^2),\\[1ex]
        \label{proof: besselderivatives: eq3}
        &\frac{\partial^{\ell_1}}{\partial\xi^{\ell_1}}\widehat\chi_
{i,n}(\xi)=\bigO(1/n^2),\qquad
        \frac{\partial^{\ell_1}}{\partial\xi^{\ell_1}}\chi_{i,n}(\xi)
=\bigO(1),
    \end{align}
    as $n\to\infty$, uniformly for $\xi,\eta$ in compact subsets of
    $\mathbb{C}$. From (\ref{proof: besselderivatives: eq1})--(\ref
{proof: besselderivatives: eq3}) it
    now follows that (\ref{proof: besselderivaties: sufficiency}) 
holds for
    $i=1,2$.

    As in the proof of \cite[Lemma 6.1]{v6} one can
    show, by writing $L_n^{-1}(\eta_n)L_n(\xi_n)-I$ as a contour
    integral, that
    \[
        \frac{\partial^{\ell_1+\ell_2}}{\partial\xi^{\ell_1}
\partial\eta^{\ell_2}}
        \frac{L_n^{-1}(\eta_n)L_n(\xi_n)-I}{\xi-\eta}=\bigO(1/n),
    \]
    as $n\to\infty$, uniformly for $\xi,\eta$ in bounded subsets of
    $(0,\infty)$. This together with (\ref{proof: besselderivatives:
    eq3}) then proves (\ref{proof: besselderivaties: sufficiency}) for 
$i=3$, as well.
    Hence, the Proposition is proven.
\end{proof}

\begin{proposition}\label{proposition: universality hard edge:
integrals Kn}
    As $n\to\infty$, uniformly for $\xi,\eta$ in bounded subsets of
    $(0,\infty)$,
    \begin{align}
        \label{proposition: universality hard edge:
integrals Kn: eq1}
        & \int_0^\xi \frac{1}{\nu_n^2}K_n(\tilde
        s^{(n)},\tilde\eta^{(n)})ds=\int_0^\xi K_J(s,\eta)ds
        +\bigO\left(\frac{\xi^{\frac{\alpha}{2}+1}\eta^{\frac{\alpha}{2}}}{n}\right),
        \\[1ex]
        \label{proposition: universality hard edge:
integrals Kn: eq2}
        & \int_\xi^\eta \frac{1}{\nu_n^2}K_n(\tilde
        s^{(n)},\tilde\eta^{(n)})ds=\int_\xi^\eta K_J(s,\eta)ds
        +\bigO\left(\frac{\eta^{\frac{\alpha}{2}}}{n}\right).
    \end{align}
\end{proposition}

\begin{proof}
    This is immediate from Proposition \ref{proposition: universality hard edge:
derivates Kn}.
\end{proof}

\begin{proposition}\label{proposition: epsilonPhi}
    There exists $0<\tau=\tau(m,\alpha)<1$ such that as (even) $n\to\infty$,
    \begin{align}
        \varepsilon\Phi_1(+\infty) &=\frac{1}{2}\sqrt\frac{\beta_n}{n}\left[\frac{1}{\sqrt m}\,{\bf a}
            -{\bf e} +\bigO(n^{-\tau})\right],
        \\[2ex]
        \varepsilon\Phi_2(+\infty) &=\frac{1}{2}\sqrt\frac{\beta_n}{n}\left[\frac{1}{\sqrt m}\,{\bf a}
            +{\bf e} +\bigO(n^{-\tau})\right],
    \end{align}
    where {\bf a} and {\bf e} are $m$-dimensional row vectors given by,
    \begin{equation}\label{definition: a and e}
        {\bf a}=\left(1,\ldots ,1,\sqrt\frac{m}{2m-1}\right),\qquad {\bf e}=(0,\ldots ,0,1).
    \end{equation}
\end{proposition}

\begin{proof}
    Fix $j\in\mathbb{Z}$ and let $r=1,2$. From Lemma \ref{lemma: single integrals phin}
    and Proposition \ref{proposition1: double integrals out bulk} we have
    \begin{align*}
        \int_0^\infty\phi_{n+j}(x)dx &=\sqrt{\beta_{n+j}}\int_0^\infty\hat\phi_{n+j}(x)dx
            =\sqrt\frac{\beta_{n+j}}{n+j}\left(\frac{1}{\sqrt m}+\bigO(n^{-\tau})\right) \\[1ex]
            &= \sqrt\frac{\beta_n}{n}\left(\frac{1}{\sqrt m}+\bigO(n^{-\tau})\right),
    \end{align*}
    and from Lemma \ref{lemma: single integrals psin} we have for $n$ even,
    \[
        \int_0^\infty\psi_r(x)dx =\sqrt{\beta_n}\int_0^\infty\hat\psi_r(x)dx=\sqrt\frac{\beta_n}{n}
        \left(\frac{1}{\sqrt{2m-1}}+(-1)^r+\bigO(n^{-\tau})\right),
    \]
    for some $0<\tau<1$. Since $\varepsilon\Phi_r(+\infty)=\frac{1}{2}\int_0^\infty\Phi_r(x)dx$ this
    proves the Proposition.
\end{proof}

\begin{proposition}\label{proposition: hard universality: Phi}
    Uniformly for $\xi$ in bounded subsets of $(0,\infty)$, as $n\to\infty$
    \begin{align}
        \label{proposition: hard universality: Phi: eq1}
        & \frac{1}{\nu_n^2}\Phi_1(\tilde\xi^{(n)})=-\frac{1}{2}\sqrt\frac{\beta_n}{n}
            \left[\frac{J_{\alpha+1}(\sqrt \xi)}{\sqrt \xi} \cdot {\bf e}
            +\bigO\left(\frac{\xi^{\frac{\alpha}{2}}}{n}\right) \right],
        \\[2ex]
        \label{proposition: hard universality: Phi: eq2}
        & \frac{1}{\nu_n^2}\Phi_2(\tilde\xi^{(n)})=-\frac{1}{2}\sqrt\frac{\beta_n}{n}
            \left[\left(\frac{J_{\alpha+1}(\sqrt \xi)}{\sqrt \xi}
                -\frac{2\alpha}{\xi}J_\alpha(\sqrt \xi) \right)\cdot {\bf e}
            +\bigO\left(\frac{\xi^{\frac{\alpha}{2}}}{n}\right) \right].
    \end{align}
\end{proposition}

\begin{proof}
    The Proposition follows from equations (\ref{lemma: asymptotics psi: bessel:
    eq1}) and (\ref{lemma: asymptotics psi: bessel: eq2}), and from the fact that for every $j\in\mathbb Z$,
    \begin{align*}
        \frac{1}{\nu_n^2}\phi_{n+j}(\tilde \xi^{(n)}) &=
            \frac{\beta_n}{4\tilde c_n n^2}
            \phi_{n+j}\left(\beta_{n+j}\frac{\xi}{4 \tilde c_n n^2}\frac{\beta_n}{\beta_{n+j}}\right)=
            \frac{\beta_n}{4\tilde c_n n^2}
            \frac{1}{\sqrt{\beta_{n+j}}}\hat\phi_{n+j}
            \left(\frac{\xi}{4 \tilde c_n n^2}\frac{\beta_n}{\beta_{n+j}}\right)
            \\[1ex] &=\bigO\left(\sqrt\frac{\beta_n}{n}
            \frac{\xi^{\frac{\alpha}{2}}}{n}\right).
    \end{align*}
    In the last equality we have used
    (\ref{lemma: asymptotics phin: bessel}).
\end{proof}

\begin{proposition} \label{Pinthard}
    Uniformly for $\xi,\eta$ in bounded subsets of $(0,\infty)$, as $n\to\infty$
    \begin{align}
        \label{corollary: hard universality: intPhi: eq1}
        & \int_0^{\tilde\eta^{(n)}}\Phi_1(s)ds=
        -\sqrt{\frac{\beta_n}{n}}\left[\int_0^{\sqrt\eta}J_{\alpha+1}(s)ds \cdot {\bf e}
            +\bigO\left(\frac{\eta^{\frac{\alpha}{2}+1}}{n}\right) \right],
        \\[2ex]
        \label{corollary: hard universality: intPhi: eq2}
        & \int_0^{\tilde\eta^{(n)}}\Phi_2(s)ds=-\sqrt{\frac{\beta_n}{n}}\left[\int_0^{\sqrt \eta}\left(
        J_{\alpha+1}(s)-\frac{2\alpha}{s}J_\alpha(s)\right)ds \cdot {\bf e}
            +\bigO\left(\frac{\eta^{\frac{\alpha}{2}+1}}{n}\right) \right],
        \\[2ex]
        \label{corollary: hard universality: intPhi: eq3}
        & \int_{\tilde\xi^{(n)}}^{\tilde\eta^{(n)}}\Phi_1(s)ds=
            -\sqrt{\frac{\beta_n}{n}}\left[\int_{\sqrt\xi}^{\sqrt\eta}J_{\alpha+1}(s)ds \cdot {\bf e}
            +\bigO\left(\frac{1}{n}\right) \right].
    \end{align}
\end{proposition}

\begin{proof}
    This is immediate from Proposition \ref{proposition: hard universality: Phi}.
\end{proof}

\begin{proposition}\label{corollary: hard universality: intPhi epsilon}
    There exists $0<\tau=\tau(m,\alpha)<1$ such that, uniformly for $\eta$ in
    bounded subsets of $(0,\infty)$, as $n\to\infty$, $n$ even,
    \begin{equation}
        \int_0^{\tilde\eta^{(n)}}\Phi_1(s)ds-\varepsilon\Phi_1(+\infty)+\varepsilon\Phi_2(+\infty)
        = \sqrt{\frac{\beta_n}{n}}\left[\int_{\sqrt\eta}^\infty
                J_{\alpha+1}(s)ds \cdot {\bf e}
                +\bigO(n^{-\tau})
                \right],
    \end{equation}
    \begin{multline}
        \int_0^{\tilde\eta^{(n)}}\Phi_2(s)ds-\varepsilon\Phi_2(+\infty)+\varepsilon\Phi_1(+\infty)
        \\[1ex]
            = \sqrt{\frac{\beta_n}{n}}\left[\int_{\sqrt\eta}^\infty\left(
                J_{\alpha+1}(s)-\frac{2\alpha}{s}J_\alpha(s)\right)ds \cdot {\bf e}
                +\bigO(n^{-\tau})
                \right].
    \end{multline}
\end{proposition}

\begin{proof}
    This follows from equations (\ref{corollary: hard universality: intPhi: eq1})
    and (\ref{corollary: hard universality: intPhi: eq2}), from Proposition \ref{proposition: epsilonPhi},
    and from the facts that $\int_0^{\infty}J_{\alpha+1}(s)ds=1$ and
    $\int_0^\infty\frac{\alpha}{s}J_\alpha(s)ds=1$, see \cite[(11.4.16)
 and (11.4.17)]{AbramowitzStegun}.
\end{proof}

Now we have the necessary ingredients to prove Theorem
\ref{theorem: universality hard edge} for the cases $\beta=1,4$.

\begin{proof}[Proof of Theorem \ref{theorem: universality hard edge}(ii).]

\medskip

\noindent \textsc{The $(1,1)$- and $(2,2)$-entry:} By
(\ref{definition: conjugate Kn}), (\ref{definition: Kn4}) and
(\ref{convenient formula S4: hard edge}), we have
\begin{multline*}
    \frac{2}{\nu_n^2}\left[K_{\frac{n}{2},4}^{(\nu_n)}(\tilde\xi^{(n)},\tilde\eta^{(n)})\right]_{11}
    =\frac{1}{\nu_n^2}S_{\frac{n}{2},4}(\tilde\xi^{(n)},\tilde\eta^{(n)}) \\[1ex]
    =
    \frac{1}{\nu_n^2}K_n(\tilde\xi^{(n)},\tilde\eta^{(n)})
    -\frac{1}{\nu_n^2}\Phi_2(\tilde\xi^{(n)})A_{21}\int_0^{\tilde\eta^{(n)}}\Phi_1(s)^tds
        -\frac{1}{\nu_n^2}\Phi_2(\tilde\xi^{(n)})G_{11}\int_0^{\tilde\eta^{(n)}}\Phi_2(s)^tds.
\end{multline*}
The asymptotics of the first term of the right hand side of the
latter equation have been determined in part (i) of the theorem.
From  (\ref{proposition: hard universality: Phi: eq2}),
(\ref{corollary: hard universality: intPhi: eq1}), and the facts
that ${\bf e}A_{21}{\bf e}^t=-\frac{1}{2}\frac{n}{\beta_n}$ (which
follows from equation (\ref{determination A21})) and
$A_{21}=\bigO(\frac{n}{\beta_n})$ (see Lemma \ref{lemma:
A21matrix}), we obtain
\begin{align*}
    &
    \frac{1}{\nu_n^2}\Phi_2(\tilde\xi^{(n)})A_{21}\int_0^{\tilde\eta^{(n)}}\Phi_1(s)^tds
    \\[1ex]
    &\qquad\qquad= \frac{1}{2}
            \left[\left(\frac{J_{\alpha+1}(\sqrt \xi)}{\sqrt \xi}
                -\frac{2\alpha}{\xi}J_\alpha(\sqrt \xi) \right)\cdot {\bf e}
            +\bigO\left(\frac{\xi^{\frac{\alpha}{2}}}{n}\right) \right]\frac{\beta_n}{n} A_{21}
            \\[1ex]
    &\hspace{8cm} \times\left[\int_0^{\sqrt\eta}J_{\alpha+1}(s)ds \cdot {\bf
    e}^t
            +\bigO\left(\frac{\eta^{\frac{\alpha}{2}+1}}{n}\right) \right]
            \\[1ex]
    &\qquad\qquad= -\frac{1}{4}\left(\frac{J_{\alpha+1}(\sqrt\xi)}{\sqrt\xi}
    -\frac{2\alpha}{\xi}J_\alpha(\sqrt\xi)
    \right)\int_0^{\sqrt\eta}J_{\alpha+1}(s)ds
    +\bigO\left(\frac{\xi^{\frac{\alpha}{2}-1}\eta^{\frac{\alpha}{2}+1}}{n}\right).
\end{align*}
From (\ref{proposition: hard universality: Phi: eq2}),
(\ref{corollary: hard universality: intPhi: eq2}), and the facts
that ${\bf e} G_{11} {\bf e}^t=0$ (which follows from the skew
symmetry of $G_{11}$, see Lemma \ref{lemma: simplification Widom})
and $G_{11}=\bigO(\frac{n}{\beta_n})$ (see Corollary
\ref{corollary: asymptotics Gmatrices}), we have
\begin{align*}
    & \frac{1}{\nu_n^2}\Phi_2(\tilde\xi^{(n)})G_{11}\int_0^{\tilde\eta^{(n)}}\Phi_2(s)^tds \\[1ex]
    &\qquad\qquad=
    \left[\bigO(\xi^{\frac{\alpha}{2}-1})\cdot\textbf{e}+
    \bigO\left(\frac{\xi^{\frac{\alpha}{2}}}{n}\right)\right]
    \frac{\beta_n}{n} G_{11}
    \left[\bigO(\eta^{\frac{\alpha}{2}})\cdot\textbf{e}^t
    +\bigO\left(\frac{\eta^{\frac{\alpha}{2}+1}}{n}\right)\right]
        \\[1ex]
        &\qquad\qquad=\bigO\left(\frac{\xi^{\frac{\alpha}{2}-1}\eta^{\frac{\alpha}{2}}}{n}\right).
\end{align*}
We conclude that
\begin{multline} \label{t6.0}
    \frac{2}{\nu_n^2}\left[K_{\frac{n}{2},4}^{(\nu_n)}(\tilde\xi^{(n)},\tilde\eta^{(n)})\right]_{11}
    \\[1ex]
    = K_J(\xi,\eta)+\frac{1}{4}\left(\frac{J_{\alpha+1}(\sqrt\xi)}{\sqrt\xi}
    -\frac{2\alpha}{\xi}J_\alpha(\sqrt\xi)
    \right)\int_0^{\sqrt\eta}J_{\alpha+1}(s)ds+
    \bigO\left(\frac{\xi^{\frac{\alpha}{2}-1}\eta^{\frac{\alpha}{2}}}{n}\right).
\end{multline}

\medskip

\noindent \textsc{The $(1,2)$-entry}: Again by (\ref{convenient formula S4:
hard edge}) we have,
\[
    (-\frac{\partial}{\partial y}S_{\frac{n}{2},4})(x,y)
       =-\frac{\partial}{\partial y}K_n(x,y)+\Phi_2(x)A_{21}\Phi_1(y)^t
        +\Phi_2(x)G_{11}\Phi_2(y)^t.
\]
As for the $(1,1)$- and $(2,2)$-entry, we obtain
from Propositions \ref{proposition: universality hard edge:
derivates Kn} and \ref{proposition: hard universality: Phi},
\begin{align} \label{t6.01}
    \nonumber
    &\frac{2}{\nu_n^2}\left[K_{\frac{n}{2},4}^{(\nu_n)}(\tilde\xi^{(n)},\tilde\eta^{(n)})\right]_{12} =
        \frac{1}{\nu_n^4}
         (-\frac{\partial}{\partial y}S_{\frac{n}{2},4})(\tilde\xi^{(n)},\tilde\eta^{(n)})
    \\[2ex]
    \nonumber
    &\quad= -\frac{\partial}{\partial \eta}\left(\frac{1}{\nu_n^2}K_n(\tilde\xi^{(n)},\tilde\eta^{(n)})\right)
    +\frac{1}{\nu_n^4}\Phi_2(\tilde\xi^{(n)})A_{21}\Phi_1(\tilde\eta^{(n)})^t
        +\frac{1}{\nu_n^4}\Phi_2(\tilde\xi^{(n)})G_{11}\Phi_2(\tilde\eta^{(n)})^t
    \\[2ex]
    &\quad=-\frac{\partial}{\partial\eta}K_J(\xi,\eta)
        -\frac{1}{8}\left(\frac{J_{\alpha+1}(\sqrt\xi)}{\sqrt\xi}-\frac{2\alpha}{\xi}J_\alpha(\sqrt\xi) \right)
        \frac{J_{\alpha+1}(\sqrt\eta)}{\sqrt\eta}
        +\bigO\left(\frac{\xi^{\frac{\alpha}{2}-1}\eta^{\frac{\alpha}{2}-1}}{n}\right).
\end{align}

\medskip

\noindent \textsc{The $(2,1)$-entry:} Using relation
$ \displaystyle
    (\varepsilon S_{\frac{n}{2},4})(x,y) = \int_0^x S_{\frac{n}{2},4}(s,y)ds
$ of Proposition \ref{proposition: 2.0}, we obtain from
(\ref{convenient formula S4: hard edge}),
\begin{multline}
    (\varepsilon S_{\frac{n}{2},4})(x,y) =\int_0^x K_n(s,y)ds-\int_0^x\Phi_2(s)ds A_{21}\int_0^y\Phi_1(s)^tds\\[1ex]
        -\int_0^x \Phi_2(s)ds G_{11}\int_0^y\Phi_2(s)^tds.
\end{multline}
Therefore, we obtain from (\ref{proposition: universality hard
edge: integrals Kn: eq1}), (\ref{corollary: hard universality:
intPhi: eq1}) and (\ref{corollary: hard universality: intPhi:
eq2}) in the same way as before,
\begin{align} \label{t6.02}
    \nonumber
    &\frac{2}{\nu_n^2}\left[K_{\frac{n}{2},4}^{(\nu_n)}(\tilde\xi^{(n)},\tilde\eta^{(n)})\right]_{21}=
        (\varepsilon
        S_{\frac{n}{2},4})(\tilde\xi^{(n)},\tilde\eta^{(n)})
        =\int_0^{\xi}\frac{1}{\nu_n^2}K_n(\tilde s^{(n)},\tilde\eta^{(n)})ds \\[2ex]
    \nonumber
        &\qquad\qquad\qquad-\int_0^{\tilde\xi^{(n)}}\Phi_2(s)ds A_{21}\int_0^{\tilde\eta^{(n)}}\Phi_1(s)^tds
        -\int_0^{\tilde\xi^{(n)}} \Phi_2(s)ds G_{11}\int_0^{\tilde\eta^{(n)}}\Phi_2(s)^tds
        \\[2ex]
        &=\int_0^\xi K_J(s,\eta)ds+\frac{1}{2}\int_0^{\sqrt \xi}\left(
        J_{\alpha+1}(s)-\frac{2\alpha}{s}J_\alpha(s)\right)ds
        \int_0^{\sqrt\eta}J_{\alpha+1}(s)ds
        +\bigO\left(\frac{\xi^{\frac{\alpha}{2}}\eta^{\frac{\alpha}{2}}}{n}\right).
\end{align}
This concludes the proof of the second part of the Theorem.
\end{proof}

\begin{proof}[Proof of Theorem \ref{theorem: universality hard edge}(iii).]

\medskip

\noindent \textsc{The $(1,1)$- and $(2,2)$-entry}: 
Using $\varepsilon\Phi_1(+\infty)=\bigO(\sqrt\frac{\beta_n}{n})=\varepsilon\Phi_2(+\infty)$
(see Proposition \ref{proposition: epsilonPhi}),
$A_{12}=\bigO(\frac{n}{\beta_n})$ (see Lemma \ref{lemma:
A21matrix}), $\widehat C_{22}^{-1}=\bigO(1)$ (see Corollary
\ref{corollary: control inverse}) and (\ref{proposition: hard universality: Phi:
eq1}), we obtain the following estimate for the last term in 
(\ref{convenient formula S1: hard edge})
\[
    \frac{1}{\nu_n^2}\Phi_1(\tilde\xi^{(n)})A_{12}\widehat C_{22}^{-1}
    \left[\bigO(n^{-\tau})\varepsilon\Phi_1(+\infty)^t+\bigO(n^{-\tau})\varepsilon\Phi_2(+\infty)^t\right]
    =\bigO(\xi^{\frac{\alpha}{2}}n^{-\tau}).
\]
By (\ref{definition: conjugate Kn}), (\ref{definition: Kn1}),
(\ref{convenient formula S1: hard edge}), 
Proposition \ref{proposition: universality hard edge:
derivates Kn},
equation (\ref{proposition: hard universality: Phi: eq1}), and
Proposition \ref{corollary: hard universality: intPhi epsilon} we
then derive in the same way as before (note that also $\widehat
G_{11}$ is skew symmetric, see Lemma \ref{lemma: simplification
Widom}(ii), and that also ${\bf e}A_{12}{\bf
e}^t=-\frac{1}{2}\frac{n}{\beta_n}$)
\begin{align}
    \nonumber
    & \frac{1}{\nu_n^2}\left[K_{n,1}^{(\nu_n)}(\tilde\xi^{(n)},\tilde\eta^{(n)})\right]_{11}
    = \frac{1}{\nu_n^2}S_{n,1}(\tilde\xi^{(n)},\tilde\eta^{(n)}) \\[2ex]
    \nonumber
    & \qquad= \frac{1}{\nu_n^2}K_n(\tilde\xi^{(n)},\tilde\eta^{(n)}) \\[1ex]
    &\nonumber\qquad \qquad-\frac{1}{\nu_n^2}\Phi_1(\tilde\xi^{(n)})A_{12}
        \left(\int_0^{\tilde\eta^{(n)}}\Phi_2(s)^tds
        -\varepsilon\Phi_2(+\infty)^t+\varepsilon\Phi_1(+\infty)^t\right)
        \\[1ex]
    \nonumber
    &\qquad\qquad
        -\frac{1}{\nu_n^2}\Phi_1(\tilde\xi^{(n)}) \widehat G_{11}\left(\int_0^{\tilde\eta^{(n)}}\Phi_1(s)^tds
        -\varepsilon\Phi_1(+\infty)^t+\varepsilon\Phi_2(+\infty)^t\right)+\bigO(\xi^{\frac{\alpha}{2}}n^{-\tau})
    \\[2ex]
    & \qquad= K_J(\xi,\eta)-\frac{1}{4}\frac{J_{\alpha+1}(\sqrt\xi)}{\sqrt\xi}
    \int_{\sqrt\eta}^\infty\left(
                J_{\alpha+1}(s)-\frac{2\alpha}{s}J_\alpha(s)\right)ds
                +\bigO(\xi^{\frac{\alpha}{2}}n^{-\tau}).
\end{align}

\medskip

\noindent \textsc{The $(1,2)$-entry}: 
Equation (\ref{convenient formula S1: hard edge}) gives
\[
    (-\frac{\partial}{\partial y}S_{n,1})(x,y) =
    -\frac{\partial}{\partial y}K_n(x,y)+\Phi_1(x)A_{12}\Phi_2(y)^t
        +\Phi_1(x) \widehat G_{11} \Phi_1(y)^t.
\]
As before we then obtain from Propositions
\ref{proposition: universality hard edge: derivates Kn} and
\ref{proposition: hard universality: Phi},
\begin{align}
    \nonumber
    &\frac{1}{\nu_n^2}\left[K_{n,1}^{(\nu_n)}(\tilde\xi^{(n)},\tilde\eta^{(n)})\right]_{12} =
        \frac{1}{\nu_n^4}\left(-\frac{\partial S_{n,1}}{\partial y}\right)(\tilde\xi^{(n)},\tilde\eta^{(n)})
    \\[2ex]
    \nonumber
    &\qquad = -\frac{\partial}{\partial \eta}\left(\frac{1}{\nu_n^2}K_n(\tilde\xi^{(n)},\tilde\eta^{(n)})\right)
    +\frac{1}{\nu_n^4}\Phi_1(\tilde\xi^{(n)})A_{12}\Phi_2(\tilde\eta^{(n)})^t
        +\frac{1}{\nu_n^4}\Phi_1(\tilde\xi^{(n)})\widehat
        G_{11}\Phi_1(\tilde\eta^{(n)})^t\\[2ex]
    &\qquad =-\frac{\partial}{\partial\eta}K_J(\xi,\eta)
        -\frac{1}{8} \frac{J_{\alpha+1}(\sqrt\xi)}{\sqrt\xi}
        \left(\frac{J_{\alpha+1}(\sqrt\eta)}{\sqrt\eta}-\frac{2\alpha}{\eta}J_\alpha(\sqrt\eta)\right)
        +\bigO\left(\frac{\xi^{\frac{\alpha}{2}}\eta^{\frac{\alpha}{2}-1}}{n}\right).
\end{align}

\medskip

\noindent \textsc{The $(2,1)$-entry}: As for the $(1,1)$-entry we first derive  
\[
    \int_{\tilde\xi^{(n)}}^{\tilde\eta^{(n)}}\Phi_1(s)ds A_{12} \widehat C_{22}^{-1}
    \left[\bigO(n^{-\tau})\varepsilon\Phi_1(+\infty)^t+\bigO(n^{-\tau})\varepsilon\Phi_2(+\infty)^t\right]
    =\bigO(n^{-\tau}),
\]
using (\ref{corollary: hard universality: intPhi: eq3}) instead of (\ref{proposition: hard universality: Phi:
eq1}).
With  $(\varepsilon S_{n,1})(x,y)=-\int_x^y S_{n,1}(s,y)ds$ (see Proposition \ref{proposition: 2.0})
we obtain from
(\ref{convenient formula S1: hard edge}), (\ref{proposition:
universality hard edge: integrals Kn: eq2}), (\ref{corollary: hard
universality: intPhi: eq3}) and Proposition \ref{corollary: hard
universality: intPhi epsilon}, in the same way as before,
\begin{align}
    \nonumber
    &
    \frac{1}{\nu_n^2}\left[K_{n,1}^{(\nu_n)}(\tilde\xi^{(n)},\tilde\eta^{(n)})\right]_{21}=
    (\varepsilon S_{n,1})(\tilde\xi^{(n)},\tilde\eta^{(n)}) -\frac{1}{2}\sgn(\xi-\eta) \\[2ex]
    &\nonumber
    \qquad=
    -\int_\xi^\eta \frac{1}{\nu_n^2} K_n(\tilde
    s^{(n)},\tilde\eta^{(n)})ds \\[1ex]
    &\nonumber\qquad\qquad\qquad
    +\int_{\tilde\xi^{(n)}}^{\tilde\eta^{(n)}} \Phi_1(s)ds
    A_{12}\left(\int_0^{\tilde\eta^{(n)}}\Phi_2(s)^tds-\varepsilon\Phi_2(+\infty)^t+\varepsilon\Phi_1(+\infty)^t\right)
    \\[1ex]
    \nonumber
    &\qquad\qquad\qquad +\int_{\tilde\xi^{(n)}}^{\tilde\eta^{(n)}}
        \Phi_1(s)ds \widehat G_{11}\left(\int_0^{\tilde\eta^{(n)}}\Phi_1(s)^tds
        -\varepsilon\Phi_1(+\infty)^t+\varepsilon\Phi_2(+\infty)^t\right)
        \\[1ex]
    &\nonumber \qquad\qquad\qquad -\frac{1}{2}\sgn(\xi-\eta)+\bigO(n^{-\tau}) \\[2ex]
    &\nonumber\qquad= - \int_\xi^\eta
    K_J(s,\eta)ds+\frac{1}{2}\int_{\sqrt\xi}^{\sqrt\eta}J_{\alpha+1}(s)ds\int_{\sqrt\eta}^\infty\left(
                J_{\alpha+1}(s)-\frac{2\alpha}{s}J_\alpha(s)\right)ds
                \\[1ex]
                &\qquad\qquad\qquad -\frac{1}{2}\sgn(\xi-\eta)+\bigO(n^{-\tau}).
\end{align}
This completes the proof of Theorem \ref{theorem: universality
hard edge}.
\end{proof}

\begin{proof}[Proof of Corollary \ref{corollary:hard}(b).]

\medskip

\noindent
{\em The case $\beta= 2$.} This result can already be found in \cite{v6},
see also \cite{KV2}. Nevertheless we follow \cite[Subsection 2.2]{DeiftGioev2}
and present a somewhat different argument which is also useful for orthogonal and
symplectic ensembles.

Using the representation of gap probabilities by Fredholm determinants,
 we have
the following expression for the distribution of the smallest eigenvalue $\lambda_1(M)$,
\begin{equation} \label{t6.1}
\mathbb P_{n,2} \left( \lambda_1(M) \le \frac{s}{\nu_n^2} \right) = 1 - \det 
\left( I - \hat{K}_{n,2}|_{L^2((0,s])} \right) \, , 
\end{equation}
where $\hat{K}_{n,2}$ denotes the integral operator with kernel 
\begin{equation*}
\hat{K}_{n,2}(\xi, \eta) = 
\frac{1}{\nu_n^2}K_n(\tilde\xi^{(n)},\tilde\eta^{(n)}) \, .
\end{equation*} 
We now prove that (\ref{t6.1}) converges to 
$1 - \det \left( I - K_{J}|_{L^2((0,s])} \right)$. As the trace class 
determinant is continuous with respect to the trace class norm it suffices
to prove that 
\begin{equation*}
\Delta_n := \hat{K}_{n,2} - K_J
\end{equation*} 
converges to zero in trace class norm when considered as an integral
operator on $L^2((0,s])$. Denoting $H_n := H_{n, 1} + H_{n, 2} + H_{n, 3}$
we obtain from (\ref{eqn:6.23}), (\ref{proof: besselderivaties: sufficiency})
that
\begin{equation*}
\Delta_n(\xi, \eta) = \xi^{\frac{\alpha}{2}} \eta^{\frac{\alpha}{2}} H_n(\xi, \eta) \quad {\textrm{and}} 
\quad \frac{\partial^{k+j}}{\partial\xi^k\partial\eta^j}
        H_n(\xi,\eta)=\bigO(1/n)
\end{equation*}
for $\xi$, $\eta$ in bounded subsets of $(0,\infty)$.
Following \cite{DeiftGioev2} we formally write $\Delta_n$ as a product of two integral operators
\begin{equation} \label{t6.5}
\Delta_n = F_1 \cdot F_2 \equiv
\left(\xi^{-\varepsilon} \frac{1}{D+I} 
\right) \cdot
\left( (D+I) \xi^{\frac{\alpha}{2}+\varepsilon} \eta^{\frac{\alpha}{2}} H_n
\right) \, ,
\end{equation}
where $\varepsilon \in \mathbb R$ and $D$ denotes differentiation. We 
may think of $\frac{1}{D+I}$ as shorthand for the integral operator
\begin{equation} \label{t6.6}
\left( \frac{1}{D+I} f \right) (\xi) := \int_0^\xi e^{\eta-\xi} f(\eta)\, d\eta \, .
\end{equation}
Indeed, integration by parts then yields  
\begin{equation*}
\frac{1}{D+I} (f' + f) = f \qquad \mbox{for all } f \in C^1(\mathbb R_+) \cap 
C^0 ([0,\infty)) \mbox{ with } f(0) = 0 \, .
\end{equation*}
Thus decomposition (\ref{t6.5}) with the interpretation of (\ref{t6.6})
is valid whenever $\frac{\alpha}{2}+\varepsilon > 0$.
$F_1$ and $F_2$ can then be written as integral operators with kernels
\begin{align*}
& F_1(\xi, \eta) = \xi^{-\varepsilon} e^{\eta-\xi} {\bf 1}_{ \{\eta < \xi\} } \, ,\\
& F_2(\xi, \eta) = \xi^{\frac{\alpha}{2} + \varepsilon -1} \eta^{\frac{\alpha}{2}}
\left(\frac{\alpha}{2} + \varepsilon + \xi + \xi \frac{\partial}{\partial \xi} \right)H_n(\xi, \eta)
=\bigO\left(\frac{\xi^{\frac{\alpha}{2} + \varepsilon -1} \eta^{\frac{\alpha}{2}}}{n}\right) \, ,
\end{align*}
uniformly for $\xi$, $\eta \in (0, s]$. Assuming in addition that $\frac{1-\alpha}{2} < \varepsilon < \frac{1}{2}$
we see that $F_1$ and $F_2$ are both Hilbert--Schmidt operators on $L^2((0, s])$, because their respective
kernels lie in $L^2((0,s]\times (0, s])$. Moreover, $\| F_2 \|_{HS} = \bigO(1/n)$ which in turn implies
$\| \Delta_n \|_1 = \bigO(1/n)$, where
$\| \cdot \|_{HS}$ denotes the Hilbert--Schmidt norm
and $\| \cdot \|_{1}$ denotes the trace norm for operators acting on $L^2((0, s])$. 
This completes the proof for unitary ensembles. 

\medskip

\noindent
{\em The case $\beta= 4$.}
A slight modification of the derivation in \cite[Section 8]{TracyWidom}, which is described in 
\cite[Subsection 2.2.3]{DeiftGioev2}, provides the following representation for the distribution
of the smallest eigenvalue $\lambda_1(M)$,
\begin{equation} \label{t6.10}
\mathbb P_{\frac{n}{2},4} \left( \lambda_1(M) \le \frac{s}{\nu_n^2} \right) = 1 - 
\sqrt{\det \left( I - \hat{K}_{\frac{n}{2},4}|_{L^2((0,s])^2} \right)} \, , 
\end{equation} 
where $\hat{K}_{\frac{n}{2},4}$ denotes the integral operator with kernel 
\begin{equation*}
\hat{K}_{\frac{n}{2},4}(\xi, \eta) = 
\frac{1}{\nu_n^2} g(\xi) K_{\frac{n}{2},4}^{(\nu_n)}(\tilde\xi^{(n)},\tilde\eta^{(n)})g(\eta)^{-1},
\qquad g(\xi) = \begin{pmatrix} \xi^\delta & 0 \\ 0 & \xi^{-\delta}
\end{pmatrix}
\end{equation*}  
For the derivation of (\ref{t6.10}) one needs to ensure that both $\xi^{-\delta} \sqrt{w(\xi)}$ and
$\xi^\delta \frac{d}{d\xi}\sqrt{w(\xi)}$ belong to $L^2((0, s])$. These conditions are satisfied if
$1-\alpha < 2 \delta < 1 + \alpha$. From considerations which will become clear below we further 
restrict the choice of $\delta$. From now on we assume that $\delta$ is a fixed number with
max$(0, \frac{1-\alpha}{2}) < \delta < \frac{1}{2}$. 
Our goal is to prove that (\ref{t6.10})
converges as $n \to \infty$ ($n$ even) to 
\begin{equation*}
1 - \sqrt{\det \left( I - g(\xi) K^{(4)}(\xi, \eta) g(\eta)^{-1}|_{L^2((0,s])^2} \right)} \, .
\end{equation*}
Using again the continuity of the trace class determinant with respect to trace class norm it 
suffices to prove that each entry of 
\begin{equation*}
\Delta_n (\xi, \eta) := g(\xi) \left(
\frac{1}{\nu_n^2} K_{\frac{n}{2},4}^{(\nu_n)}(\tilde\xi^{(n)},\tilde\eta^{(n)}) - K^{(4)}(\xi, \eta)
\right) g(\eta)^{-1}
\end{equation*}
converges to zero in trace class norm when considered as an integral operator on $L^2((0, s])$.
As in \cite{DeiftGioev2} we split $\Delta_n = \Delta_n^{(1)} + \Delta_n^{(2)}$,
 where the first term
refers to the Christoffel--Darboux part and the latter corresponds to the correction term. For example,
for the 11-entry we have 
\begin{align*}
2\left[\Delta_n^{(1)} (\xi, \eta)
\right]_{11} &= \xi^\delta \eta^{-\delta} \left[
\frac{1}{\nu_n^2}K_n(\tilde\xi^{(n)},\tilde\eta^{(n)}) - K_J(\xi, \eta)
\right] \, ,  \\[1ex]
2\left[\Delta_n^{(2)} (\xi, \eta)
\right]_{11} &= \xi^\delta \eta^{-\delta} 
\left[
-\frac{1}{\nu_n^2}\Phi_2(\tilde\xi^{(n)})A_{21}\int_0^{\tilde\eta^{(n)}}\Phi_1(s)^tds
        -\frac{1}{\nu_n^2}\Phi_2(\tilde\xi^{(n)})G_{11}\int_0^{\tilde\eta^{(n)}}\Phi_2(s)^tds
\right.\\[1ex]
& \qquad \qquad - \left.
\frac{1}{4}\left(\frac{J_{\alpha+1}(\sqrt\xi)}{\sqrt\xi}
    -\frac{2\alpha}{\xi}J_\alpha(\sqrt\xi)
    \right)\int_0^{\sqrt\eta}J_{\alpha+1}(s)ds
\right] \, .
\end{align*}
Since $0 < \delta < \frac{1}{2}$ one can prove the trace norm convergence 
$[\Delta_n^{(1)}]_{11} \to 0$ in exactly the same way as $\Delta_n \to 0$ was
proven in the case $\beta=2$. In order to treat $[\Delta_n^{(2)}]_{11}$ we first
observe that the rank of this operator is bounded by $m+1$ for all $n$. We may therefore
estimate the trace norm by the Hilbert--Schmidt norm (cf. \cite[(2.7)]{DeiftGioev2})
$\| [\Delta_n^{(2)}]_{11}\|_1 \le \sqrt{m+1} \|[\Delta_n^{(2)}]_{11} \|_{HS}$.
The above proof of part (ii) of Theorem \ref{theorem: universality hard edge} (see 
(\ref{t6.0}) and above) shows
\begin{equation*}
\left[\Delta_n^{(2)} (\xi, \eta)
\right]_{11} = \bigO \left(
\frac{\xi^{\frac{\alpha}{2}-1 + \delta} \eta^{\frac{\alpha}{2}-\delta}}{n}
\right) \, .
\end{equation*}
This implies $\|[\Delta_n^{(2)}]_{11} \|_{HS}=\bigO(1/n)$, because both exponents
$\frac{\alpha}{2}-1 + \delta$ and $\frac{\alpha}{2}-\delta$ 
are larger than $-\frac{1}{2}$ by the choice of $\delta$. This completes the
proof that $[\Delta_n]_{11}$ converges to zero in trace norm, and also proves
the corresponding result for $[\Delta_n]_{22}$, because $[\Delta_n]_{22}$ is 
the adjoint of the operator $[\Delta_n]_{11}$ acting on $L^2((0, s])$. 

Applying the same method of proof to the 12-entry we obtain 
$[\Delta_n]_{12} = [\Delta_n^{(1)}]_{12} + [\Delta_n^{(2)}]_{12}$
where the correction part satisfies 
$[\Delta_n^{(2)} (\xi, \eta)]_{12} = \bigO \left(
\frac{\xi^{\frac{\alpha}{2}-1 + \delta} \eta^{\frac{\alpha}{2}-1+\delta}}{n}
\right)
$ by (\ref{t6.01}) and 
$2 [\Delta_n^{(1)}]_{12} = F_1 \cdot F_2$ can be written as a composition of
integral operators with kernels
\begin{align*}
& \!\!\!\!\!\!\!F_1(\xi, \eta) = -\xi^{\delta-\varepsilon} e^{\eta-\xi} {\bf 1}_{ \{\eta < \xi\} } \, ,\\[1ex]
& \!\!\!\!\!\!\!F_2(\xi, \eta) = \xi^{\frac{\alpha}{2} + \varepsilon -1} \eta^{\frac{\alpha}{2}+\delta-1}
\left(\frac{\alpha}{2} + \varepsilon + \xi + \xi \frac{\partial}{\partial \xi} \right)
\left( \frac{\alpha}{2} +\eta \frac{\partial}{\partial \eta} \right)H_n(\xi, \eta)
=\bigO\left(\frac{\xi^{\frac{\alpha}{2} + \varepsilon -1} \eta^{\frac{\alpha}{2}+\delta-1}}{n}\right).
\end{align*}
Choosing $\frac{1-\alpha}{2} < \varepsilon < \frac{1}{2}$ we ensure
 that $F_1$, $F_2$, 
$[\Delta_n^{(2)}]_{12}$ are Hilbert--Schmidt with 
$\| F_2 \|_{HS} = \bigO(1/n)$ and $\|[\Delta_n^{(2)}]_{12} \|_{HS}=\bigO(1/n)$. As the 
rank of $[\Delta_n^{(2)}]_{12}$ is bounded above by $m+1$ we have proven the 
trace class convergence of $[\Delta_n]_{12}$ to $0$.

Finally we turn to the 21-entry. From (\ref{t6.02}) we learn 
$[\Delta_n^{(2)}]_{21} = \bigO \left(
\frac{\xi^{\frac{\alpha}{2}- \delta} \eta^{\frac{\alpha}{2}-\delta}}{n}
\right)
$ and 
$2 [\Delta_n^{(1)}]_{21} = F_1 \cdot F_2$ with kernels
\begin{align*}
& F_1(\xi, \eta) = \xi^{-\delta} e^{\eta-\xi} {\bf 1}_{ \{\eta < \xi\} } \, ,\\[1ex]
& F_2(\xi, \eta) = \left(
\xi^{\frac{\alpha}{2}} H_n(\xi, \eta) + \int_0^\xi t^{\frac{\alpha}{2}} H_n(t, \eta) dt \right)
\eta^{\frac{\alpha}{2}-\delta}
=\bigO\left(\frac{\eta^{\frac{\alpha}{2}-\delta}}{n}\right) \, .
\end{align*}
The choice of $\delta$ ensures that $F_1$, $F_2$, 
$[\Delta_n^{(2)}]_{21}$ are Hilbert--Schmidt with 
$\| F_2 \|_{HS} = \bigO(1/n)$, $\|[\Delta_n^{(2)}]_{21} \|_{HS}=\bigO(1/n)$ and 
rank of $[\Delta_n^{(2)}]_{21} \le m+1$. This completes the proof 
for the symplectic
case.

\medskip

\noindent
{\em The case $\beta= 1$.}
We choose max$(0, \frac{1-\alpha}{2}) < \delta < \frac{1}{2}$ and 
$g(\xi) = \begin{pmatrix} \xi^\delta & 0 \\ 0 & \xi^{-\delta} \end{pmatrix}$ as above.
Following \cite[Section 9]{TracyWidom}, \cite[Subsection 2.2.3]{DeiftGioev2} we may express the
distribution of the smallest eigenvalue $\lambda_1(M)$ for even values of $n$ by
\begin{equation} \label{t6.15}
\mathbb P_{n,1} \left( \lambda_1(M) \le \frac{s}{\nu_n^2} \right) = 1 - 
\sqrt{\mbox{ det}_2 \left( I - \hat{K}_{n, 1}|_{L^2((0,s])^2} \right)} \, , 
\end{equation} 
where  
\begin{equation*}
\hat{K}_{n, 1}(\xi, \eta) = 
\frac{1}{\nu_n^2} g(\xi) K_{n, 1}^{(\nu_n)}(\tilde\xi^{(n)},\tilde\eta^{(n)})g(\eta)^{-1} \, ,
\end{equation*}  
and the regularized 2-determinant $\det_2$ is defined by 
$\det_2 (I+A) \equiv \det\left((I+A)e^{-A}\right)e^{tr(A_{11}+A_{22})}$ for $2 \times 2$ block operators
$A = (A_{ij})_{i,j=1,2 }$ with $A_{11}$, $A_{22}$ in trace class and $A_{12}$, $A_{21}$ Hilbert--Schmidt
(cf. \cite[below Corollary 1.2]{DeiftGioev2}, \cite{Simon}). Define
\begin{equation*}
\Delta_n (\xi, \eta) := g(\xi) \left(
\frac{1}{\nu_n^2} K_{n, 1}^{(\nu_n)}(\tilde\xi^{(n)},\tilde\eta^{(n)}) - K^{(1)}(\xi, \eta)
\right) g(\eta)^{-1} \, .
\end{equation*}
In order to prove the convergence of (\ref{t6.15}) to 
\begin{equation*}
1 - \sqrt{\mbox{ det}_2 \left( I - g(\xi) K^{(1)}(\xi, \eta) g(\eta)^{-1}|_{L^2((0,s])^2} \right)} \, ,
\end{equation*}
it suffices to show that the diagonal blocks $[\Delta_n]_{11}$, $[\Delta_n]_{22}$ converge to zero
in trace class  and that the off-diagonals $[\Delta_n]_{12}$, $[\Delta_n]_{21}$ converge to zero
in Hilbert--Schmidt norm. The convergence of the diagonal blocks is proven in exactly the same way as
in the case $\beta = 4$. For the off-diagonals we learn from Theorem \ref{theorem: universality hard edge}(iii)
that
\begin{equation*}
\left[\Delta_n (\xi, \eta)\right]_{12} = \bigO \left(
\frac{\xi^{\frac{\alpha}{2} + \delta} \eta^{\frac{\alpha}{2}-1+\delta}}{n^\tau}
\right), \qquad  \qquad
[\Delta_n (\xi, \eta)]_{21} = \bigO \left(
\frac{\xi^{- \delta} \eta^{-\delta}}{n^\tau}
\right) \, .
\end{equation*}
The choice of $\delta$ ensures $\|[\Delta_n]_{12} \|_{HS}=\bigO(1/n)$ and 
$\|[\Delta_n]_{21} \|_{HS}=\bigO(1/n)$, completing the proof for orthogonal
ensembles. Statement (b) of Corollary \ref{corollary:hard} is now established.  
\end{proof}

\subsection{The soft edge of the spectrum}

The proof of Theorem \ref{theorem: universality soft edge} is similar to the proofs of Theorem
\ref{theorem: universality hard edge} and \cite[Theorem
1.1]{DeiftGioev2}. Instead of the property ${\bf e} A_{21} {\bf
e}^t=-\frac{1}{2}\frac{n}{\beta_n}$, which was used to prove
universality at the hard edge, we will need at the soft edge the
following (quite remarkable) fact.

\begin{proposition}\label{proposition: algebraic formula soft edge}
    Let ${\bf a}$ be the $m$-dimensional row vector given by {\rm (\ref{definition: a and e})}.
    As $n\to\infty$,
    \begin{equation}
        {\bf a} A_{21} {\bf
        a}^t={\bf a}A_{12}{\bf a}^t=-\frac{n}{\beta_n}\left(\frac{m}{2}+\bigO(n^{-1/m})\right).
    \end{equation}
\end{proposition}

\begin{proof}
    Since $A_{12}=A_{21}^t$, see (\ref{th1.6}), we have ${\bf a} A_{21} {\bf
        a}^t={\bf a}A_{12}{\bf a}^t$. 
Further, from Lemma \ref{lemma: A21matrix} we have,
    \[
        {\bf a} A_{21} {\bf a}^t =-\frac{n}{\beta_n}\left({\bf a}Y{\bf
        a}^t+\bigO(n^{-1/m})\right),\qquad\mbox{where } Y=\begin{pmatrix}
            Q & 0 \\
            0 & \frac{1}{2}
        \end{pmatrix}.
    \]
    Here, $Q$ is the $(m-1)\times(m-1)$-matrix with entries
    $Q(i,j)=c_{i+j-1}$, where $c_\ell$ is given by (\ref{definition: cl}).
    With the notation $d_k=\sum_{j=k+1}^{m-1}c_j$ as in
    the beginning of Section \ref{subsection: proof invertibility: third part},
    we obtain from (\ref{definition: a and e})
    and Proposition \ref{proposition: c1 sum dj},
    \[
        {\bf a}Y{\bf
        a}^t=\sum_{k=0}^{m-1}d_k+\frac{1}{2}\frac{m}{2m-1}=\frac{m}{2}.
    \]
    This proves the Proposition.
\end{proof}

Furthermore, instead of Propositions
\ref{proposition: universality hard edge: derivates
Kn}-\ref{corollary: hard universality: intPhi epsilon} 
we will need the
following two Propositions.

\begin{proposition}\label{proposition: universality soft edge: derivates integrals Kn}
    {\rm (cf.~\cite[(3.8) and (3.56)]{DeiftGioev2})}
    There exists $c>0$ such that, uniformly for
    $\xi,\eta\in[L_0,\infty)$,  as $n\to\infty$
    \begin{align}
        \label{proposition: universality soft edge: derivates integrals Kn: eq1}
        & \frac{\partial^{k+j}}{\partial\xi^k\partial\eta^j}
        \left[\frac{1}{\lambda_n^2}K_n(\xi^{(n)},\eta^{(n)})\right]=
        \frac{\partial^{k+j}}{\partial\xi^k\partial\eta^j}K_\Ai(\xi,\eta)+\bigO(n^{-1/3})e^{-c\xi}e^{-c\eta},
        \\[2ex]
        \label{proposition: universality soft edge: derivates integrals Kn: eq2}
        &\int_\xi^\infty
        \frac{1}{\lambda_n^2}K_n(s^{(n)},\eta^{(n)})ds=\int_\xi^\infty
        K_\Ai(s,\eta)ds+\bigO(n^{-1/3})e^{-c\xi}e^{-c\eta},
        \\[2ex]
        \label{proposition: universality soft edge: derivates integrals Kn: eq3}
        &\int_\xi^\eta
        \frac{1}{\lambda_n^2}K_n(s^{(n)},\eta^{(n)})ds=\int_\xi^\eta
        K_\Ai(s,\eta)ds+\bigO(n^{-1/3})e^{-c\min(\xi,\eta)}e^{-c\eta}.
    \end{align}
\end{proposition}

\begin{proof}
    The proof of (\ref{proposition: universality soft edge: derivates integrals Kn: eq1})
    can be given by either following the path of the
proof of \cite[(3.8)]{DeiftGioev2}
    or by adjusting the arguments of the proof of Proposition 
    \ref{proposition: universality hard edge: derivates Kn} making efficient use of the
    formulae presented in \cite{v6}.
    Estimates (\ref{proposition: universality soft edge: derivates integrals Kn: eq2}) and
    (\ref{proposition: universality soft edge: derivates integrals Kn: eq3}) are immediate from
    (\ref{proposition: universality soft edge: derivates integrals Kn: eq1}) with $k=j=0$.
\end{proof}


\begin{proposition}\label{proposition: soft universality}
    {\rm (cf.~\cite[Proposition 4.1]{DeiftGioev2})}
    Let $j=1,2$. There exists $\tau>0$ and $c>0$ such that,
 uniformly for
    $\xi\in[L_0,\infty)$, as $n\to\infty$,
    \begin{align}
        \label{proposition: soft universality: eq1}
        & \frac{1}{\lambda_n^2}\Phi_j(\xi^{(n)})=\frac{1}{\sqrt
        m}\sqrt\frac{\beta_n}{n}\left[\Ai(\xi)\cdot{\bf
        a}+\bigO\left(\frac{e^{-c\xi}}{n^\tau}\right)\right], \\[2ex]
        \label{proposition: soft universality: eq2a}
        & \int_{\xi^{(n)}}^{\eta^{(n)}}\Phi_j(s)ds=\frac{1}{\sqrt
        m}\sqrt\frac{\beta_n}{n}\left[\int_\xi^\eta \Ai(s)ds\cdot{\bf
        a}+\bigO\left(\frac{e^{-c\min (\xi, \eta)}}{n^\tau}\right)\right], \\[2ex]        
        \label{proposition: soft universality: eq2}
        & \int_{\xi^{(n)}}^\infty\Phi_j(s)ds=\frac{1}{\sqrt
        m}\sqrt\frac{\beta_n}{n}\left[\int_\xi^\infty \Ai(s)ds\cdot{\bf
        a}+\bigO\left(\frac{e^{-c\xi}}{n^\tau}\right)\right], \\[2ex]
        \label{proposition: soft universality: eq3}
        &
        \int_{\xi^{(n)}}^\infty\Phi_j(s)ds-\varepsilon\Phi_1(+\infty)-\varepsilon\Phi_2(+\infty)=
        -\frac{1}{\sqrt m}\sqrt\frac{\beta_n}{n}\left[\int_{-\infty}^\xi\Ai(s)ds\cdot{\bf a}+
        \bigO(n^{-\tau})\right].
    \end{align}
\end{proposition}

\begin{proof}
    Using Lemmas \ref{lemma: asymptotics phin}, 
\ref{lemma: asymptotics psi: airy} and
    \ref{lemma: asymptotics psi: exponential},    
    the proof of (\ref{proposition: soft universality: eq1}) is
    similar to the proof of \cite[(4.4)]{DeiftGioev2}.
    Estimate
    (\ref{proposition: soft universality: eq2}) is immediate from (\ref{proposition: soft universality:
    eq1}), and estimate (\ref{proposition: soft universality:
    eq3}) follows from (\ref{proposition: soft universality: eq2}),
    Proposition \ref{proposition: epsilonPhi} and the fact that
    $\int_{-\infty}^\infty\Ai(s)ds=1$.
\end{proof}

We have now the necessary ingredients to prove our Theorem for the soft edge.

\begin{proof}[Proof of Theorem \ref{theorem: universality soft edge}.]
    (i) The result for the $\beta=2$ case is proven in \cite{v6} and follows also from
    (\ref{proposition: universality soft edge: derivates integrals Kn: eq1})
    with $k=j=0$.

    (ii) The proof of the second part of the theorem (the case $\beta=4$) is similar to the
    proofs of Theorem \ref{theorem: universality hard edge}(ii) and
\cite[Theorem 1.1: case $\beta=4$]{DeiftGioev2}. 
\medskip

\noindent \textsc{The $(1,1)$- and $(2,2)$-entry}: By
(\ref{convenient formula S4: soft edge}), (\ref{definition: Kn4}) and  
(\ref{definition: conjugate Kn}) we have
\begin{multline*}
    \frac{2}{\lambda_n^2}\left[K_{\frac{n}{2},4}^{(\lambda_n)}(\xi^{(n)},\eta^{(n)})\right]_{11}
    =\frac{1}{\lambda_n^2}S_{\frac{n}{2},4}(\xi^{(n)},\eta^{(n)}) \\[1ex]
    =
    \frac{1}{\lambda_n^2}K_n(\xi^{(n)},\eta^{(n)})
    +\frac{1}{\lambda_n^2}\Phi_2(\xi^{(n)})A_{21}\int_{\eta^{(n)}}^\infty\Phi_1(s)^tds
        +\frac{1}{\lambda_n^2}\Phi_2(\xi^{(n)})G_{11}\int_{\eta^{(n)}}^\infty\Phi_2(s)^tds.
\end{multline*}
The asymptotics of the first term on the right hand side of the
latter equation have been determined in part (i). From
(\ref{proposition: soft universality: eq1}), (\ref{proposition:
soft universality: eq2}), Proposition \ref{proposition: algebraic
formula soft edge} and the facts that
$A_{21}=\bigO(\frac{n}{\beta_n})$,  $|\Ai(\xi)|\leq C e^{-\xi}$
and $|\int_\eta^\infty Ai(s)ds|\leq C e^{-\eta}$ for
$\xi,\eta\in[L_0,\infty)$ and $C>0$ some constant, we have
\begin{align*}
    & \frac{1}{\lambda_n^2}\Phi_2(\xi^{(n)})A_{21}\int_{\eta^{(n)}}^\infty\Phi_1(s)^tds\\[1ex]
    &\qquad\qquad = \left[\Ai(\xi)\cdot{\bf a}+\bigO\left(\frac{e^{-c\xi}}{n^\tau}\right)\right]
    \frac{1}{m}\frac{\beta_n}{n}A_{21}
    \left[\int_\eta^\infty\Ai(s)ds\cdot{\bf a}^t+\bigO\left(\frac{e^{-c\eta}}{n^\tau}\right)\right]
    \\[1ex]
    &\qquad\qquad
    =-\frac{1}{2}\Ai(\xi)\int_\eta^\infty\Ai(s)ds+\bigO\left(n^{-\tau}\right)e^{-c\xi}e^{-c\eta}.
\end{align*}
Since $G_{11}$ is skew symmetric, see Lemma \ref{lemma:
simplification Widom}, we have ${\bf a} G_{11} {\bf a}^t=0$. Using
in addition (\ref{proposition: soft universality: eq1}),
(\ref{proposition: soft universality: eq2}) and the facts that
$G_{11}=\bigO(\frac{n}{\beta_n})$ (see Corollary \ref{corollary:
asymptotics Gmatrices}), $|\Ai(\xi)|\leq C e^{-\xi}$ and
$|\int_\eta^\infty Ai(s)ds|\leq C e^{-\eta}$ for
$\xi,\eta\in[L_0,\infty)$, we have,
\[
    \frac{1}{\lambda_n^2}\Phi_2(\xi^{(n)})G_{11}\int_{\eta^{(n)}}^\infty\Phi_2(s)^tds=
        \bigO\left(n^{-\tau}\right)e^{-c\xi}e^{-c\eta}.
\]
We conclude that,
\begin{equation}
    \frac{2}{\lambda_n^2}\left[K_{\frac{n}{2},4}^{(\lambda_n)}(\xi^{(n)},\eta^{(n)})\right]_{11}
    = K_\Ai(\xi,\eta)-\frac{1}{2}\Ai(\xi)\int_\eta^\infty\Ai(s)ds+\bigO\left(n^{-\tau}\right)e^{-c\xi}e^{-c\eta}.
\end{equation}

\medskip

\noindent \textsc{The $(1,2)$-entry}: We conclude from 
(\ref{convenient formula S4: soft edge}) that
\[
    (-\frac{\partial}{\partial y}S_{\frac{n}{2},4})(x,y)=
        -\frac{\partial}{\partial y}K_n(x,y)+\Phi_2(x)A_{21}\Phi_1(y)^t
        +\Phi_2(x)G_{11}\Phi_2(y)^t.
\]
Using (\ref{definition: Kn4}),  
(\ref{definition: conjugate Kn}), (\ref{proposition:
universality soft edge: derivates integrals Kn: eq1}),
(\ref{proposition: soft universality: eq1}) and Proposition
\ref{proposition: algebraic formula soft edge}, we obtain
\begin{align}
    \nonumber
    &\frac{2}{\lambda_n^2}\left[K_{\frac{n}{2},4}^{(\lambda_n)}(\xi^{(n)},\eta^{(n)})\right]_{12} =
        \frac{1}{\lambda_n^4}(-\frac{\partial}{\partial y}S_{\frac{n}{2},4})(\xi^{(n)},\eta^{(n)})
    \\[2ex]
    \nonumber
    &\qquad= -\frac{\partial}{\partial \eta}\left(\frac{1}{\lambda_n^2}K_n(\xi^{(n)},\eta^{(n)})\right)
    +\frac{1}{\lambda_n^4}\Phi_2(\xi^{(n)})A_{21}\Phi_1(\eta^{(n)})^t
        +\frac{1}{\lambda_n^4}\Phi_2(\xi^{(n)})G_{11}\Phi_2(\eta^{(n)})^t
    \\[2ex]
    &\qquad=-\frac{\partial}{\partial\eta}K_\Ai(\xi,\eta)
        -\frac{1}{2}\Ai(\xi)\Ai(\eta)+\bigO\left(\frac{e^{-c\xi}e^{-c\eta}}{n^\tau}\right).
\end{align}

\medskip

\noindent \textsc{The $(2,1)$-entry}: We employ
$
    (\varepsilon S_{\frac{n}{2},4})(x,y) = -\int_x^\infty S_{\frac{n}{2},4}(s,y)ds
$ of Proposition \ref{proposition: 2.0} and derive
from (\ref{convenient formula S4: soft edge}) that
\begin{multline}
    (\varepsilon S_{\frac{n}{2},4})(x,y) = -\int_x^\infty K_n(s,y)ds
        -\int_x^\infty\Phi_2(s)ds A_{21}\int_y^\infty\Phi_1(s)^tds\\[1ex]
        -\int_x^\infty \Phi_2(s)ds G_{11}\int_y^\infty\Phi_2(s)^tds.
\end{multline}
As above, we obtain from (\ref{definition: Kn4}),   
(\ref{definition: conjugate Kn}), (\ref{proposition:
universality soft edge: derivates integrals Kn: eq2}),
(\ref{proposition: soft universality: eq2}) and Proposition
\ref{proposition: algebraic formula soft edge},
\begin{align}
    \nonumber
    &\frac{2}{\lambda_n^2}\left[K_{\frac{n}{2},4}^{(\lambda_n)}(\xi^{(n)},\eta^{(n)})\right]_{21}=
        (\varepsilon
        S_{\frac{n}{2},4})(\xi^{(n)},\eta^{(n)})
        =-\int_\xi^\infty\frac{1}{\lambda_n^2}K_n(s^{(n)},\eta^{(n)})ds \\[2ex]
    \nonumber
        &\qquad\qquad\qquad-\int_{\xi^{(n)}}^\infty\Phi_2(s)ds A_{21}\int_{\eta^{(n)}}^\infty\Phi_1(s)^tds
        -\int_{\tilde\xi^{(n)}}^\infty \Phi_2(s)ds G_{11}\int_{\tilde\eta^{(n)}}^\infty \Phi_2(s)^tds
        \\[2ex]
        &=-\int_\xi^\infty K_\Ai(s,\eta)ds+\frac{1}{2}\int_\xi^\infty\Ai(s)ds
        \int_\eta^\infty\Ai(s)ds
        +\bigO\left(n^{-\tau}\right)e^{-c\xi}e^{-c\eta}.
\end{align}

(iii) The proof of the third part of the theorem is similar to the
proofs of Theorem \ref{theorem: universality hard edge}(iii) and
\cite[Theorem 1.1: case $\beta=1$]{DeiftGioev2}. One starts with
formula (\ref{convenient formula S1: soft edge}). Using (\ref{definition: Kn1}),
Proposition \ref{proposition: 2.0} together with Propositions
\ref{proposition: universality soft edge: derivates integrals Kn},
\ref{proposition: soft universality}, and
\ref{proposition: algebraic formula soft edge},
the same arguments as described in the proof of 
\ref{theorem: universality hard edge}(iii), prove the result.
However, one needs to use some identities
for Airy functions (\cite[(2.3)]{DeiftGioev2} and $\int_{-\infty}^\infty \Ai(s) \, ds\, = 1$) in order to convince oneself that 
\begin{align*}
&-\int_{\xi}^\eta K_\Ai(s,\eta)ds - \frac{1}{2}  \int_{\xi}^\eta \Ai(s)\,ds\,\int_{-\infty}^\eta \Ai(s)\,ds\, \\=&
 -\int_\xi^\infty
        K_\Ai(s,\eta)ds-\frac{1}{2}\int_\xi^\eta\Ai(s)ds+\frac{1}{2}\int_\xi^\infty\Ai(s)ds\int_\eta^\infty\Ai(s)ds
\end{align*}
which is needed to verify that the limit of the $(2,1)$-entry agrees with the one stated in the theorem.  

\end{proof}

\subsection{Universality in the bulk of the spectrum}

The proof of this theorem is similar to the proof of \cite[Theorem
1.1]{DeiftGioev}. We need the following two Propositions.

\begin{proposition}\label{proposition: universality bulk: asymptotics}
    Let $j=1,2$. As $n\to\infty$, uniformly for $\xi,\eta$ in compact subsets of
    $\mathbb{R}$ and $x$ in compact subsets of $(0,1)$,
    \begin{align}
        \label{proposition: universality bulk: asymptotics: eq1}
        & \frac{1}{q_n^2}\Phi_j\left(\beta_n
        x+\frac{\xi}{q_n^2}\right)=\bigO\left(\frac{\sqrt{\beta_n}}{n}\right),
        \\[2ex]
        \label{proposition: universality bulk: asymptotics: eq2}
        &
        \varepsilon\Phi_j\left(\beta_nx+\frac{\xi}{q_n^2}\right)=\bigO\left(\sqrt\frac{\beta_n}{n}\right),
        \\[2ex]
        \label{proposition: universality bulk: asymptotics: eq3}
        &\int_{\beta_n x+\frac{\xi}{q_n^2}}^{\beta_n
        x+\frac{\eta}{q_n^2}}\Phi_j(s)ds=\bigO\left(\frac{\sqrt{\beta_n}}{n}\right).
    \end{align}
\end{proposition}

\begin{proof}
    Let $k\in\mathbb Z$. By (\ref{definition: phinhat psirhat}), (\ref{notation:
    bulk}),
    Proposition \ref{proposition1: double integrals out bulk} and
    Lemma
    \ref{lemma: asymptotics phin}(ii) we have,
 uniformly for $\xi$ in compact subsets of
    $\mathbb{R}$ and $x$ in compact subsets of $(0,1)$, as $n\to\infty$
    \[
        \frac{1}{q_n^2}\phi_{n+k}\left[\beta_n
        x+\frac{\xi}{q_n^2}\right]=\frac{\beta_n}{n\omega_n(x)}\frac{1}{\sqrt{\beta_{n+k}}}\hat\phi_{n+k}
        \left[\frac{\beta_n}{\beta_{n+k}}\left(x+\frac{\xi}{n\omega_n(x)}\right)\right]=
        \bigO\left(\frac{\sqrt{\beta_n}}{n}\right).
    \]
    Further, with $j=1,2$, we have by (\ref{definition: phinhat
    psirhat}), (\ref{notation:
    bulk}) and Lemma \ref{lemma: asymptotics psi: bulk},
    \[
        \frac{1}{q_n^2}\psi_j\left(\beta_nx+\frac{\xi}{q_n^2}\right)=\frac{\sqrt{\beta_n}}{n\omega_n(x)}
        \hat\psi_j\left(x+\frac{\xi}{n\omega_n(x)}\right)=\bigO\left(\frac{\sqrt{\beta_n}}{n}\right).
    \]
    We now have proven (\ref{proposition: universality bulk: asymptotics:
    eq1}). Similarly, (\ref{proposition: universality bulk: asymptotics:
    eq2}) follows from (\ref{lemma: single integrals phin: eq2}) and (\ref{lemma: single integrals psin:
    eq2}). Finally (\ref{proposition: universality bulk: asymptotics:
    eq3}) is immediate from (\ref{proposition: universality bulk: asymptotics:
    eq1}).
\end{proof}

\begin{proposition}\label{proposition: universality bulk: derivatives integrals Kn}
    Uniformly for $\xi,\eta$ in compact subsets of
    $\mathbb{R}$ and $x$ in compact subsets of $(0,1)$, as $n\to\infty$ 
    \begin{align}
        \label{proposition: universality bulk: derivatives integrals Kn: eq1}
        & \frac{\partial^{k+j}}{\partial\xi^k\partial\eta^j}
        \left[\frac{1}{q_n^2}K_n\left(\beta_n x+\frac{\xi}{q_n^2},\beta_n
        x+\frac{\eta}{q_n^2}\right)\right]=\frac{\partial^{k+j}}{\partial\xi^k\partial\eta^j}K_\infty(\xi-\eta)
        +\bigO\left(\frac{1}{n}\right),
        \\[2ex]
        \label{proposition: universality bulk: derivatives integrals Kn: eq2}
        & -\int_{\beta_n x+\frac{\xi}{q_n^2}}^{\beta_n
        x+\frac{\eta}{q_n^2}} K_n\left(s,\beta_n
        x+\frac{\eta}{q_n^2}\right)ds=\int_0^{\xi-\eta}K_\infty(s)ds+\bigO\left(\frac{1}{n}\right).
    \end{align}
\end{proposition}

\begin{proof}
     It is straightforward to modify the proof of Proposition
    \ref{proposition: universality hard edge: derivates Kn} to derive the desired result. 
\end{proof}

\begin{proof}[Proof of Theorem \ref{theorem: universality bulk}.]
    (i) The case $\beta=2$ has been proven in \cite[Theorem 2.8(a)]{v6}.

    (ii) We only consider the case $\beta=1$. The case $\beta=4$ is proved in a completely analogous fashion.

    \noindent\textsc{The $(1,1)$- and $(2,2)$-entry:} Since, by (\ref{definition: conjugate Kn})
    and (\ref{definition: Kn1}),
    \[
        \left[K_{n,1}^{(q_{n,1})}(x,y)\right]_{11}=S_{n,1}(x,y)
    \]
    we obtain from (\ref{proposition: kernels: eq2}), (\ref{theorem: universality bulk: case beta=2}),
    (\ref{proposition: universality bulk: asymptotics: eq1}),
    (\ref{proposition: universality bulk: asymptotics: eq2}) and the fact that
    $A_{12}=\bigO\left(\frac{n}{\beta_n}\right)=\widehat G_{11}$ (see Lemma \ref{lemma: A21matrix} and
    Corollary \ref{corollary: asymptotics Gmatrices}) and $q_{n,1}=q_n$,
    \begin{align}
        \nonumber
        & \frac{1}{q_{n,1}^2}\left[K_{n,1}^{(q_{n,1})}\left(\beta_n
        x+\frac{\xi}{q_{n,1}^2},\beta_n x+\frac{\eta}{q_{n,1}^2}\right)\right]_{11} \\[2ex]
        \nonumber
        &\qquad\qquad\qquad = \frac{1}{q_n^2}K_n\left(\beta_n x+\frac{\xi}{q_n^2},
        \beta_n x+\frac{\eta}{q_n^2}\right)+\bigO\left(\frac{\sqrt{\beta_n}}{n}\right)
        \bigO\left(\frac{n}{\beta_n}\right)\bigO\left(\sqrt\frac{\beta_n}{n}\right)
        \\[2ex]
        &\qquad\qquad\qquad = K_\infty(\xi-\eta)+\bigO\left(n^{-1/2}\right).
    \end{align}

    \medskip

    \noindent\textsc{The $(1,2)$-entry:} Since, by (\ref{definition: conjugate Kn})
    and (\ref{definition: Kn1}),
    $
        \left[K_{n,1}^{(q_{n,1})}(x,y)\right]_{12}
        =-\frac{1}{q_{n,1}^2}\frac{\partial}{\partial y}S_{n,1}(x,y),
    $
    we obtain from (\ref{proposition: kernels: eq2}), (\ref{proposition: universality bulk: derivatives integrals Kn:
    eq1}), (\ref{proposition: universality bulk: asymptotics: eq1}) and the facts that
    $A_{12}=\bigO\left(\frac{n}{\beta_n}\right)=\widehat G_{11}$ and $q_{n,1}=q_n$,
    \begin{align}
        \nonumber
        & \frac{1}{q_{n,1}^2}\left[K_{n,1}^{(q_{n,1})}\left(\beta_n
        x+\frac{\xi}{q_{n,1}^2},\beta_n x+\frac{\eta}{q_{n,1}^2}\right)\right]_{12} \\[2ex]
        \nonumber
        &\qquad\qquad\qquad
        = -\frac{\partial}{\partial\eta}\left[\frac{1}{q_n^2}K_n\left(\beta_n x+\frac{\xi}{q_n^2},
        \beta_n
        x+\frac{\eta}{q_n^2}\right)\right]+\bigO\left(\frac{\sqrt{\beta_n}}{n}\right)
        \bigO\left(\frac{n}{\beta_n}\right)\bigO\left(\frac{\sqrt{\beta_n}}{n}\right)
        \\[2ex]
        &\qquad\qquad\qquad = -\frac{\partial}{\partial\eta}K_\infty(\xi-\eta)+\bigO\left(\frac{1}{n}\right).
    \end{align}
    Since
    $-\frac{\partial}{\partial\eta}K_\infty(\xi-\eta)=\frac{\partial}{\partial\xi}K_\infty(\xi-\eta)$,
    this proves the convergence of the $(1,2)$-entry.

    \medskip

    \noindent\textsc{The $(2,1)$-entry:} We use the formula 
        $(\varepsilon S_{n,1})(x,y)=-\int_x^y S_{n,1}(s,y)ds$ of Proposition \ref{proposition: 2.0}
    (in contrast to the edge cases, one should use the same formula also for $\beta=4$) and arrive via
    (\ref{definition: conjugate Kn})
    and (\ref{definition: Kn1}) at
    \[
        \left[K_{n,1}^{(q_{n,1})}(x,y)\right]_{21}=q_{n,1}^2 \left[(\varepsilon
        S_{n,1})(x,y) - \frac{1}{2} \sgn (x-y) \right]= -q_{n,1}^2 \left[ 
        \int_x^y S_{n,1}(s,y)ds + \frac{1}{2} \sgn (x-y) \right].
    \]
    This together with (\ref{proposition: kernels: eq2}),
    (\ref{proposition: universality bulk: derivatives integrals Kn: eq2}),
    (\ref{proposition: universality bulk: asymptotics: eq2}),
    (\ref{proposition: universality bulk: asymptotics: eq3}) and the facts that
    $A_{12}=\bigO\left(\frac{n}{\beta_n}\right)=\widehat G_{11}$ and $q_{n,1}=q_n$ yields
    \begin{align}
        \nonumber
        & \frac{1}{q_{n,1}^2}\left[K_{n,1}^{(q_{n,1})}\left(\beta_n
        x+\frac{\xi}{q_{n,1}^2},\beta_n x+\frac{\eta}{q_{n,1}^2}\right)\right]_{21} \\[2ex]
        \nonumber
        &\qquad\qquad\qquad
        = -\int_{\beta_n x+\frac{\xi}{q_n^2}}^{\beta_n
        x+\frac{\eta}{q_n^2}} K_n\left(s,\beta_n
        x+\frac{\eta}{q_n^2}\right)ds- \frac{1}{2} \sgn (\xi-\eta)
        +\bigO\left(\frac{1}{n}\right)
        \\[2ex]
        &\qquad\qquad\qquad = \int_0^{\xi-\eta}K_\infty(s)ds - \frac{1}{2} \sgn (\xi-\eta)
        +\bigO\left(\frac{1}{n}\right).
    \end{align}
    This completes the proof in the $\beta=1$ case.
\end{proof}

\end{document}